\documentclass[letterpaper,twocolumn,10pt]{article}
\usepackage{usenix,epsfig,endnotes}
\usepackage{authblk}
\usepackage{url,graphicx}
\usepackage[cmex10]{amsmath}
\usepackage{enumerate}
\usepackage[ruled,vlined]{algorithm2e}
\usepackage[font=small, labelfont=bf]{caption}
\usepackage{subcaption}
\usepackage{titlesec}
\usepackage{rotating}
\usepackage{hyperref}

\titlespacing*{\section} {0pt}{5pt}{0pt}
\titlespacing*{\subsection} {0pt}{4pt}{0pt}
\titlespacing*{\subsubsection} {0pt}{3pt}{0pt}

\newcommand{\myparatight}[1]{\smallskip\noindent{\bf {#1}:}~}

\DeclareMathOperator{\sign}{sign}

\begin{document}

\title{SybilFrame: A Defense-in-Depth Framework for Structure-Based Sybil Detection}

\author[1]{\rm Peng Gao}
\author[2]{\rm Neil Zhenqiang Gong}
\author[3]{\rm Sanjeev Kulkarni}
\author[4]{\rm Kurt Thomas}
\author[5]{\rm Prateek Mittal}

\affil[1, 3, 5]{Department of Electrical Engineering, Princeton University}
\affil[2]{Computer Science Division, University of California, Berkeley}
\affil[4]{Google}

\affil[1, 3, 5]{\textit {\{pgao,pmittal,kulkarni\}@princeton.edu}}
\affil[2]{\textit {neilz.gong@berkeley.edu}}
\affil[4]{\textit {kurtthomas@google.com}}

\maketitle

\begin{abstract}
Sybil attacks are becoming increasingly widespread, and pose a significant threat to online social systems; a single adversary can inject multiple colluding identities in the system to compromise security and privacy. Recent works have leveraged the use of social network-based trust relationships to defend against Sybil attacks. However, existing defenses are based on oversimplified assumptions, which do not hold in real world social graphs.

In this work, we propose SybilFrame, a defense-in-depth framework for mitigating the problem of Sybil attacks when the oversimplified assumptions are relaxed. Our framework is able to incorporate prior information about users and edges in the social graph. We validate our framework on synthetic and 
real world network topologies, including a large-scale Twitter dataset with 20M nodes and 265M edges, and demonstrate that our scheme performs an order of magnitude better than previous structure-based approaches. 
\end{abstract}

\section{Introduction}
\label{sec:intro}

Our systems today are vulnerable to Sybil attacks, in which an attacker injects multiple fake accounts into the system to compromise security and privacy~\cite{sybil}. 
Recently, the increasing popularity of online social networks have made them attractive targets for Sybil attacks. It is estimated that tens of millions of Sybil accounts exist in popular social networks such as Twitter and Facebook~\cite{Facebooksybil}~\cite{benevenuto2010detecting}. 
Attackers can leverage Sybil accounts to compromise system security via propagating social malware, as well as system privacy via learning users' private information~\cite{Facebooksybil}-\cite{Thomas11-imc}

An important thread of research proposes to mitigate Sybil attacks using social network-based trust relationships. The key insight of this line of defense is that it is  hard for attackers to establish trust relationships with benign users. That is, the number of edges between benign users and Sybil identities (called attack edges) is limited. Systems such as SybilGuard~\cite{Yu06}, SybilLimit~\cite{Yu08}, SybilInfer~\cite{Danezis09}, SybilRank~\cite{sybilrank}, and SybilBelief~\cite{Gong13} exploit the limited number of attack edges to detect Sybil identities using graph-theoretic techniques.

While these systems as well as related works have pioneered the use of social network structure for Sybil defense, the actual deployment of these ideas in real world networks remains controversial. Yang et al.~\cite{Yang11-sybil} showed that network structure-based Sybil defenses failed in identifying Sybil 
accounts in RenRen, a popular social network in China. This is because structure-based defense mechanisms make assumptions of \emph{strong trust} relationships between users, such that the number of attack edges is limited~\cite{Yu06}-\cite{Gong13}. These assumptions do not hold in networks with weak trust 
relationships, which enables an adversary to create a large number of attack edges. Ghosh et al.~\cite{Ghosh12} showed that on Twitter, a link farming phenomenon is wide spread and poisonous, in which certain benign accounts blindly accept follow requests. Thus, in such weak-trust social networks,
previous structure-based Sybil defenses have limited applicability and performance.

In this paper, we focus on the problem of mitigating Sybil attacks in social networks with weak trust, i.e., when the number of attack edges is large. We propose SybilFrame, an approach that provides defense-in-depth against Sybil attacks. SybilFrame uses a multi-stage classification mechanism that is able to incorporate heterogeneous sources and types of information about the social network. In the first stage, SybilFrame leverages fine grained local information about 
users and edges in the social network to design classifiers for predicting whether users or edges are benign or malicious. In the second stage, SybilFrame combines information from local classifiers with global structural properties of social networks (even ones with weak trust properties). 
Our approach leverages the results of local classification about users and edges as prior probabilities in a pairwise Markov Random Field model~\cite{Cross83}, 
and uses Loopy Belief Propagation~\cite{Murphy99} to make probabilistic inferences.

We experimentally evaluate the performance of SybilFrame using both synthetic and Facebook 
network topologies. 
We show that local node classifiers that are better than random (e.g., false positive/negative rates as high as $40\%$), can significantly improve the Sybil detection accuracy when combined with global structural information. Similarly, local edge classifiers with even a small predictive capability, provide synergistic 
information to global structural inference, and improve detection accuracy. Our approach is resilient to seed targeting attacks and a high number of attack edges which are common in social networks with weak trust.

We test SybilFrame on a large scale Twitter dataset with over $20M$ nodes and $265M$ edges. We obtain information about which accounts in this dataset were suspended by Twitter, and use this as ground truth for Sybil attacks. This dataset is typical of social networks with weak social trust, as the attacker has more than $18M$ attack edges for about $145,000$ Sybil identities. Even in this challenging setting with very large number of attack edges, SybilFrame is able to detect 51\% Sybil identities with $4.2\%$ false positives, with an overall accuracy of $95.4\%$.
In contrast, state-of-the-art approaches such as SybilBelief predict all nodes to be Sybil and thus completely fail on this dataset.
SybilFrame can also be used as a mechanism to rank user accounts. In the top 1K accounts ranked by SybilFrame (in increasing order of being benign), SybilFrame identifies $55\%$ Sybil accounts, which is 1-2 orders of magnitude better than state-of-the-art approaches. Furthermore, we manually examine the profile of the top 100 ranked users, of which 71 are suspended and 29 are active, and find that 24 active accounts are highly likely to be malicious. Thus, SybilFrame is able to uncover a large fraction (24/29) of suspicious accounts that Twitter fails to detect.

\section{Background}
\label{sec:background}

First, we give a formal definition of the Sybil defense problem in online social systems, and discuss state-of-the-art approaches.
Then, we introduce our design goals.

\subsection{Sybil Defense in Online Social Systems} 
Consider a network topology $G = (V, E)$, comprising a set $V$ of nodes with a set $E$ of edges. In social network topologies, a node $v \in V$ denotes a user on the network, and an edge $(u, v) \in E$ denotes a friendship relationship between two users $u$ and $v$. Here we only consider mutual relationships, hence $(u, v) \in E$ is equivalent to $(v, u) \in E$ and $G$ is an undirected graph. Every node $v \in V$ in the network is either a benign node, or a Sybil identity. %

\begin{figure}[!htp]
\vspace{-0.4cm}
\centering
\includegraphics[width=0.3\textwidth]{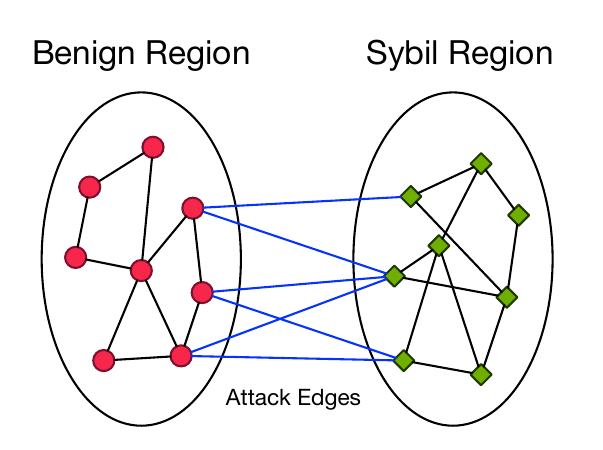}
\caption{Sybil attack problem.}
\label{fig:sybil_attack}
\vspace{-0.5cm}
\end{figure}

Figure~\ref{fig:sybil_attack} depicts the \emph{Sybil attack} problem. We denote the subnetwork containing all benign nodes to be the \emph{benign} region, and denote the subnetwork containing all Sybil nodes to be the \emph{Sybil} region. The edges that connect the benign region and the Sybil region are called \emph{attack edges}. Following the established convention in the literature, we do not impose any constraints on the size or the shape of the Sybil region. Attackers can create an unlimited number of Sybil nodes and set up edges between them arbitrarily. The main goal of Sybil defense is to design a mechanism to detect as many Sybil nodes as possible, while minimizing the number of benign nodes that are misdetected, i.e., a low false positive rate. 

\subsection{State-of-the-art Approaches}
\myparatight{Content-based approaches} Content-based approaches seek to filter Sybil accounts by 
analyzing the associated content information, such as news feeds and wall posts on Facebook and tweets and hashtags on Twitter~\cite{Thomas11-imc}. These approaches span a large category of mechanisms, including blacklisting, whitelisting, URL filtering, as well as various machine learning methods, such as Bayesian Reasoning, Support Vector Machines and Clustering~\cite{Wang10}-\cite{spam:acsac10}.
A major problem of these approaches is that attackers can mimic the behaviors of benign users and produce similar content
information, thus making content-based approaches less effective.

\myparatight{Structure-based approaches} Structure-based approaches, 
seek to exploit graph-theoretic differences between benign and Sybil identities. %
The key insight is that in a social graph where edges represent strong trust relationships between users, it is hard for attackers to set up links to benign users. %
As a result, the number of attack edges is relatively small. Such networks preserve a strong level of homophily, i.e., two linked nodes are 
likely to have similar attributes. 

SybilGuard~\cite{Yu06} and SybilLimit~\cite{Yu08}, rely on the insight that it is easy for short random walks starting from a benign user to quickly reach other benign users, while hard for random walks starting from Sybil identities to enter into the benign region. 
SybilInfer~\cite{Danezis09}, relies on random walks and a combination of Bayesian inference and Monte-Carlo sampling and aims to directly detect the bottleneck cut between benign and Sybil identities. SybilRank~\cite{sybilrank}, 
uses short random walks to distribute initial scores from a set of trusted benign seeds, and rely on the insight that benign users tend to have larger degree-normalized scores than Sybil identities. Criminal account Inference Algorithm (CIA)~\cite{Yang12-spam}, similar to SybilRank, starts random walks and distributes scores from Sybil seeded users and allows the restart from initial probability distribution with certain probability. Researchers have shown that despite considerable differences, the above schemes rely on identifying local communities around a trust node~\cite{Viswanath10}. SybilBelief~\cite{Gong13}, on the other hand, models the distribution of labels of the nodes as a pairwise Markov Random Field. Similar to SybilFrame, it adopts Loopy Belief Propagation to estimate probabilities of users being benign. 
\'{I}ntegro~\cite{integro}, is an extension to SybilRank by incorporating victim predictions using content features, thus not purely structure-based.

We note that all of the above-mentioned structure-based methods are based on two key assumptions. First, the benign region is fast mixing~\cite{mohaisen:imc10}, which presumes the existence of a well-connected, giant community structure of benign users. Second, the social network is a strong trust network, where the number of attack edges is relatively small~\cite{Viswanath10}. Given the two assumptions, these structure-based approaches have been shown to provide reliable performance.

\subsection{Assumptions vs. Reality}
We claim that the above mentioned assumptions oversimplify social network structure, and do not hold well on all real-world social graphs. 

First, benign users tend to form multiple small communities~\cite{Viswanath10} driven by different purposes (e.g., geographical location, education and career). This multi-community structure prohibits the existence of a giant community component and hence results in a longer mixing time. Mohaisen et al. \cite{mohaisen:imc10} measured the mixing time of real-world social graphs and found that the actual mixing time is longer than the theoretical anticipated value. 

Second, real-world social networks may not necessarily represent strong trust networks. Yang et al. showed that RenRen, the largest social networking platform in China, does not follow this assumption~\cite{Yang11-sybil}. Another typical example is the Twitter network. The Twitter network is a directed network, on which links are established by the action of ``follow". Unlike Facebook, users in Twitter often use a pseudonym, which makes them less serious about whom they choose to follow. 
Ghosh et al.~\cite{Ghosh12} showed that on Twitter, the notable phenomenon of link farming is wide spread, and that a majority of attack edges are farmed from a small fraction of Twitter users. Those users, the social capitalists, are benign users who are seeking to increase their social power and links by following back anyone who follows them. Even normal users, who are not as athirst for social power as social capitalists, are also likely to follow back strangers because they want to read their tweets or just by courtesy. On such weak trust social networks like Twitter, a large number of attack edges exist and the benign region may not be easily separable from the Sybil region. As a result, all of these structure-based Sybil defense mechanisms are limited in their 
performance.

\subsection{Design Goals}
We aim to design a scheme that works even when the fast-mixing and strong trust assumptions are relaxed. 
Our overall design goals are as follows:

\textbf{1) Defense-in-depth:} The scheme should provide multi-layered protection, and be robust to different attack strategies.

\textbf{2) Accuracy:} The scheme should have reliable detection accuracy when applied to a wide range of social network topologies, including both strong trust and weak trust social networks.

\textbf{3) Scalability:} The scheme should be scalable to large social networks, and be amenable to parallel deployment.

We propose SybilFrame, a defense-in-depth framework that adopts a multi-stage classification mechanism for incorporating heterogeneous sources and types of information about the social network.

\section{The SybilFrame Framework}
\label{sec:framework}

In this section, we give a detailed description of SybilFrame framework.
\subsection{Framework Overview}
\begin{figure}[!htp]
\centering
\includegraphics[width=0.49\textwidth]{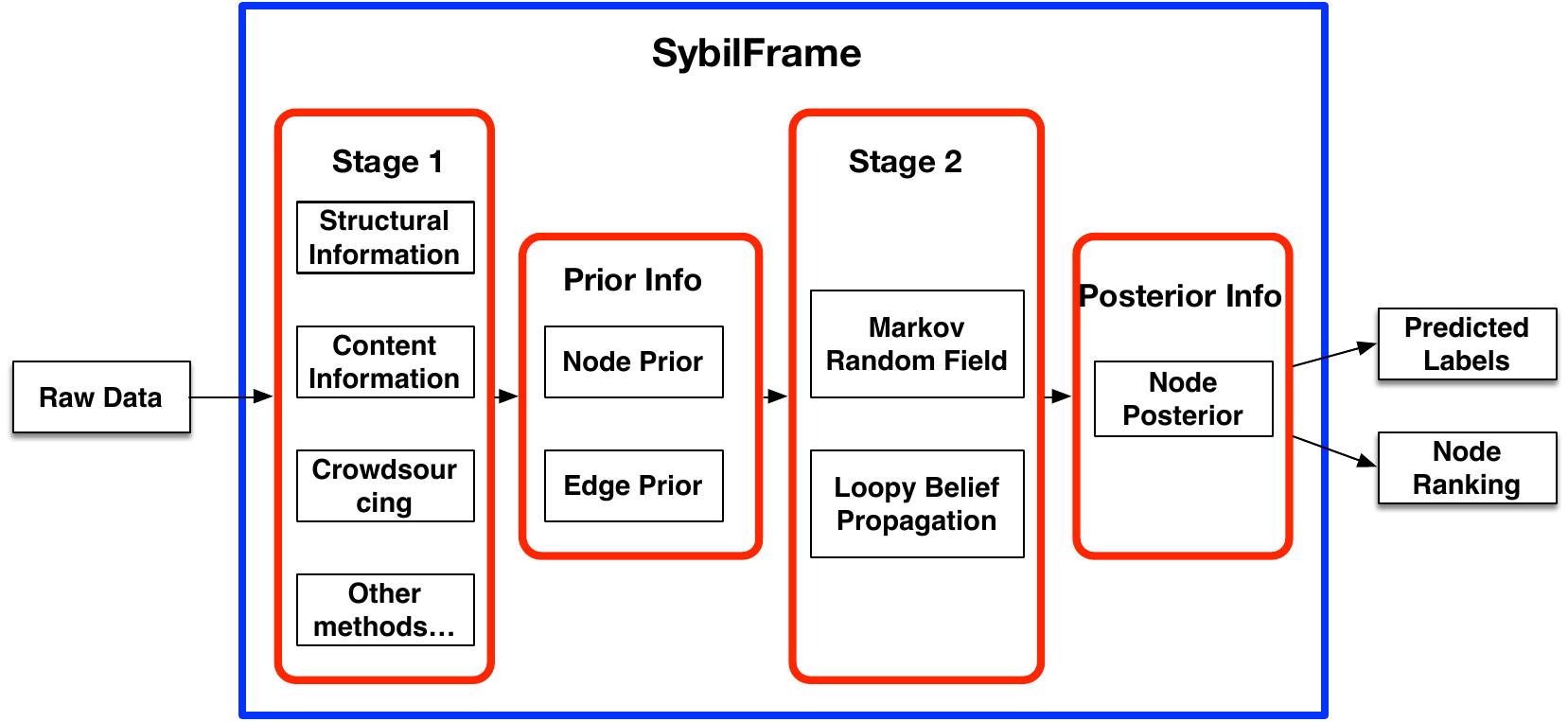}
\caption{SybilFrame framework.} 
\label{fig:sybilframe}
\vspace{-0.3cm}
\end{figure}
Figure~\ref{fig:sybilframe} shows the general framework of SybilFrame. SybilFrame is a multi-stage classification approach that leverages the attributes of an individual node and correlation between connected nodes to make a combined classification of networked data. SybilFrame has two stages of inference. Once the raw data has been fed into the framework, \emph{Stage 1} will explore the dataset and extract useful information, to compute node prior information and edge prior information (Section~\ref{subsec:prior}). This prior information, together with a small set of nodes whose labels are known, i.e., trust seeds, will be fed into \emph{Stage 2}. \emph{Stage 2} is the posterior inference layer. To represent the correlation between nodes, we model the problem as a pairwise Markov Random Field (Section~\ref{subsec:mrf}). We adopt Loopy Belief Propagation (Section~\ref{subsec:lbp}) to make inferences about the posterior information. This posterior information will then be used to classify and rank Sybil identities (Section~\ref{subsec:classify}).
\subsection{Prior Information}
\label{subsec:prior}
\emph{Stage 1} in Figure~\ref{fig:sybilframe} takes the raw dataset as input, and outputs the prior information of all nodes and edges. We now formalize our notion of priors.

For a node $v \in V$, we denote $Prior_v$ as the node prior of $v$. $Prior_v$ is a real number in the range $[0, 1]$, that quantifies the probability that node $v$ takes a benign label. The larger $Prior_v$ is, the more likely that $v$ is a benign node. Specifically, $Prior_v > 0.5$ means that $v$ is more likely to take a benign label rather than a Sybil label. Similarly, $Prior_v < 0.5$ means that $v$ is more likely to take a Sybil label, and $Prior_v = 0.5$ means that $v$ takes a benign or Sybil label with equal probability. If $v$'s label is known, then $Prior_v = 1$ for a benign trust seed, and $Prior_v = 0$ for a Sybil trust seed.

For two nodes $u$ and $v$ that are connected by an edge, we denote $Prior_{u, v}$ as the edge prior of $(u, v) \in E$. $Prior_{u, v}$ is a real number in the range $[0, 1]$, that quantifies the likelihood that node $u$ and node $v$ take the same label. The larger $Prior_{u, v}$ is, the more likely that $u$ and $v$ take the same label. Specifically, $Prior_{u, v} > 0.5$ means that $u$ and $v$ are more likely to take the same label than different labels. Similarly, $Prior_{u, v} < 0.5$ means that $u$ and $v$ are more likely to take different labels, and $Prior_{u, v} = 0.5$ means that $u$'s label has no influence on $v$'s label, and vice versa. Generally, $Prior_{u, v}$ models the level of coupling strength between $u$ and $v$. $Prior_{u, v} > 0.5$ refers to a positive coupling relationship, and $Prior_{u, v} < 0.5$ refers to a negative coupling relationship, and $Prior_{u, v} = 0.5$ means that there is no coupling between $u$ and $v$.

The derivation of node priors and edge priors is based on the dataset we are given. We can leverage heterogeneous information sources to make inferences. To compute node priors, we can leverage the structural information and explore differences of local structure between benign and Sybil nodes. We can extract useful features and build a machine learning classifier that supports probability estimates, and use these probability outputs as node priors. 
To infer edge priors, we want to assign lower scores to attack edges, and assign higher scores to edges between benign accounts. We do not care about edges between Sybil accounts, since attacker has a complete control over the Sybil region and can change it arbitrarily. This makes our approach robust to high number of attack edges and distinguishes SybilFrame from previous approaches. Since benign nodes tend to behave similarly and Sybil nodes tend to behave differently from benign nodes, a straightforward way is to explore similarities of two connected nodes under different metrics and obtain a scaled overall similarity score. This overall score can then be used as an edge prior.

We note that although we propose a structure-based scheme, as will be demonstrated and evaluated later, our framework can definitely incorporate content information. For example, we can analyze news feeds of each Facebook account and tweets of each Twitter account, and identify spam keywords and abnormal actions. We can then build a content-based classifier and compute node priors. The philosophy also works for content-based edge priors. The reason why we tend to use structural information is that it is harder for an attacker to alter the overall graph structure than mimic the content behaviors of benign users.
In Section~\ref{sec:evaluation_facebook} and Section~\ref{sec:twitter}, we will explore ways to compute node priors and edge priors on real-world, large-scale social graphs.

\subsection{Markov Random Field}
\label{subsec:mrf}
A Markov Random Field (MRF)~\cite{Cross83}, is a probabilistic graphical model over an undirected graph.
Nodes in MRF are random variables, and edges are used to model correlation between those random variables. For each node $v \in V$ on graph $G = (V, E)$, we associate it with a binary random variable $X_v$, that indicates the label of $v$. $X_v = 1$ refers to a benign label, and $X_v = -1$ refers to a Sybil label. To quantify the correlation, we use a set of functions called \emph{clique potentials}. A \emph{clique potential} is a function defined over a set of random variables, which maps any joint assignment of these random variables to a real number, which indicates how favorable this joint assignment is. Let $\Psi$ denote the set of potential functions. Specifically, if we only consider cliques comprising at most two connected nodes, $\Psi$ can be divided into the following two types of functions.
\begin{equation}
\setlength{\abovedisplayskip}{2pt}
\setlength{\belowdisplayskip}{2pt}
  \psi_v(X_v)=\begin{cases}
  	Prior_v, & \text{if }X_v = 1 \\
  	1 - Prior_v, & \text{if }X_v = -1
  \end{cases}  
\end{equation}
\begin{equation}
\setlength{\abovedisplayskip}{2pt}
\setlength{\belowdisplayskip}{2pt}
	\psi_{u, v}(X_u, X_v)=\begin{cases}
		Prior_{u,v}, & \text{if }X_uX_v = 1 \\
		1 - Prior_{u, v}, & \text{if }X_uX_v = -1
	\end{cases}  
\end{equation}

As defined in Section~\ref{subsec:prior}, $Prior_v$ is the prior information of node $v$, and $Prior_{u, v}$ is the prior information of edge $(u, v)$. We denote function $\psi_v$ as the \emph{node potential}, and function $\psi_{u, v}$ as the \emph{edge potential}. $(G, \Psi)$ then defines a \emph{pairwise Markov Random Field}.

Given a pairwise MRF $(G, \Psi)$, where $G = (V, E)$ and $\Psi = (\psi_v, \psi_{u, v})$, the full joint probability distribution is specified as
\begin{equation}
\setlength{\abovedisplayskip}{2pt}
\setlength{\belowdisplayskip}{2pt}
P(X_V) = \frac{1}{Z}\prod_{v \in V}\psi_v(X_v) \prod_{(u, v) \in E} \psi_{u, v}(X_u, X_v)
\end{equation}

Here, $X_V$ denotes a particular joint assignment of all random variables in set $V$, and $Z$ is the partition function given by
\begin{equation}
\setlength{\abovedisplayskip}{2pt}
\setlength{\belowdisplayskip}{2pt}
Z = \sum_{X_V}\prod_{v \in V}\psi_v(X_v) \prod_{(u, v) \in E} \psi_{u, v}(X_u, X_v)
\end{equation}

\subsection{Infer Posteriors}
\label{subsec:lbp}

Given the pairwise MRF $(G, \Psi)$, which contains prior information of trust seeds and other nodes and edges, for each node $v\in V$, we want to infer the posterior probability of random variable $X_v$.
\begin{equation}
\setlength{\abovedisplayskip}{2pt}
\setlength{\belowdisplayskip}{2pt}
P(X_v) = \frac{1}{Z}\sum_{X_{V\backslash v}}\prod_{s \in V}\psi_s(X_s) \prod_{(u, s) \in E} \psi_{u, s}(X_u, X_s)
\end{equation}

Exact inference is computationally difficult, and not scalable on large dataset. Therefore, we adopt \emph{Loopy Belief Propagation} to make approximate inferences. Loopy Belief Propagation~\cite{Murphy99} is an iterative process in which neighboring variables pass messages or beliefs to each other. Algorithm~\ref{alg:lbp} gives the Loopy Belief Propagation algorithm for the pairwise MRF $(G, \Psi)$.
\begin{algorithm}
\DontPrintSemicolon
\KwData{node potentials $\psi_v(X_v)$, edge potentials $\psi_{u, v}(X_u, X_v)$}
\KwResult{marginal beliefs $bel_v(X_v)$}
Initialize beliefs $bel_v(X_v) = 1$ for all nodes $v$\;
Initialize message $m_{u\rightarrow v} (X_v)= 1$ for all edges $u\rightarrow v$\;
\Repeat{
number of iterations $>$ threshold $d$}{
	Messages update $m_{u\rightarrow v} (X_v) = \sum_{X_u}\left( \psi_u(X_u)\psi_{u, v}(X_u, X_v)\prod_{s\in Nbd(u)\backslash v}m_{s \rightarrow u}(X_s) \right)$\;
	Beliefs update $bel_v(X_v) \propto \psi_v(X_v)\prod_{u\in Nbd(v)}m_{u\rightarrow v}(X_v)$\;
}
\caption{Loopy Belief Propagation Algorithm}\label{alg:lbp} 
\end{algorithm}  

We note that for social networks with loops, LBP approximates the posterior probability distribution without theoretical convergence guarantees. However, LBP has been widely used and demonstrated good results in practical applications~\cite{Murphy99}. Through our experiments, we find that setting $d$ to be 5$\sim$6  achieves good results.

\myparatight{Scalability} The complexity of LBP is $O(md)$ where $m$ is the number of edges and $d$ is the number of iterations. For sparse social networks, $O(md) = O(nd)$, where $n$ is the number of nodes. LBP is essentially parallelizable, and we will discuss related implementation issues in Section~\ref{subsec:bp_parallel}.

\subsection{Sybil Accounts Prediction and Ranking}
\label{subsec:classify}

We use posteriors obtained in Section~\ref{subsec:lbp} to predict the label of each node. For a node $v$ whose label is unknown, we predict the label $L_v$ using the following rule.
\begin{equation}
\setlength{\abovedisplayskip}{2pt}
\setlength{\belowdisplayskip}{2pt}
L_v = \sign(bel_v - 0.5)
\end{equation}
\noindent where $L_v = 1$ means that $v$ is predicted as a benign node, and $L_v = -1$ means that $v$ is predicted as a Sybil node.

We can also rank nodes in ascending order of its posterior, and produce a ranking list. Sybil nodes are likely to have lower posteriors, thus occur more in the front part. OSN operators can then go through the list from the beginning, and check a fixed number of nodes. More effective posteriors will let OSN operators detect more Sybil accounts within a certain amount of time.

\section{Security Evaluation on Synthetic Networks}
\label{sec:evaluation_synthetic}
In this section, we evaluate SybilFrame on different network structures. For comparison, we use SybilBelief~\cite{Gong13}, which
takes a similar probabilistic inference approach as SybilFrame. Since Gong et al.~\cite{Gong13} have demonstrated that SybilBelief outperforms other structure-based methods on trust networks, we limit our space here to only compare with SybilBelief. Later in Section~\ref{sec:evaluation_facebook} and Section~\ref{sec:twitter}, we will compare with other methods such as SybilLimit~\cite{Yu08}, SybilInfer~\cite{Danezis09}, and SybilRank~\cite{sybilrank}. We do not compare with \'{I}ntegro~\cite{integro} since it leverages network-specific content information for victim predictions.

\myparatight{Basic experimental setup} We adopt the Preferential Attachment (PA)~\cite{Barabasi99} model to generate both benign region and Sybil region. The size of benign region is 1000, and the size of Sybil region is 400. The average degree of both benign region and Sybil region is 10. We randomly add 1000 attack edges between the two regions. We only use 1 benign trust seed and 1 Sybil trust seed. For default node priors, we set 0.9 for benign trust seeds, and 0.1 for Sybil trust seeds, and 0.5 for others if we do not have any external priors fed in. For default edge priors, we set it to 0.9 in order to model homophily. 
We will study the impact of different factors. When we study one factor, we fix the other factors to be the same as in the basic setup, and only vary the studied one. Under each setting, we run 100 experiments. In each experiment, we randomly generate 1 benign trust seed and 1 Sybil trust seed, configure prior information, and run SybilFrame and SybilBelief. We store results of SybilFrame and SybilBelief correspondingly, and take the average over 100  experiments in the end to be our final results.

\myparatight{Evaluation metrics} Following the convention, we denote Sybil nodes as positive examples and benign nodes as negative examples. Thus, we have TP (Sybil $\rightarrow$ Sybil), TN (benign $\rightarrow$ benign), FP (benign $\rightarrow$ Sybil) and FN (Sybil $\rightarrow$ benign). We use the following four evaluation metrics:

\textbf{1) Accuracy:} $(TP+TN)/(TP+TN+FP+FN)$

\textbf{2) Number of rejected benign nodes:} $FP$

\textbf{3) Number of accepted Sybil nodes:} $FN$

\textbf{4) Area Under the Receiver Operating Characteristic Curve (AUC)~\cite{auc}:}  The probability that a randomly selected benign nodes ranks higher than a randomly selected Sybil node, given the ranking of posteriors of all nodes from the smallest to the largest.

\subsection{Influence of Node Priors}
\label{subsec:node_prior}

We want to explore SybilFrame when only incorporating external node priors. Since we are experimenting with synthetic networks, we need to figure out a way to obtain node priors that are able to model the real case. A straightforward solution is to use false positive rate (FPR) and false negative rate (FNR) to model the performance of an external node classifier. By setting up different FPR and FNR, we can generate prior scores that are able to model the level of noise, and use them to evaluate SybilFrame. Due to limited space, we list our \emph{Node Prior Generator} algorithm (Algorithm~\ref{alg:node_prior}) in Appendix~\ref{subsec:prior_generator}.
In Algorithm~\ref{alg:node_prior}, we set prior for benign/Sybil trust seeds to be 0.9/0.1, not 1/0 as discussed in Section~\ref{subsec:prior} in order to run LBP successfully.

\begin{figure}
\begin{subfigure}[H]{0.24\textwidth}
  \includegraphics[width=\linewidth]{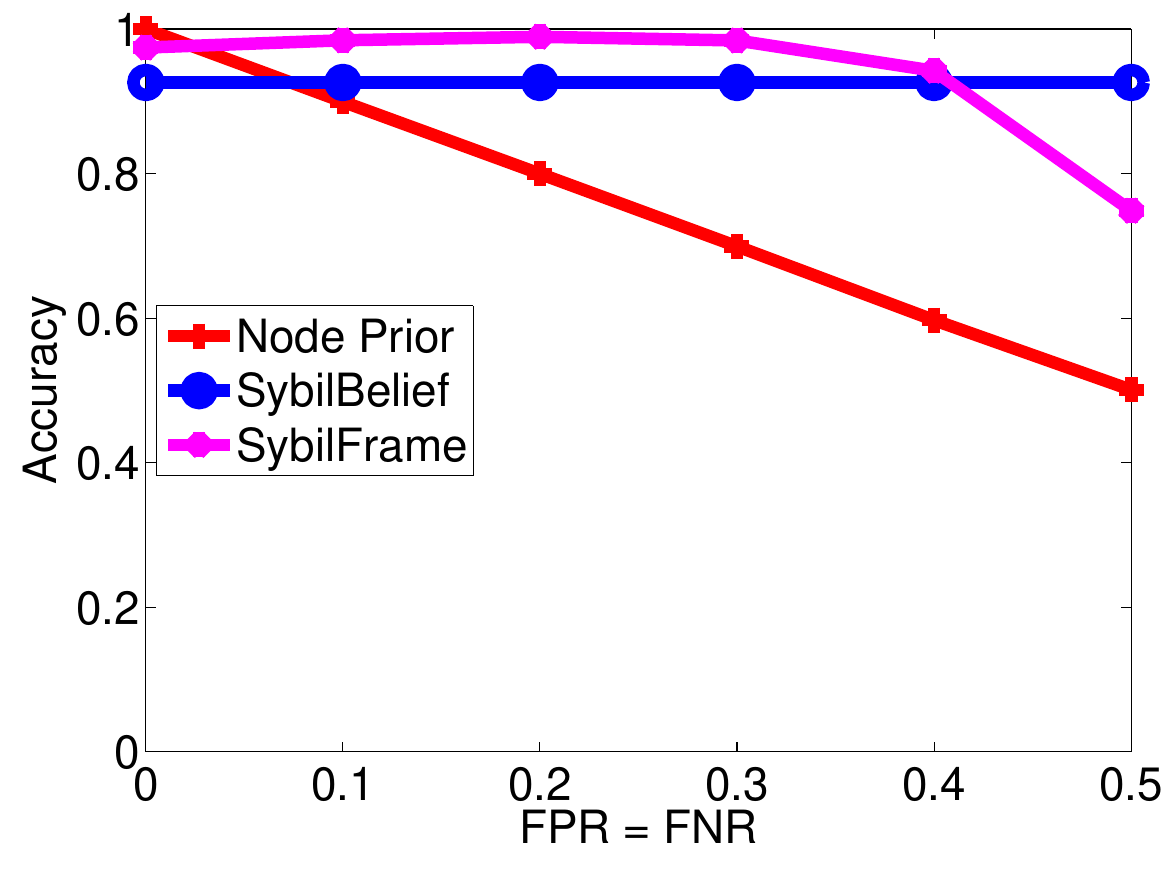}
  \caption{Accuracy}
  \label{fig:node_prior_accuracy_1}
\end{subfigure}%
\begin{subfigure}[H]{0.24\textwidth}
  \includegraphics[width=\linewidth]{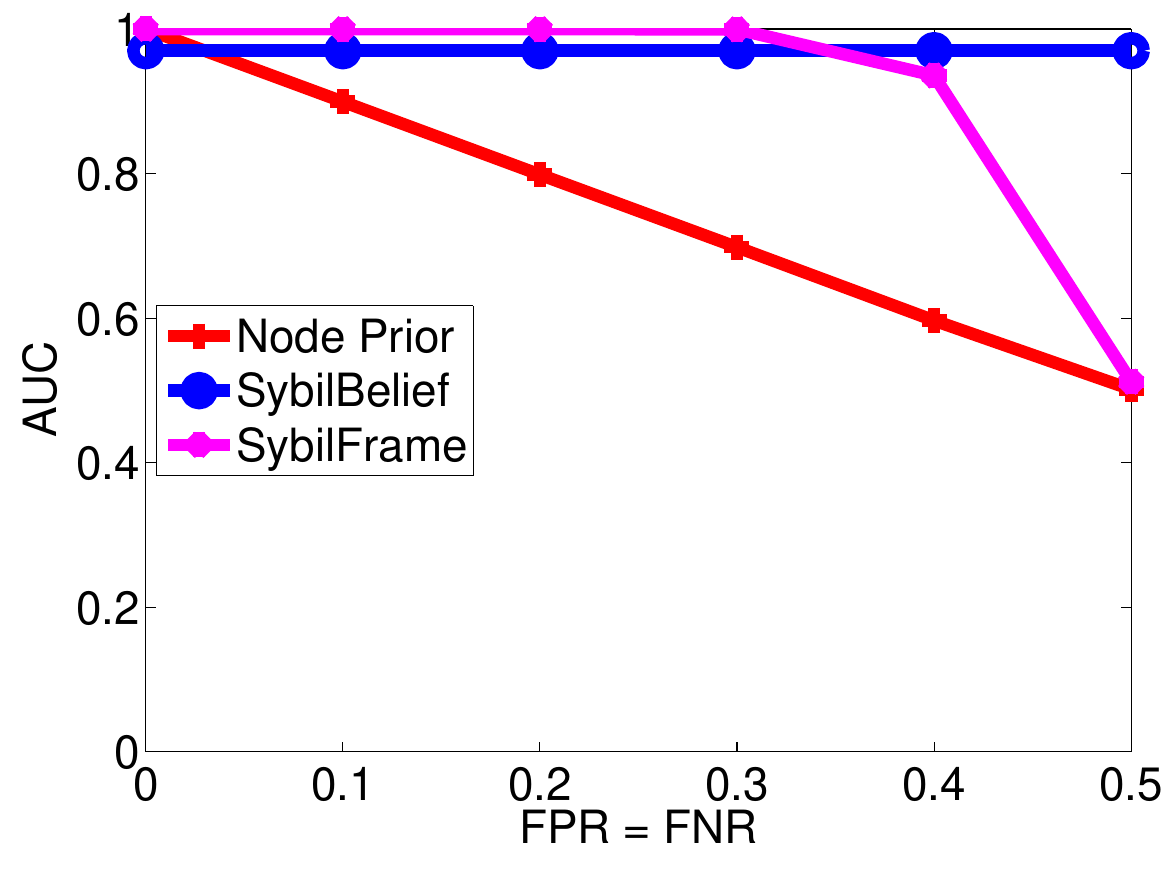}
  \caption{AUC}
  \label{fig:node_prior_auc_1}
\end{subfigure}%

\begin{subfigure}[H]{0.24\textwidth}
  \includegraphics[width=\linewidth]{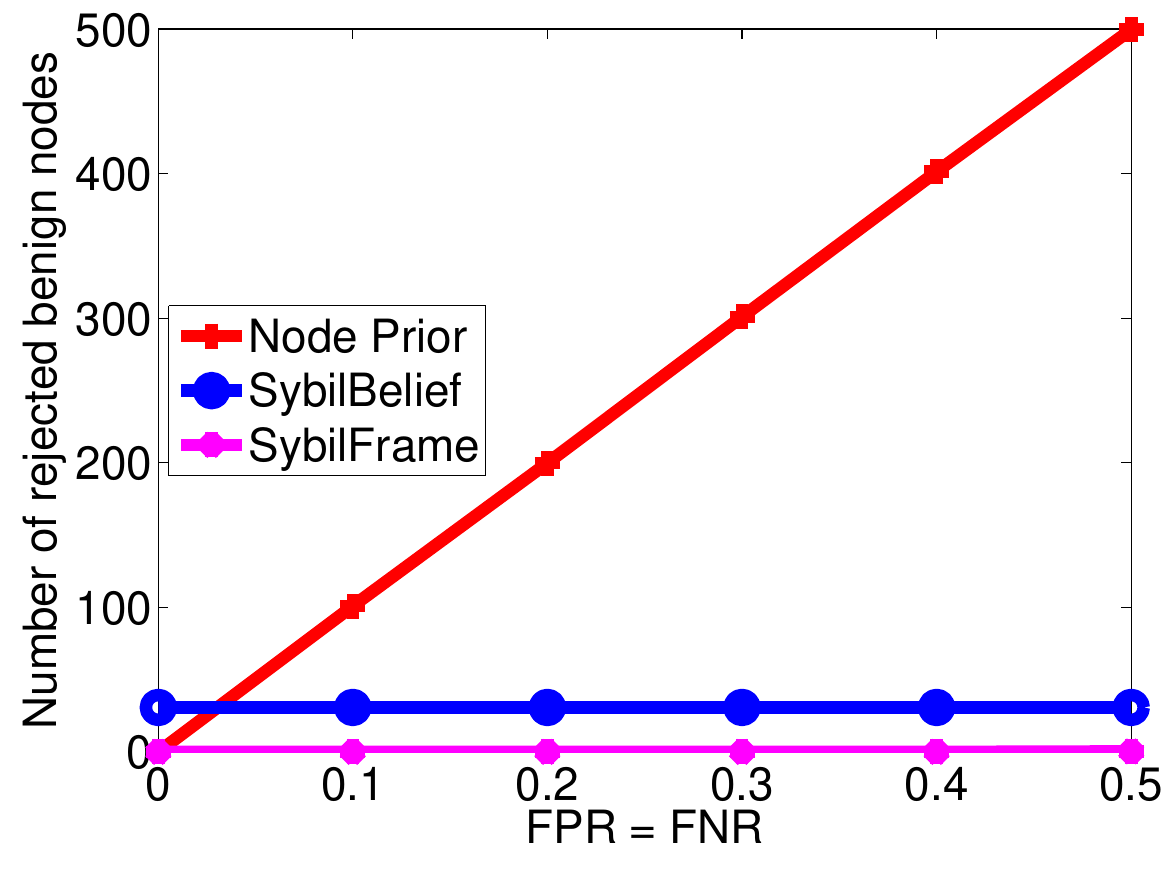}
  \caption{Rejected benign nodes}
  \label{fig:node_prior_benign_rej_1}
\end{subfigure}%
\begin{subfigure}[H]{0.24\textwidth}
  \includegraphics[width=\linewidth]{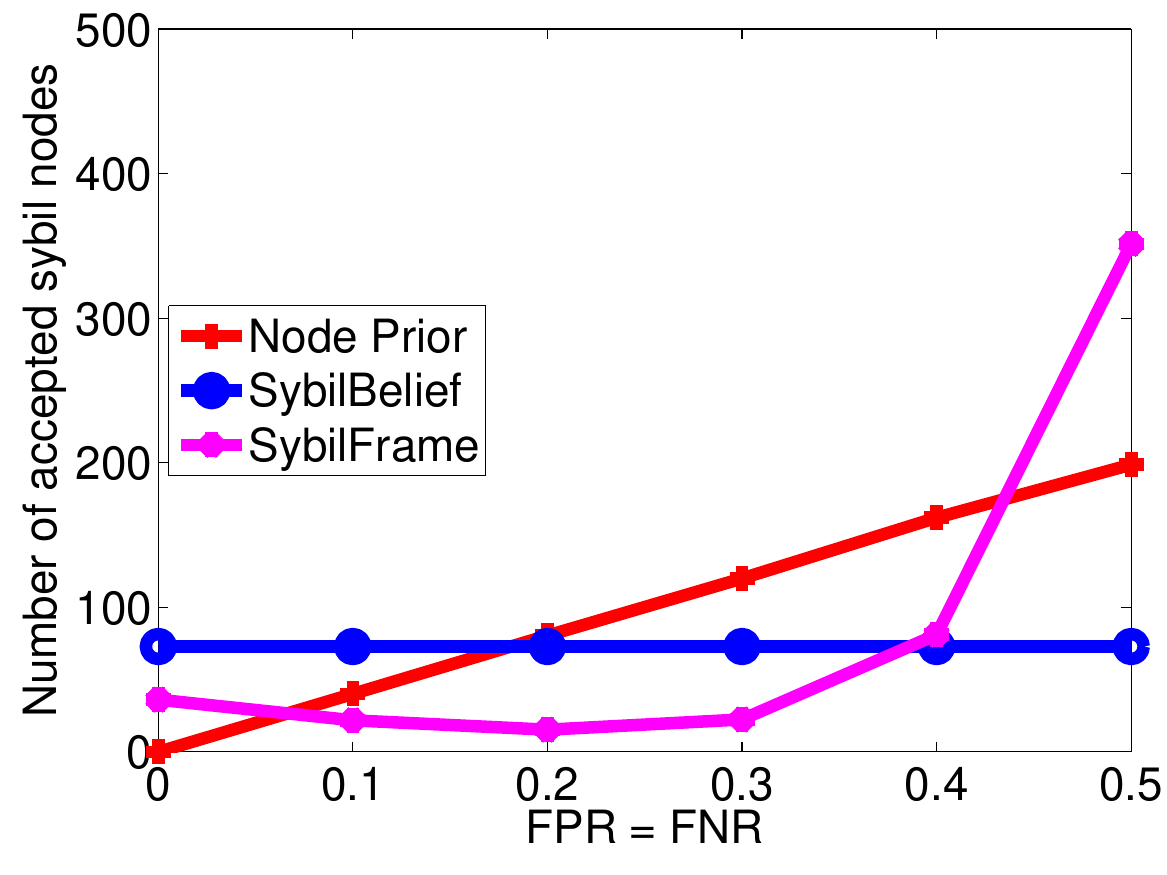}
  \caption{Accepted Sybil nodes}
  \label{fig:node_prior_sybil_acc_1}
\end{subfigure}%
\caption{
Vary FPR=FNR (node prior)}
\label{fig:node_prior_1}
\vspace{
-0.7cm}
\end{figure}

\myparatight{Varying FPR and FNR}
\label{subsubsec:fpr_fnr_f}
First, we evaluate SybilFrame given node priors with different levels of noise. We tune $FPR=FNR$ from 0 to 0.5, i.e., from perfect classification to random guess, and fix everything else as in the basic setup. In addition to comparison with SybilBelief, we also compare with the performance of external node classifier, i.e., compare with priors, and explore whether there are improvements.
Figure~\ref{fig:node_prior_1} shows the results. As $FPR=FNR$ increases, the performance of external node classifier degrades linearly. 
Besides, SybilFrame performs better than SybilBelief when $FPR=FNR\leq 0.4$, in terms of all four metrics. When $FPR=FNR\leq 0.3$, SybilFrame can achieve near optimal performance. This means that SybilFrame is resilient to prior noise with FPR and FNR as high as 40\%.

\begin{figure}
\begin{subfigure}[H]{0.24\textwidth}
  \includegraphics[width=\linewidth]{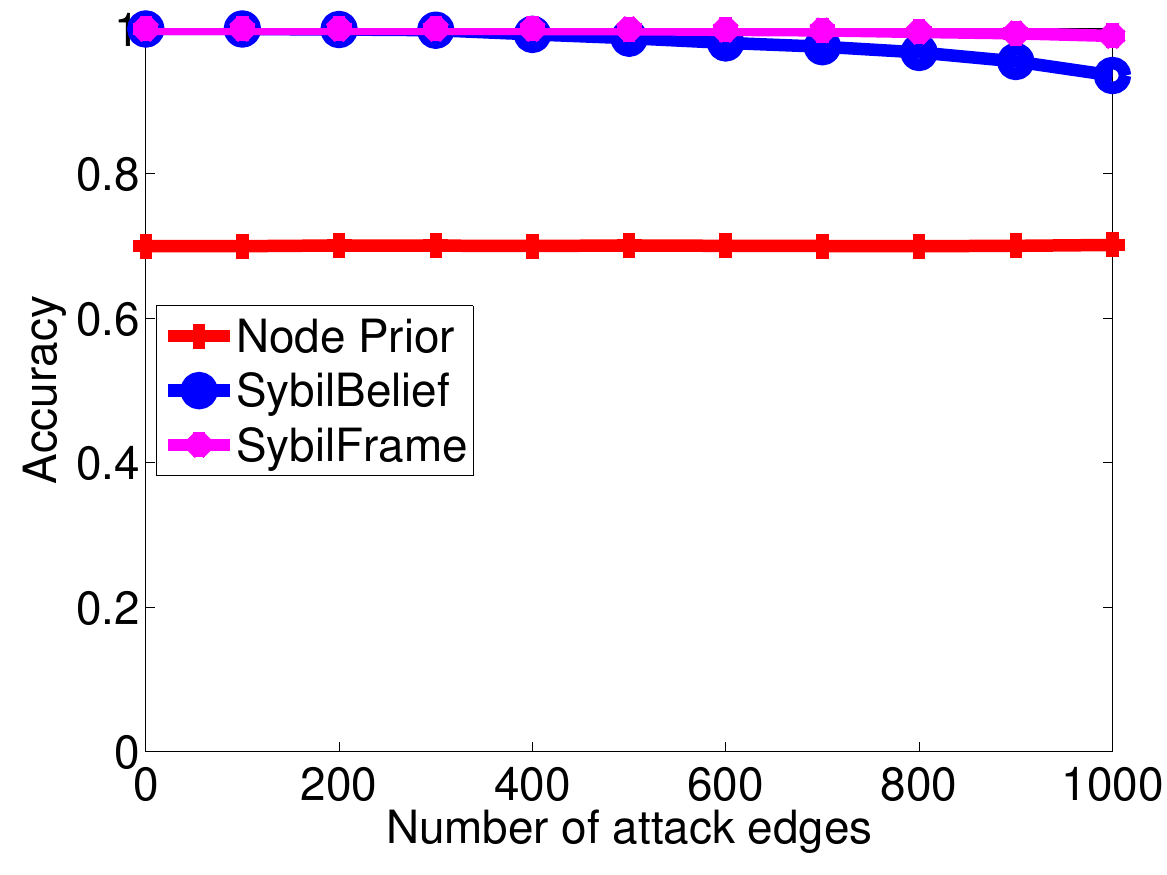}
  \caption{Accuracy}
  \label{fig:node_prior_accuracy_2}
\end{subfigure}%
\begin{subfigure}[H]{0.24\textwidth}
  \includegraphics[width=\linewidth]{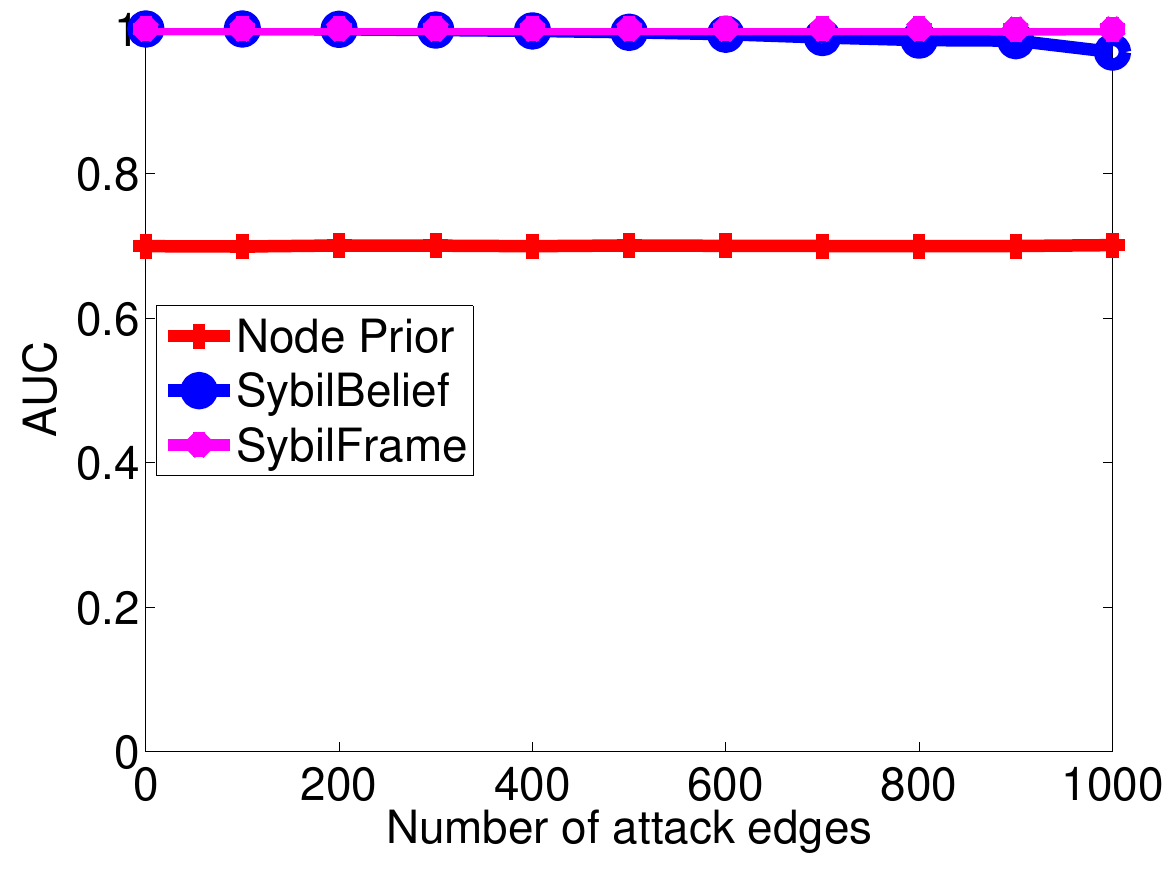}
  \caption{AUC}
  \label{fig:node_prior_auc_2}
\end{subfigure}%

\begin{subfigure}[H]{0.24\textwidth}
  \includegraphics[width=\linewidth]{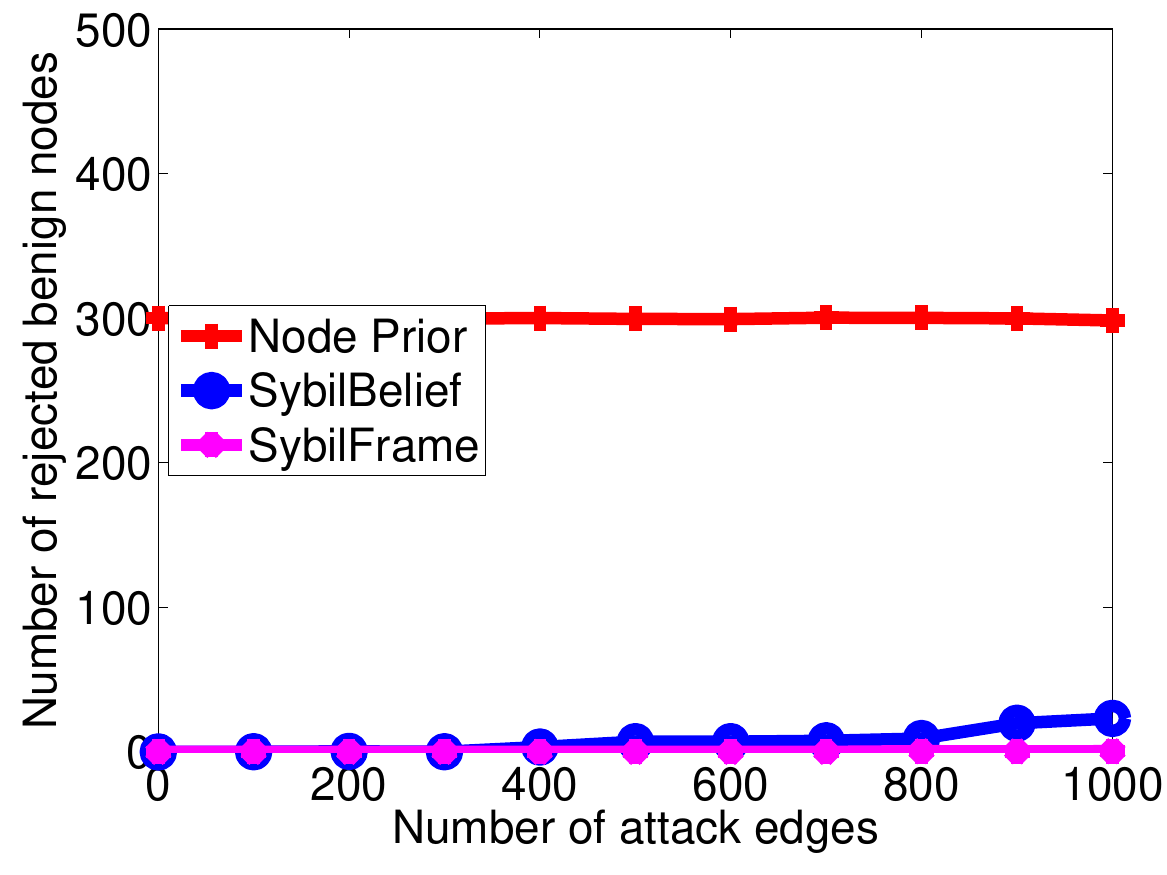}
  \caption{Rejected benign nodes}
  \label{fig:node_prior_benign_rej_2}
\end{subfigure}%
\begin{subfigure}[H]{0.24\textwidth}
  \includegraphics[width=\linewidth]{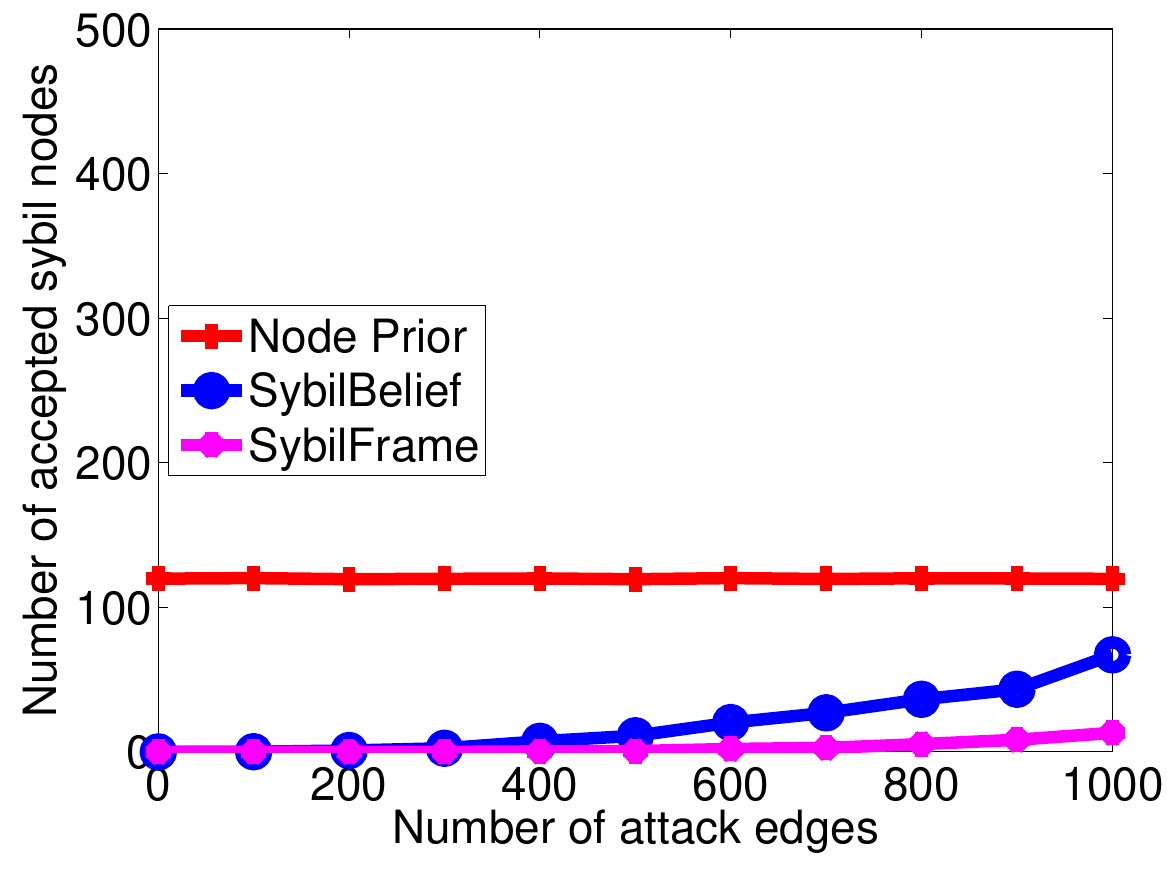}
  \caption{Accepted Sybil nodes}
  \label{fig:node_prior_sybil_acc_2}
\end{subfigure}%
\caption{
Vary the number of attack edges (node prior)}
\label{fig:node_prior_2}
\vspace{-0.7cm}
\end{figure}

\myparatight{Varying the number of attack edges}
\label{subsubsec:n_att}
Second, we evaluate SybilFrame when the number of attack edges changes. We set FPR and FNR to be 0.3, and vary the number of attack edges from 0 to 1000. Figure~\ref{fig:node_prior_2} shows the results. We find that both SybilFrame and SybilBelief have good performance with less than 200 attack edges. However, when the number of attack edges increases, SybilBelief degrades its performance while SybilFrame still has stable and near optimal detection accuracy.

\begin{figure}[!htb]
\vspace{-0.5cm}
\begin{subfigure}[H]{0.24\textwidth}
  \includegraphics[width=\linewidth]{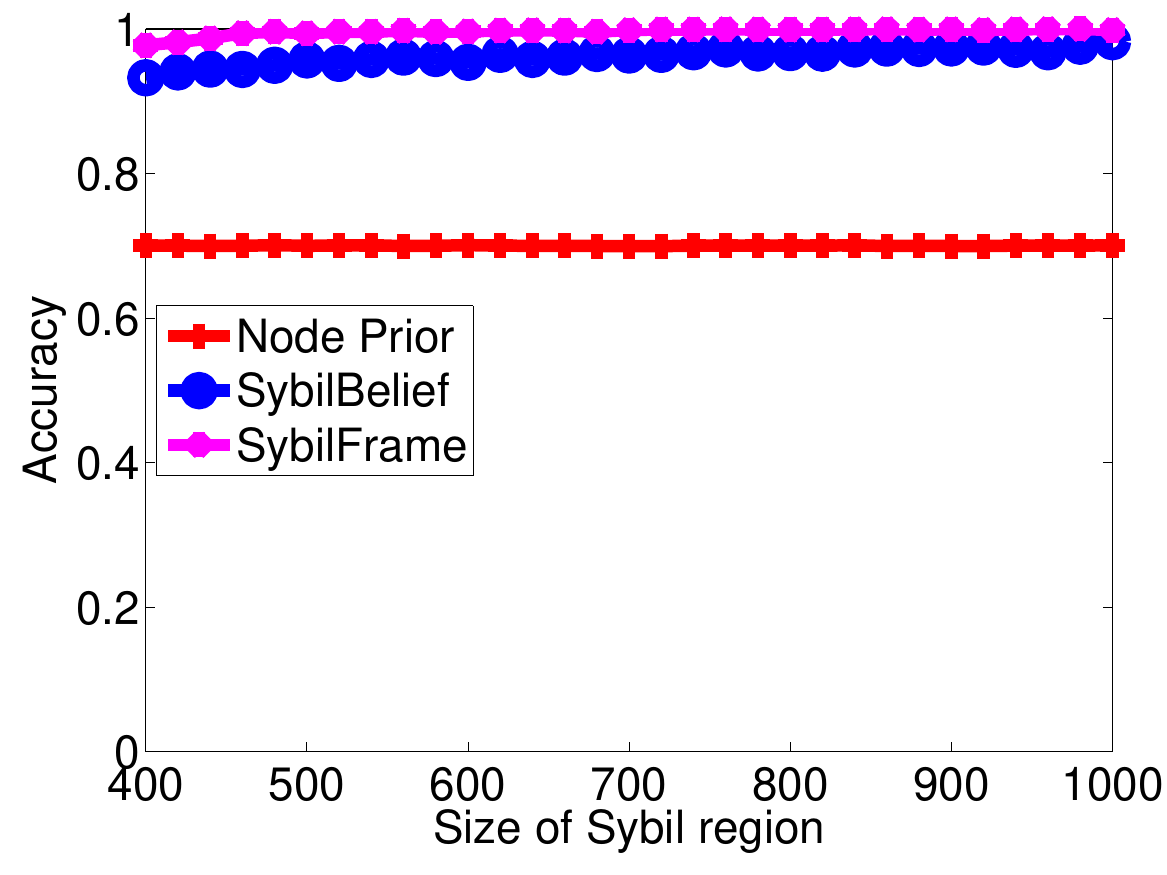}
  \caption{Accuracy}
  \label{fig:node_prior_accuracy_3}
\end{subfigure}%
\begin{subfigure}[H]{0.24\textwidth}
  \includegraphics[width=\linewidth]{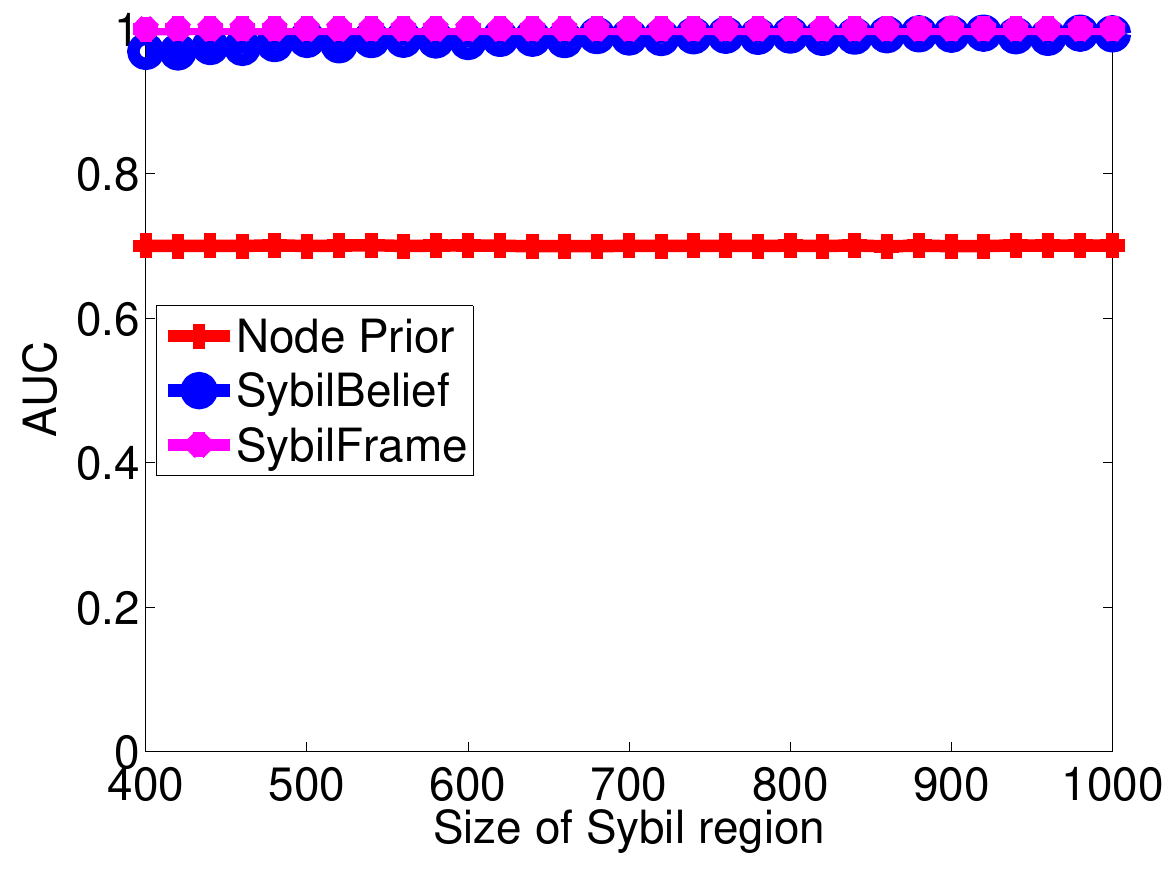}
  \caption{AUC}
  \label{fig:node_prior_auc_3}
\end{subfigure}%

\begin{subfigure}[H]{0.24\textwidth}
  \includegraphics[width=\linewidth]{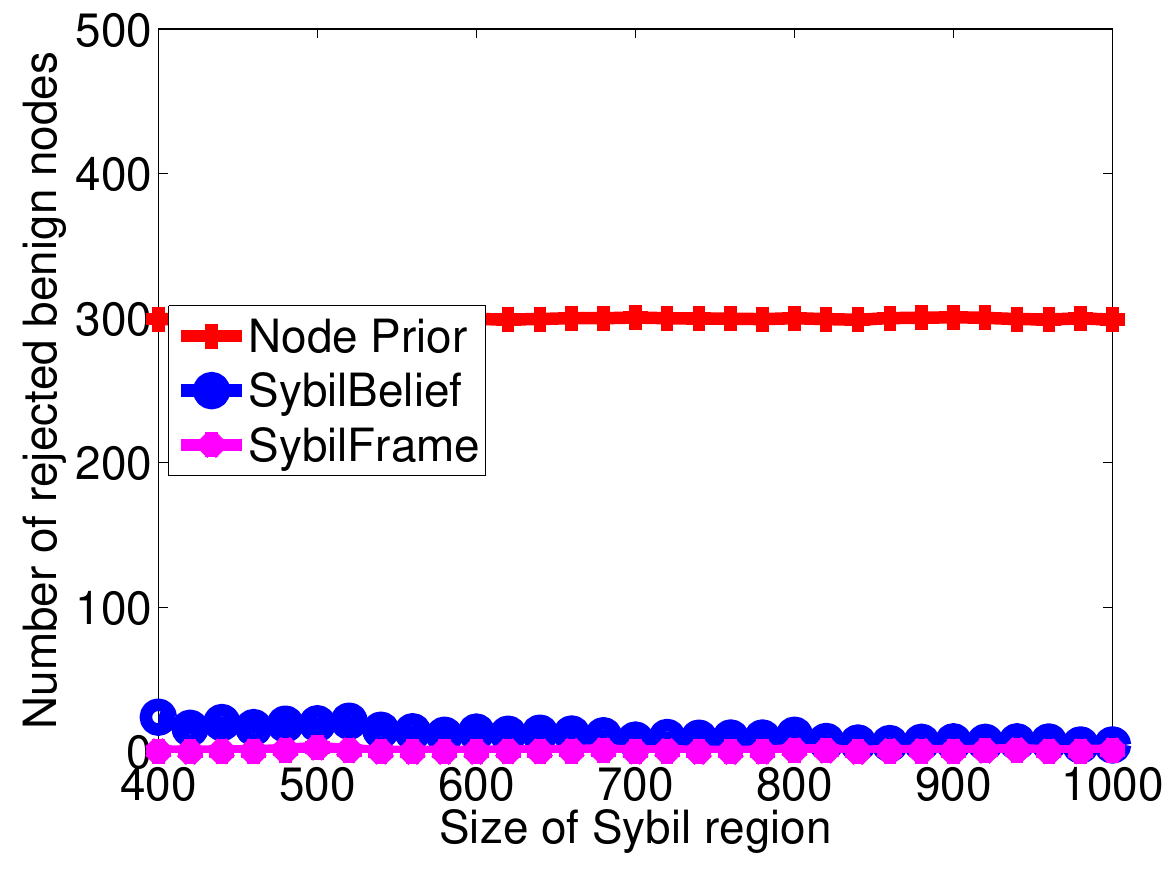}
  \caption{Rejected benign nodes}
  \label{fig:node_prior_benign_rej_3}
\end{subfigure}%
\begin{subfigure}[H]{0.24\textwidth}
  \includegraphics[width=\linewidth]{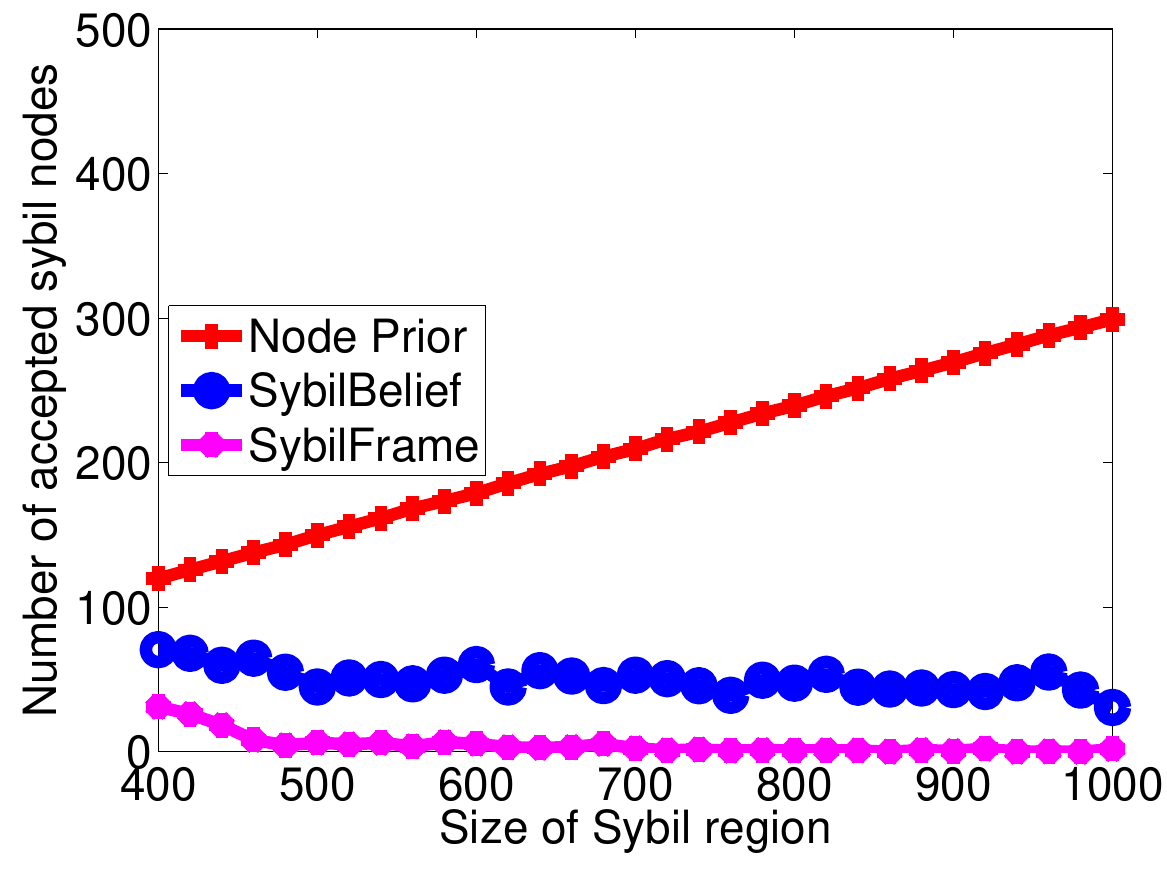}
  \caption{Accepted Sybil nodes}
  \label{fig:node_prior_sybil_acc_3}
\end{subfigure}%
\caption{
Vary the size of Sybil region (node prior)}
\label{fig:node_prior_3}
\vspace{-0.5cm}
\end{figure}

\myparatight{Varying the size of the Sybil region}
\label{subsubsec:size_of_sybil}
Furthermore, we evaluate SybilFrame when attacker changes the size of Sybil region. We will not consider the case when the Sybil region is too small, since it has limited utility to perform large-scale attacks. We set FPR and FNR to be 0.3, and vary the size of Sybil region from 400 to 1000.
Figure~\ref{fig:node_prior_3} shows the results. When there are more Sybil nodes, both SybilFrame and SybilBelief improve performance. This is because when both the benign and Sybil region are large, the internal homophily is strong enough to overcome the influence of attack edges. However, SybilFrame still performs better than SybilBelief.

\subsection{Influence of Edge Priors}
\label{subsec:edge_prior}

\begin{figure}
\begin{subfigure}[H]{0.24\textwidth}
  \includegraphics[width=\linewidth]{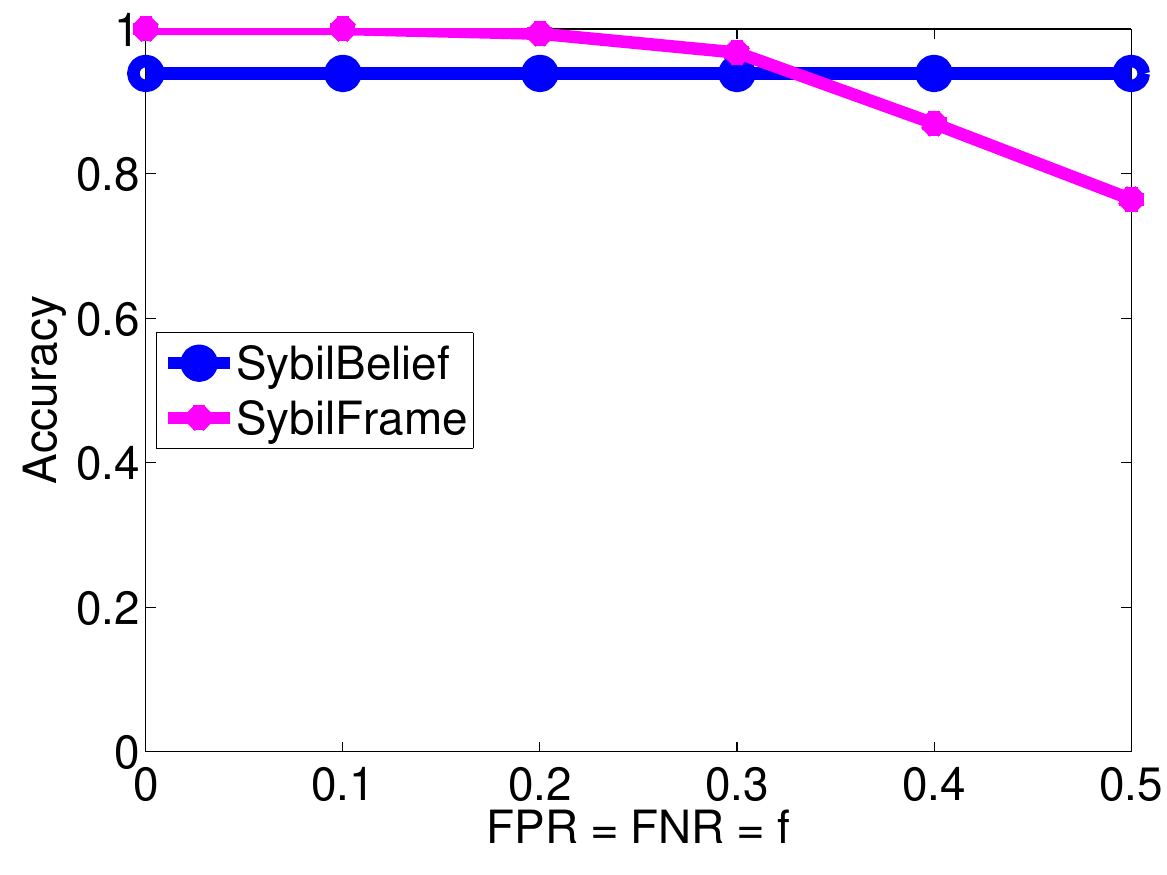}
  \caption{Accuracy}
  \label{fig:edge_prior_accuracy_1}
\end{subfigure}%
\begin{subfigure}[H]{0.24\textwidth}
  \includegraphics[width=\linewidth]{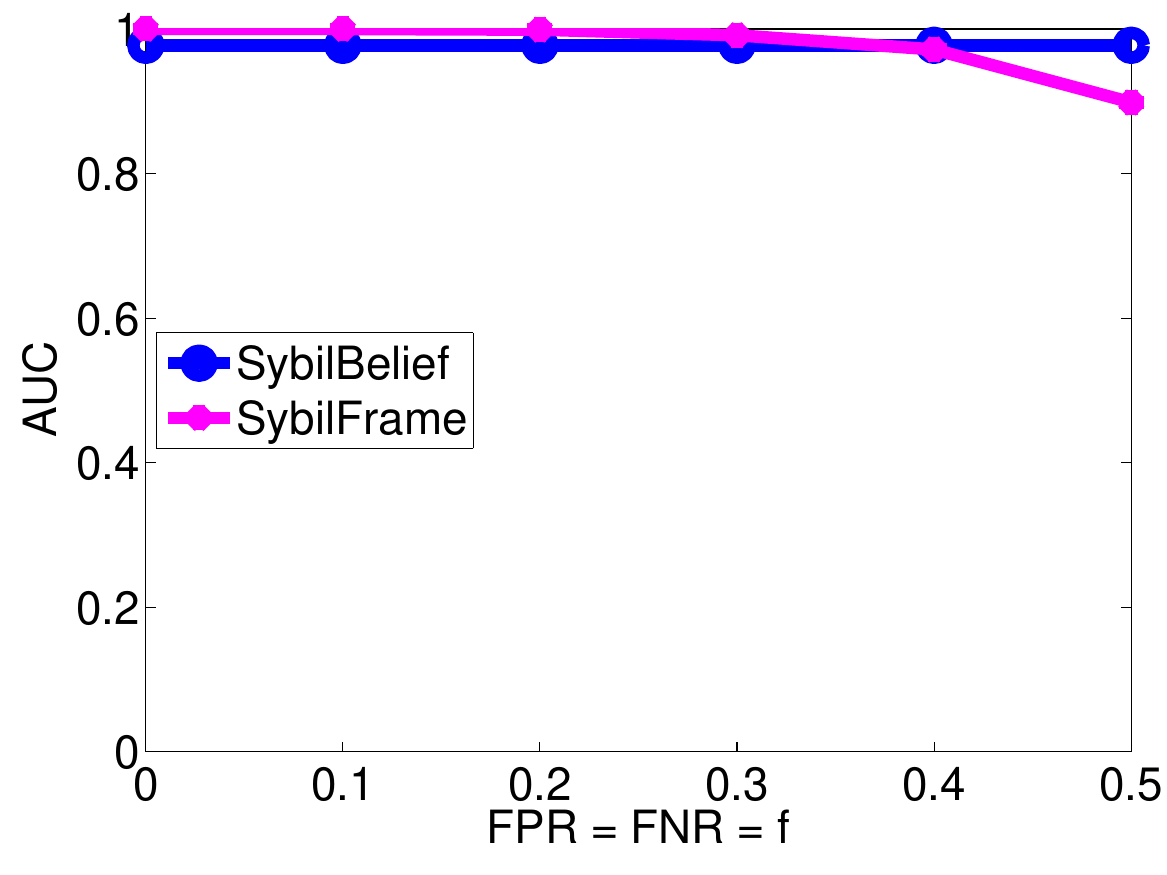}
  \caption{AUC}
  \label{fig:edge_prior_auc_1}
\end{subfigure}%

\begin{subfigure}[H]{0.24\textwidth}
  \includegraphics[width=\linewidth]{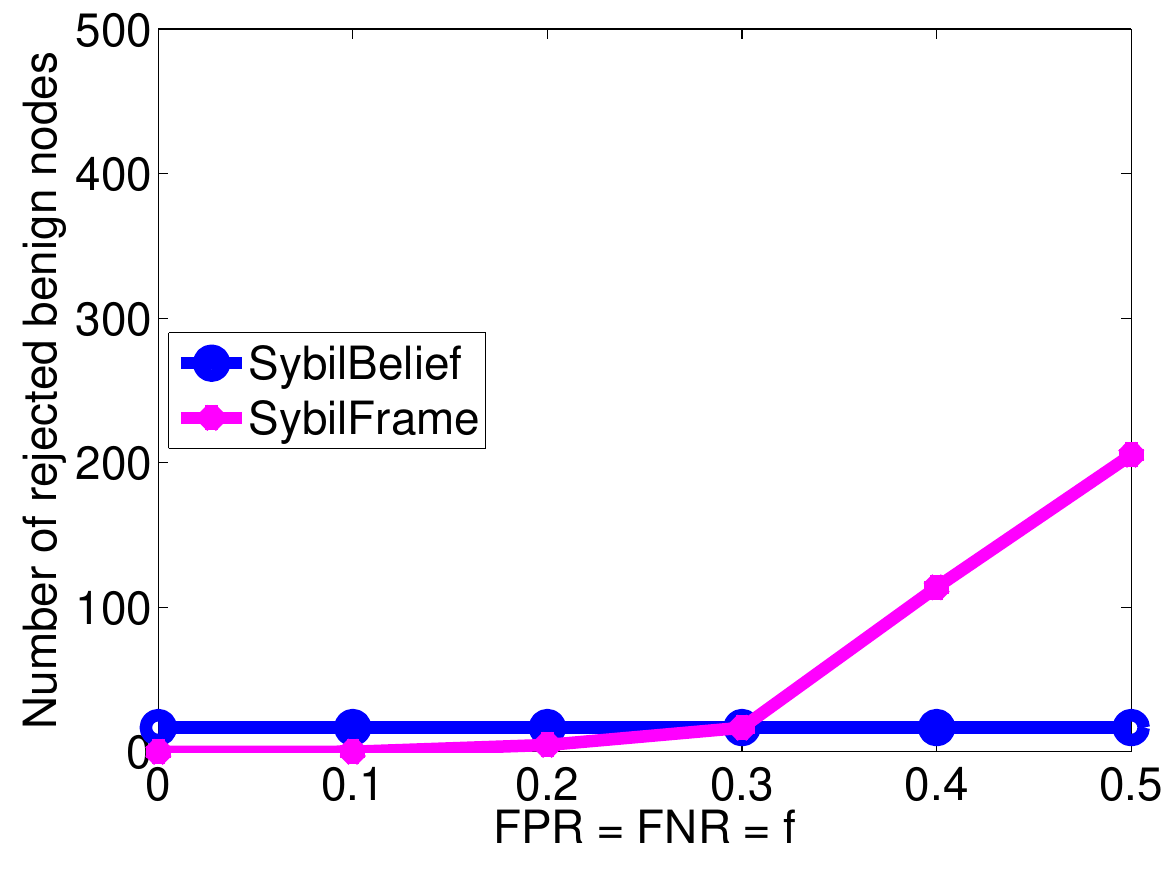}
  \caption{Rejected benign nodes}
  \label{fig:edge_prior_benign_rej_1}
\end{subfigure}%
\begin{subfigure}[H]{0.24\textwidth}
  \includegraphics[width=\linewidth]{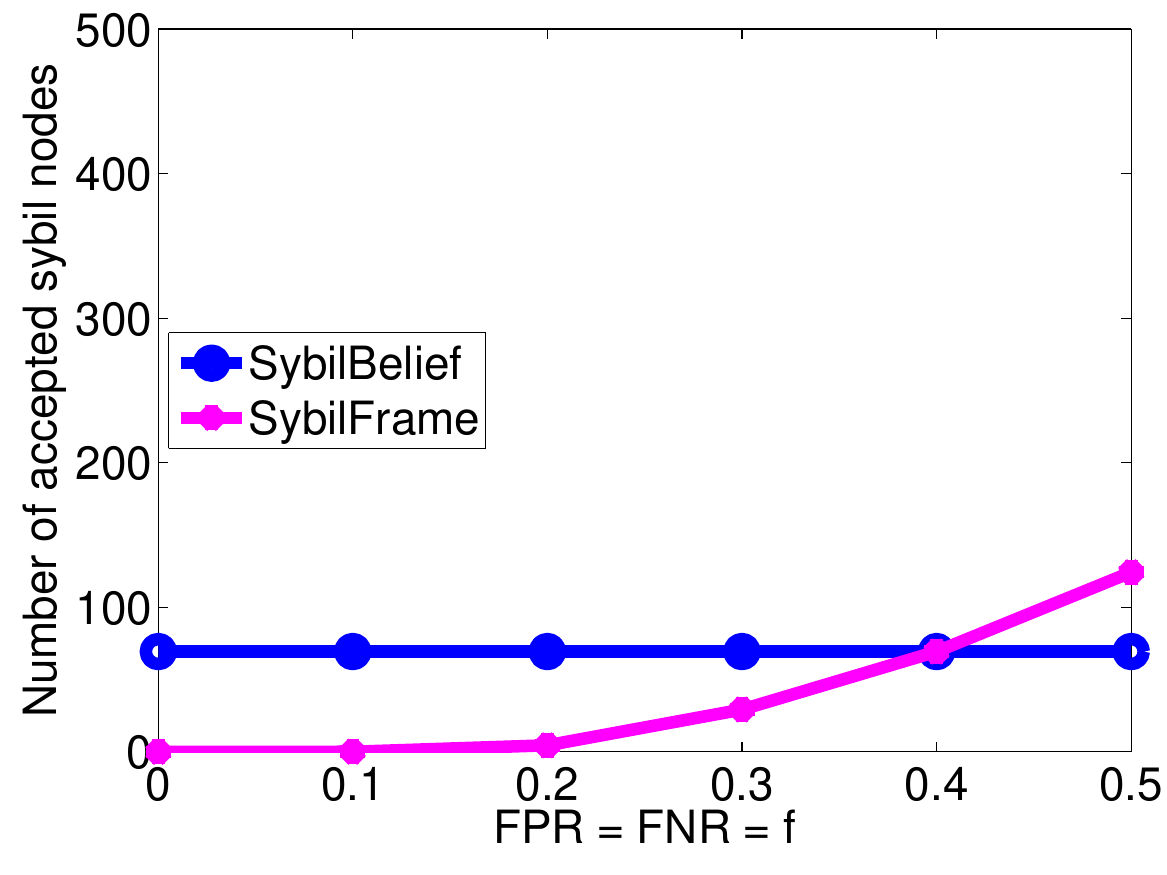}
  \caption{Accepted Sybil nodes}
  \label{fig:edge_prior_sybil_acc_1}
\end{subfigure}%
\caption{
Vary FPR=FNR (edge prior)}
\label{fig:edge_prior_1}
\vspace{-0.5cm}
\end{figure}

We want to explore SybilFrame when only incorporating external edge priors. Similarly, we use FPR and FNR to model the performance of an external edge classifier, which makes predictions of attack edges and other edges. We list our \emph{Edge Prior Generator} algorithm (Algorithm~\ref{alg:edge_prior}) in Appendix~\ref{subsec:prior_generator}.
\myparatight{Varying FPR and FNR}
\label{subsubsec:edge_fpr_fnr_f}
First, we \emph{tune $FPR=FNR$ for edge priors from 0 to 0.5}. We run SybilFrame with default node priors and compare with SybilBelief. From Figure~\ref{fig:edge_prior_1}, with $FPR=FNR\leq 0.3$, SybilFrame performs better than SybilBelief. 
Figure~\ref{fig:edge_prior_1_2} in Appendix~\ref{syn_edge_fpr0.1fpr} shows the results when we \emph{set $FPR=0.1$ and tune $FNR$ from 0 to 0.5}. As we can see, SybilFrame has good performance and outperforms SybilBelief even when $FNR$ is 0.5. This means that as long as the external edge classifier has some power to detect attack edges, incorporating edge priors into SybilFrame gives better performance.

\begin{figure}
\begin{subfigure}[H]{0.24\textwidth}
  \includegraphics[width=\linewidth]{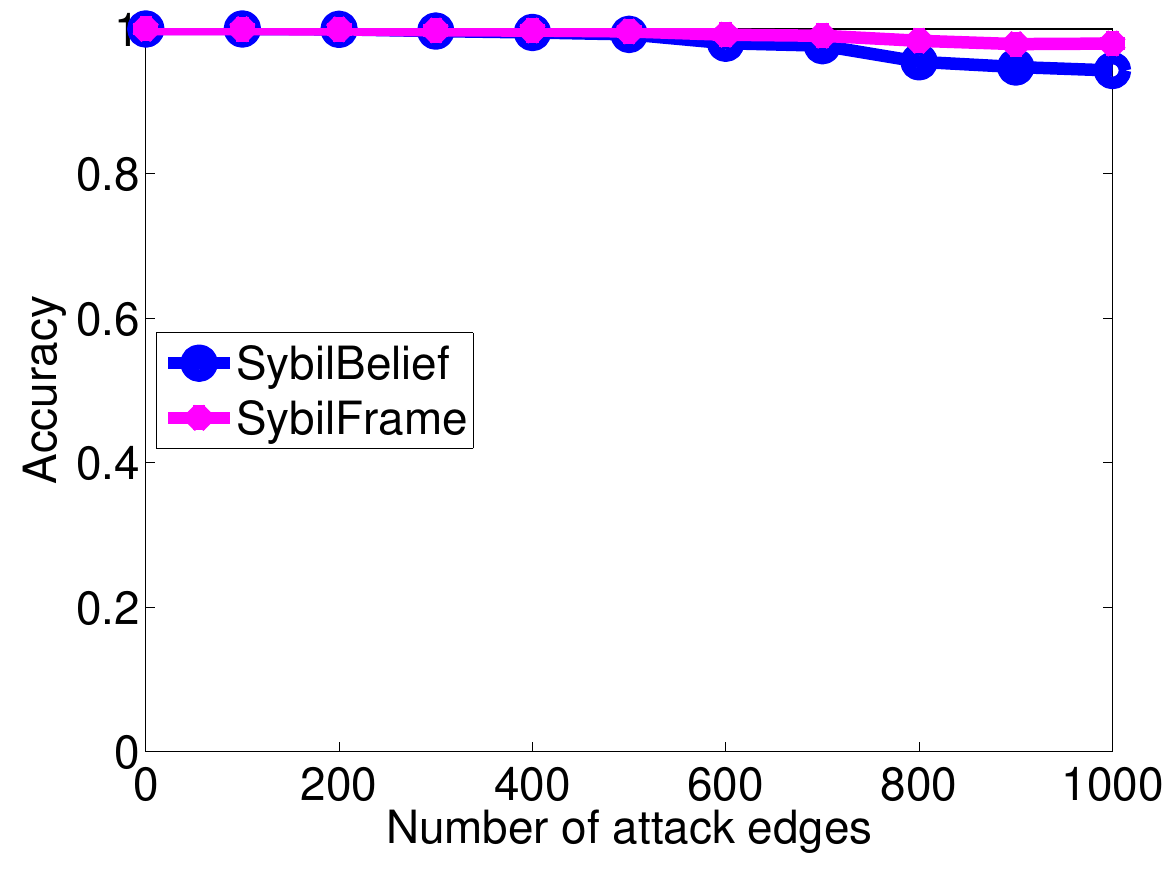}
  \caption{Accuracy}
  \label{fig:edge_prior_accuracy_2}
\end{subfigure}%
\begin{subfigure}[H]{0.24\textwidth}
  \includegraphics[width=\linewidth]{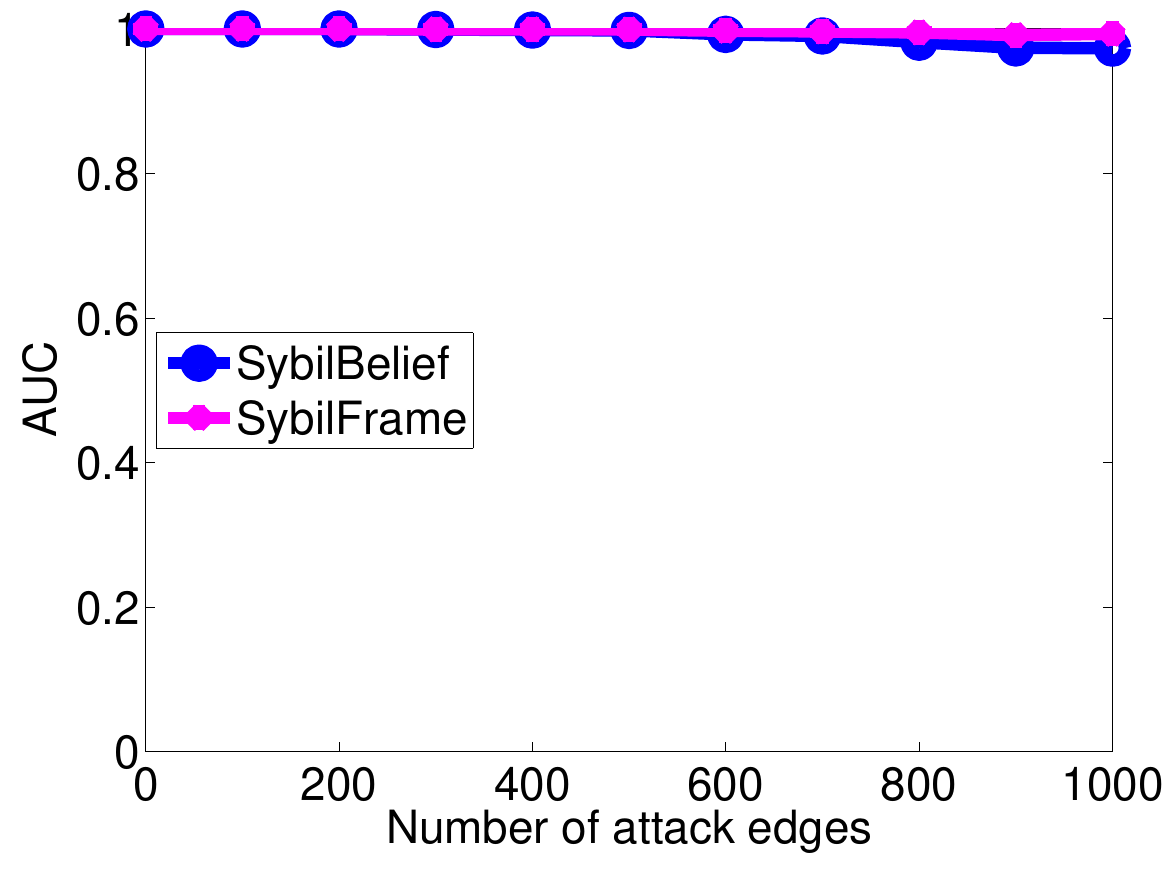}
  \caption{AUC}
  \label{fig:edge_prior_auc_2}
\end{subfigure}%

\begin{subfigure}[H]{0.24\textwidth}
  \includegraphics[width=\linewidth]{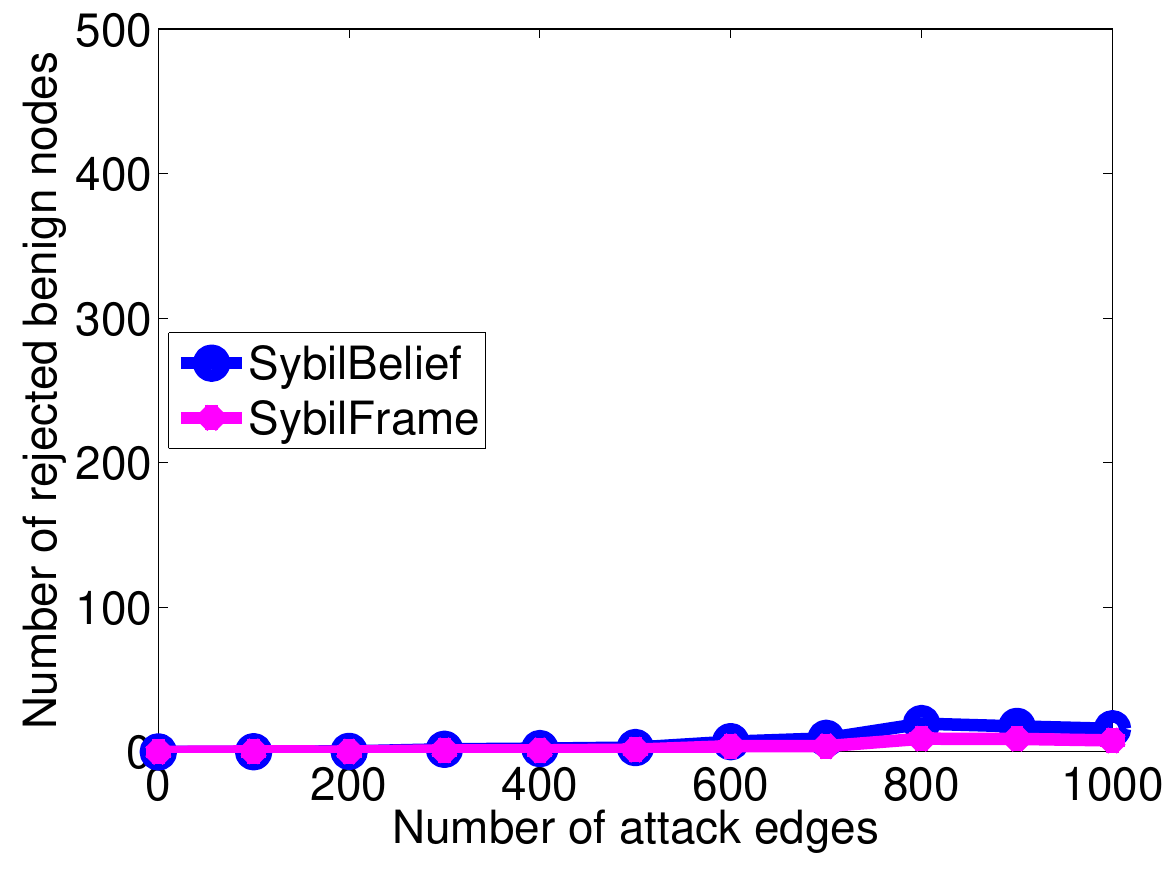}
  \caption{Rejected benign nodes}
  \label{fig:edge_prior_benign_rej_2}
\end{subfigure}%
\begin{subfigure}[H]{0.24\textwidth}
  \includegraphics[width=\linewidth]{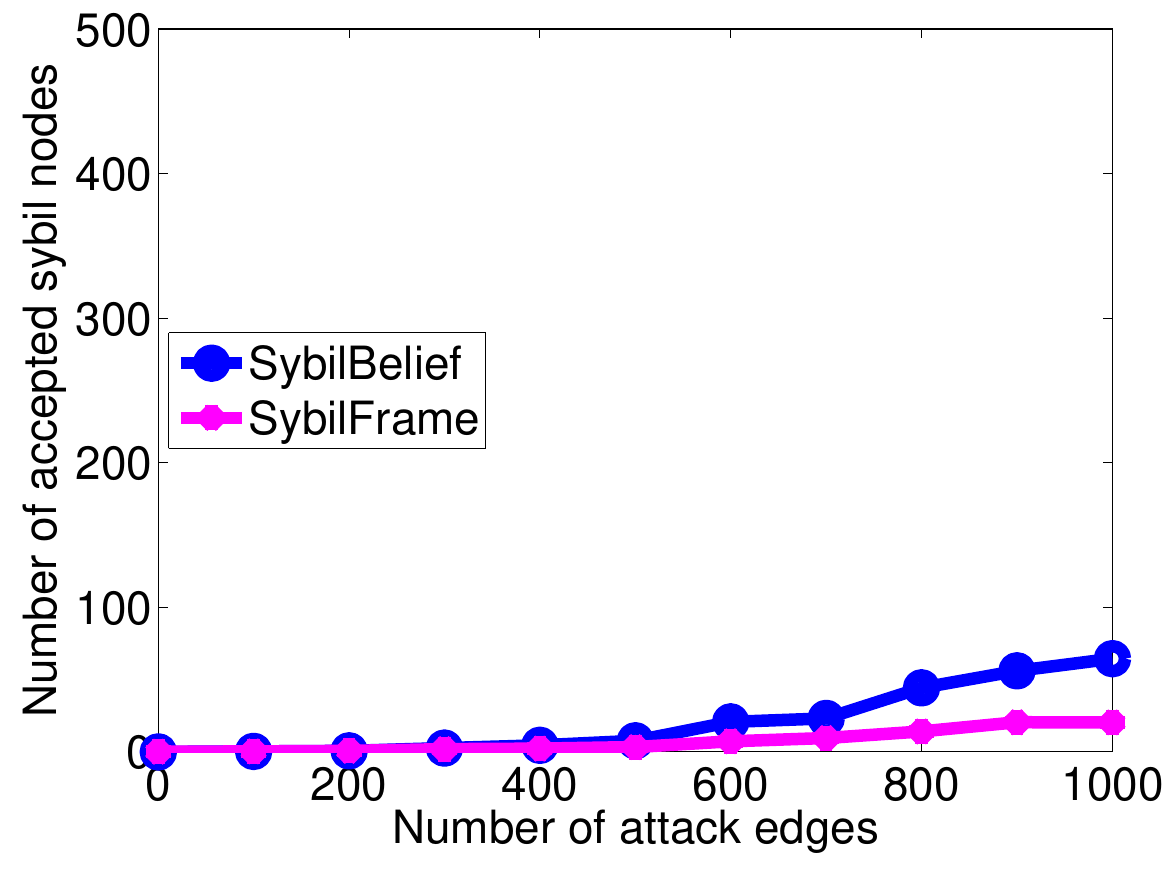}
  \caption{Accepted Sybil nodes}
  \label{fig:edge_prior_sybil_acc_2}
\end{subfigure}%
\caption{
Vary the number of attack edges (edge prior)}
\label{fig:edge_prior_2}
\vspace{-0.1cm}
\end{figure}

\myparatight{Varying the number of attack edges}
\label{subsubsec:edge_n_att}
Second, we set FPR to be 0.1 and FNR to be 0.5, and vary the number of attack edges from 0 to 1000. 
Figure~\ref{fig:edge_prior_2} shows the results. As the number of attack edges increases, both SybilFrame and SybilBelief degrade performance. However, SybilFrame still outperforms SybilBelief. Notice that the performance of SybilFrame depends the detection accuracy of external classifier. If we have a classifier with 0.1 FPR and 0.1 FNR, SybilFrame will have near optimal performance. 

\begin{figure}
\begin{subfigure}[H]{0.24\textwidth}
  \includegraphics[width=\linewidth]{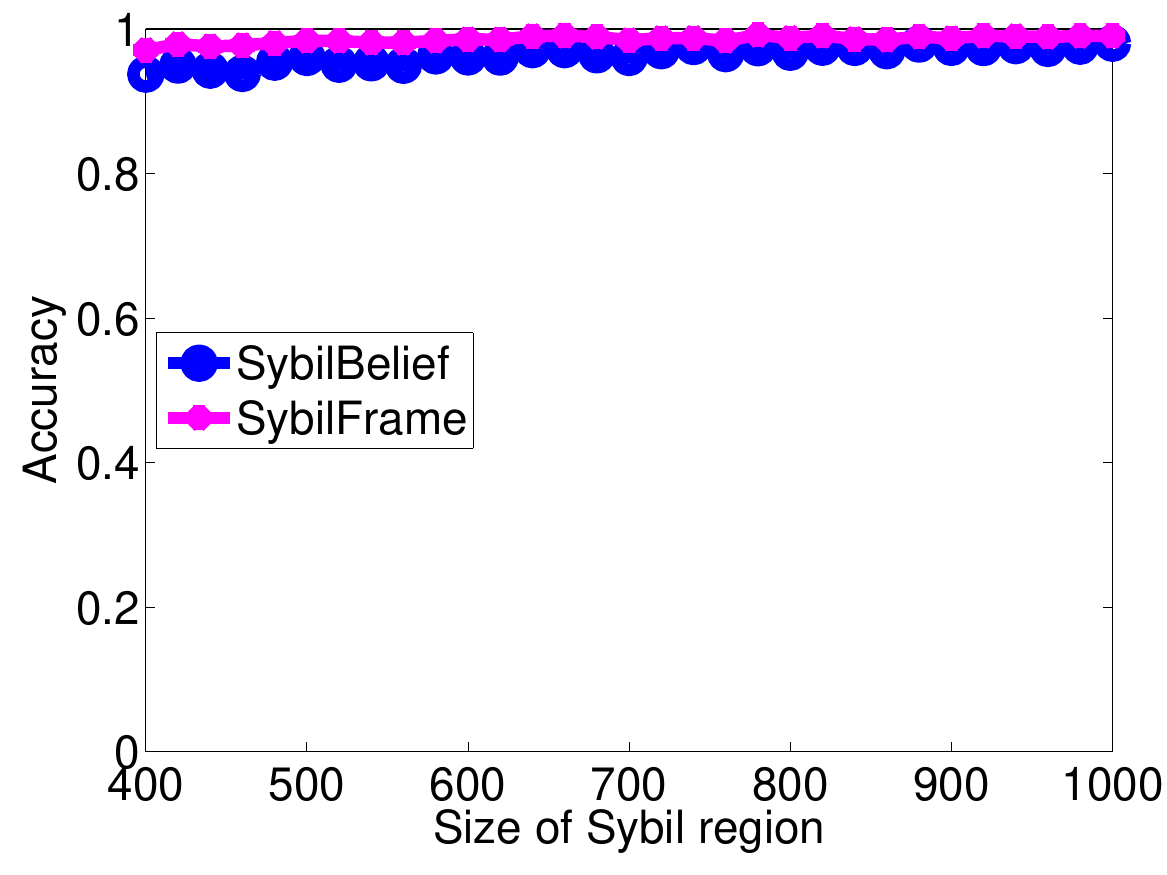}
  \caption{Accuracy}
  \label{fig:edge_prior_accuracy_3}
\end{subfigure}%
\begin{subfigure}[H]{0.24\textwidth}
  \includegraphics[width=\linewidth]{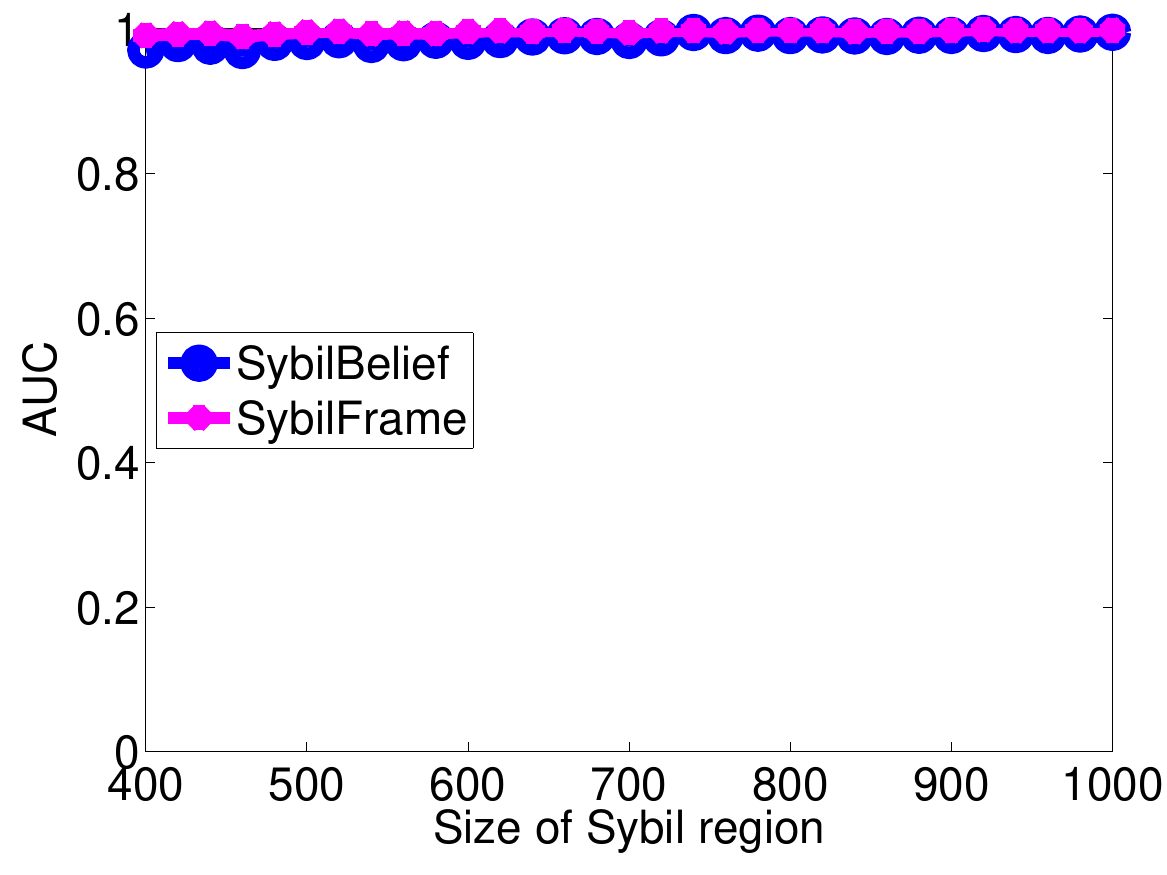}
  \caption{AUC}
  \label{fig:edge_prior_auc_3}
\end{subfigure}%

\begin{subfigure}[H]{0.24\textwidth}
  \includegraphics[width=\linewidth]{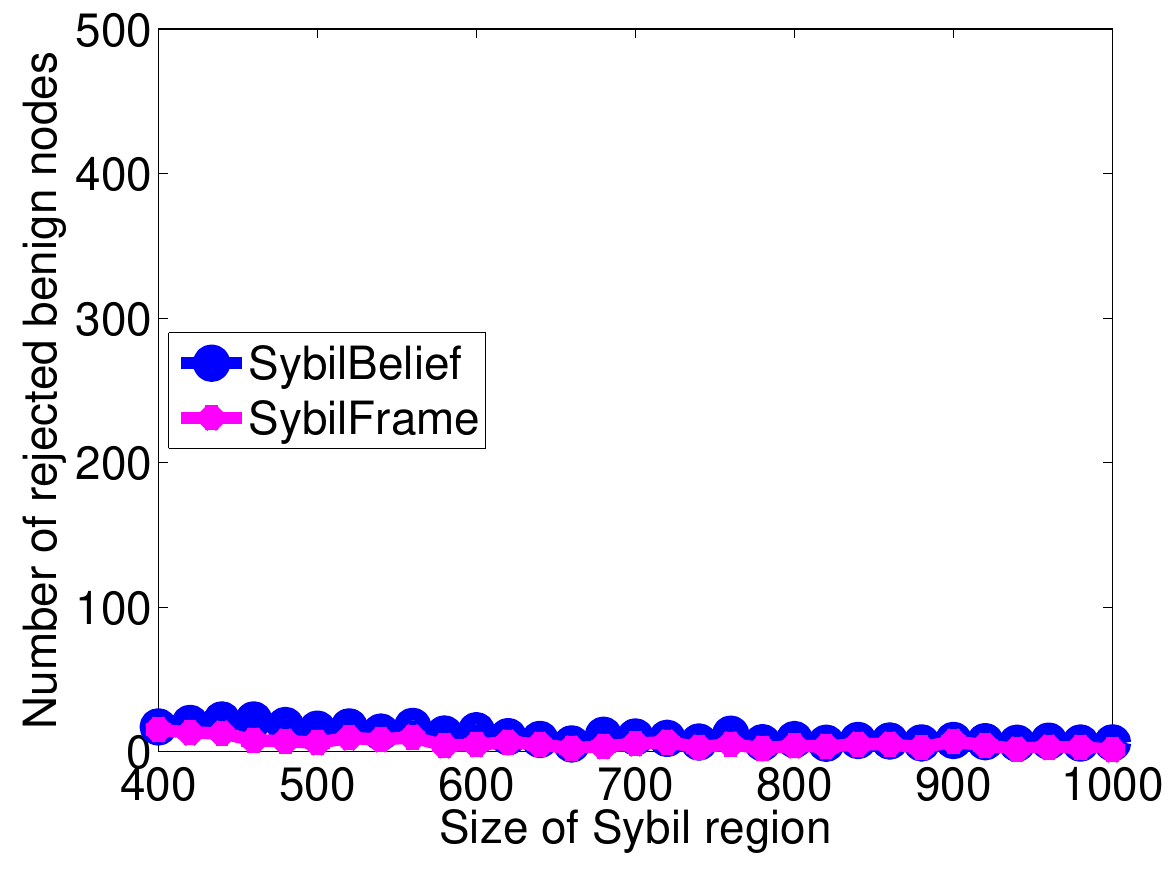}
  \caption{Rejected benign nodes}
  \label{fig:edge_prior_benign_rej_3}
\end{subfigure}%
\begin{subfigure}[H]{0.24\textwidth}
  \includegraphics[width=\linewidth]{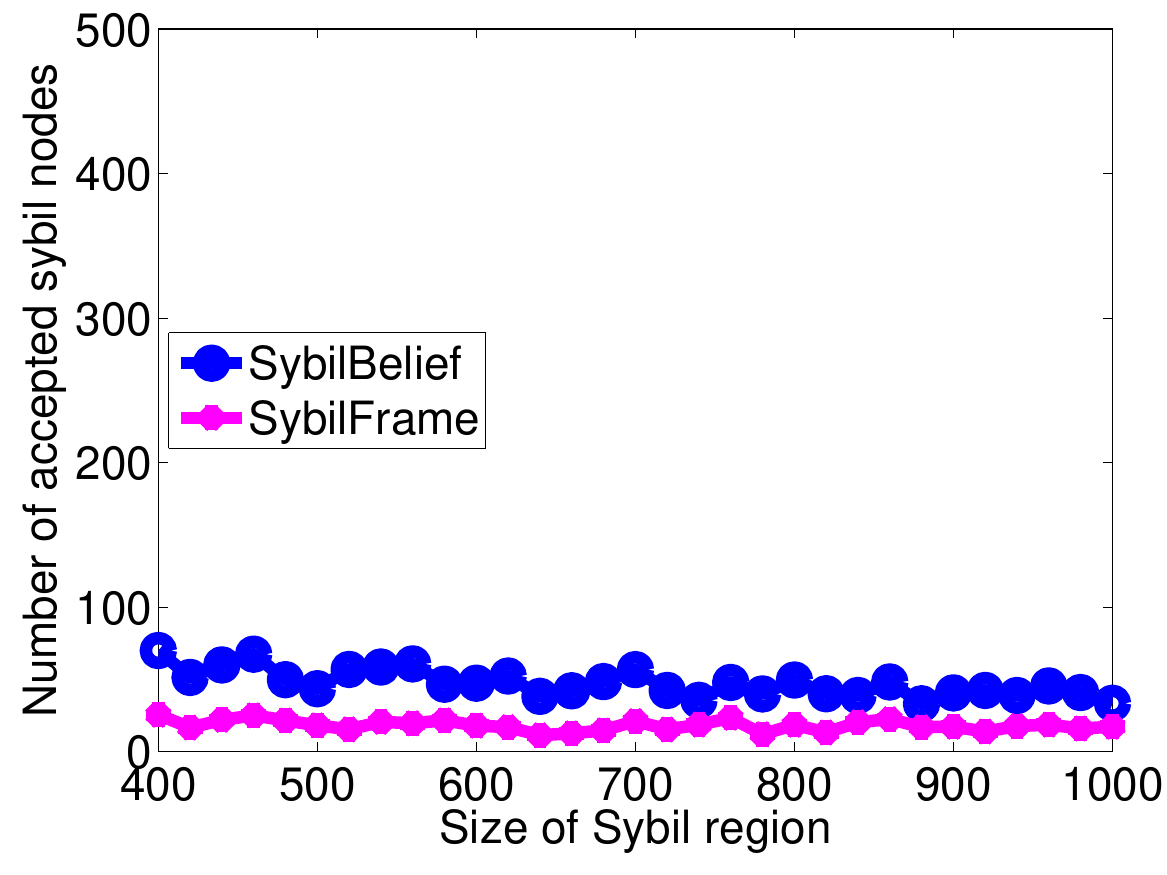}
  \caption{Accepted Sybil nodes}
  \label{fig:edge_prior_sybil_acc_3}
\end{subfigure}%
\caption{
Vary the size of Sybil region (edge prior)}
\label{fig:edge_prior_3}
\vspace{-0.6cm}
\end{figure}

\myparatight{Varying the size of the Sybil region}
\label{subsubsec:edge_size_of_sybil}
Furthermore, we evaluate SybilFrame with edge priors when attacker changes the size of the Sybil region. We set FPR to be 0.1 and FNR to be 0.5, and vary the size of Sybil region from 400 to 1000. 
From Figure~\ref{fig:edge_prior_3}, SybilFrame improves its performance when there are more Sybil nodes, and still outperforms SybilBelief.

\subsection{Resilient Against Seed Targeting Attacks}
\label{subsec:targeted_attack}

We are interested in the impact of seed targeting attacks, i.e., when the known labeled nodes are end points of attack edges. We consider the following cases:

\textbf{1) SI:} Benign (Sybil) trust seeds are not end points of attack edges.

\textbf{2) SII:} Benign (Sybil) trust seeds are end points of attack edges.
\begin{figure}[!htb]
\begin{subfigure}[H]{0.24\textwidth}
  \includegraphics[width=\linewidth]{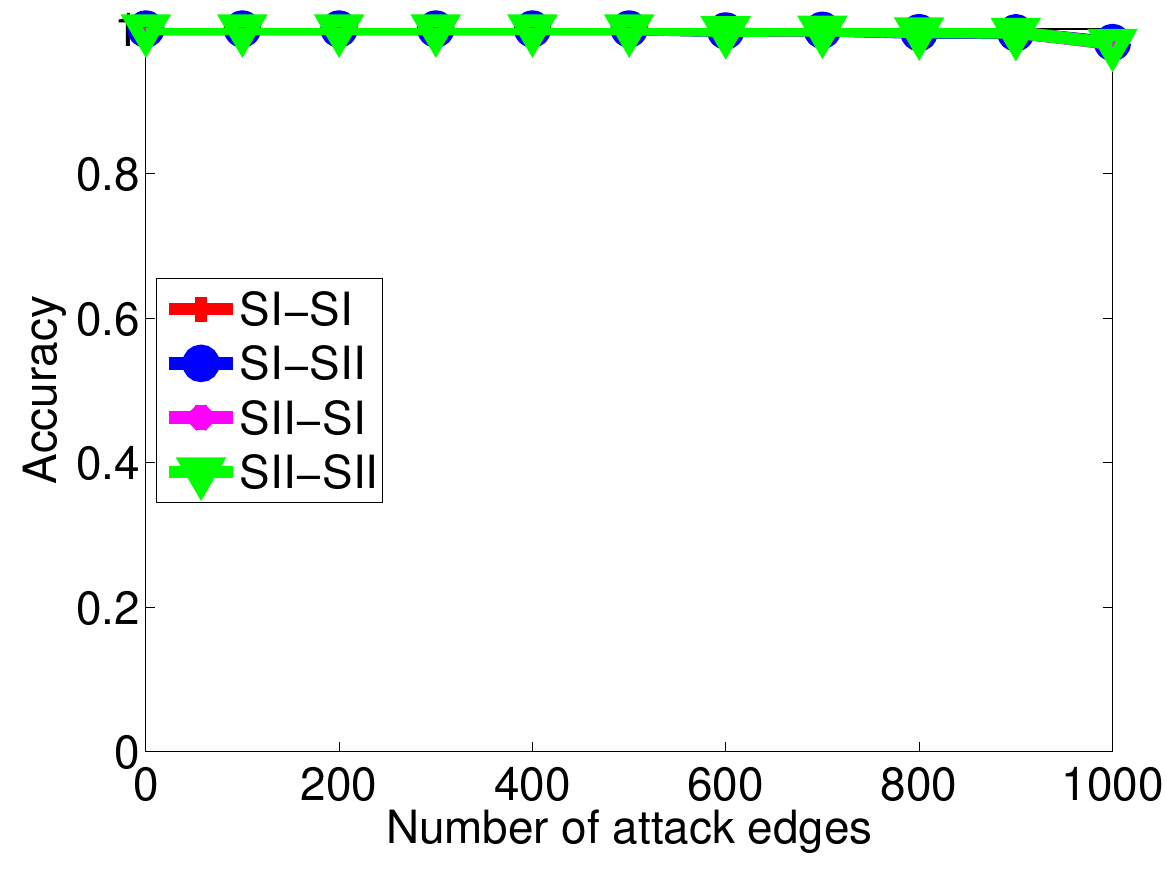}
  \caption{Accuracy (node prior)}
  \label{fig:targeted_node_accuracy}
\end{subfigure}%
\begin{subfigure}[H]{0.24\textwidth}
  \includegraphics[width=\linewidth]{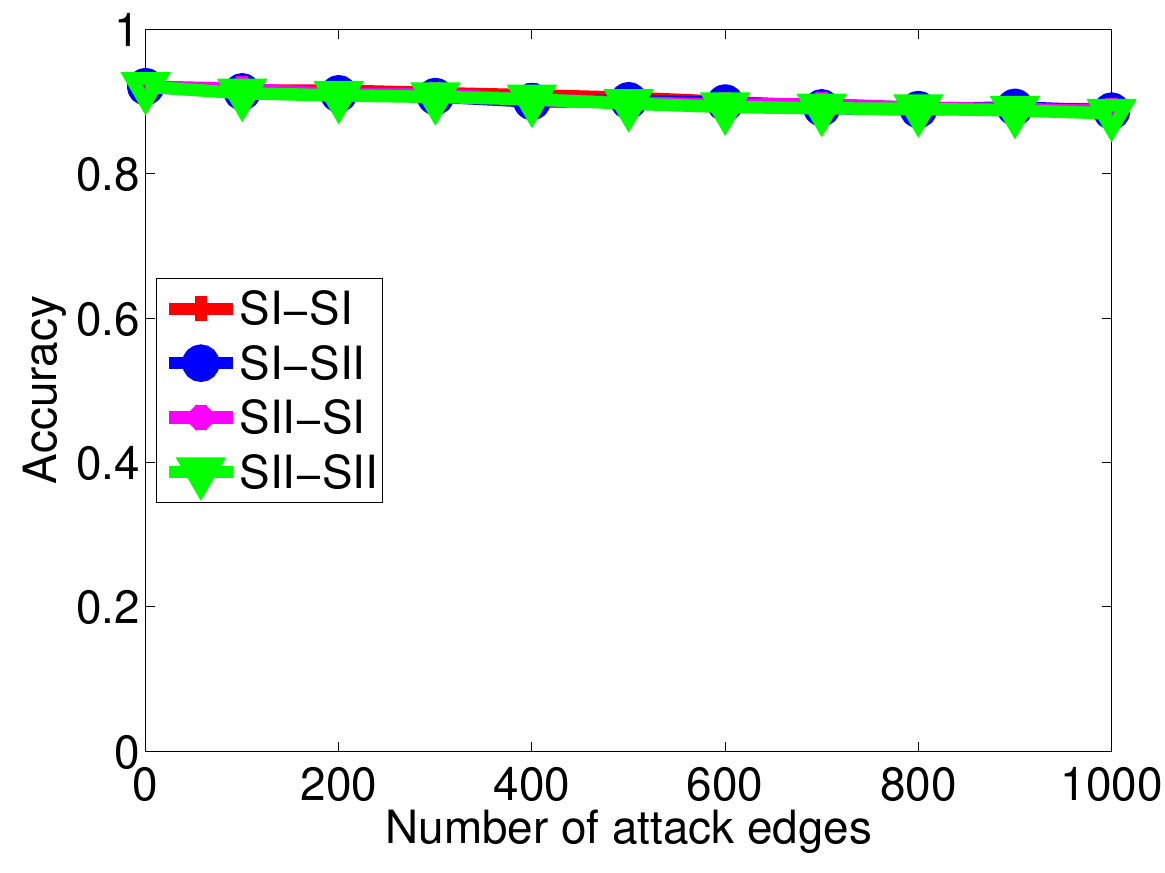}
  \caption{Accuracy (edge prior)}
  \label{fig:targeted_edge_accuracy}
\end{subfigure}%
\caption{
Accuracy of SybilFrame under seed targeting attacks. (a) Given node priors. (b) Given edge priors.
}
\label{fig:targeted_accuracy}
\vspace{-0.3cm}
\end{figure}

Figure~\ref{fig:targeted_accuracy} shows the accuracy as a function of the number of attack edges for four scenario combinations of trust seeds, in the node prior experiment (FPR=FNR=0.3) and edge prior experiment (FPR=0.1, FNR=0.5). We find that the location of trust seeds have no influence on the detection accuracy. Due to limited space, we list the results of AUC, FP and FN in Appendix~\ref{subsec:seed_targeted_appendix}. We find that SybilFrame is
resilient against seed targeting attacks, and we can simply select trust seeds uniformly at random.

\subsection{Summary}
\label{subsec:summary}

In summary, we have the following observations:

\textbf{1)} SybilFrame outputs near optimal results when incorporating node priors with FPR and FNR less than 0.3.

\textbf{2)} SybilFrame outperforms SybilBelief when incorporating node priors with FPR and FNR less than 0.4.

\textbf{3)} When incorporating edge priors, as long as the edge priors has a low FPR (0.1) and some level of FNR (less than 0.5), SybilFrame outperforms SybilBelief. 

\textbf{4)} SybilFrame is robust to different attack strategies and resilient against seed targeting attacks.

\section{Evaluation on Facebook Network}
\label{sec:evaluation_facebook}

We evaluate SybilFrame on semi-real Facebook network, and compare with state-of-the-art Sybil defense mechanisms: SybilLimit, SybilInfer, SybilRank and SybilBelief. We find that SybilFrame performs orders of magnitudes better than other methods, especially when the number of attack edges is large. Furthermore, the performance of SybilFrame is stable and near optimal.

\subsection{Dataset Description}
\label{subsec:facebook_dataset_description}
The dataset we use is the ego-Facebook dataset obtained from Stanford Network Analysis Project (SNAP)~\cite{facebookdata}. The Facebook graph contains 4,039 nodes and 88,234 edges. In this graph, nodes are Facebook accounts and edges are friendship relationships. The graph is connected and undirected, with a diameter 8 and average clustering coefficient 0.6055.

\subsection{Experimental Setup}
\label{subsec:facebook_exp_setup}

We construct the network topology as follows. We use this Facebook dataset as both the benign region and Sybil region, and randomly add attack edges between the two regions. 
We vary the number of attack edges from 1000 to 20000, and evaluate the performance of SybilFrame, as well as SybilLimit, SybilInfer, SybilRank and SybilBelief. We randomly select 1 benign trust seed and 1 Sybil trust seed, and perform the experiments 100 times and then take the average. 

\subsection{Compute Prior Information}
\label{subsec:facebook_prior}

To run SybilFrame, we need prior information. Since the benign region is identical to the Sybil region, we are not able to collect distinguishable node priors. Thus, we only explore ways to compute edge priors.

As discussed in Section~\ref{subsec:prior}, we can leverage
similarity between two connected nodes, and use it as a prior for the edge between them. Intuitively, connected benign nodes are similar and connected benign and Sybil nodes are not similar. Therefore, attack edges should have a lower score than non-attack edges. We adopt the Jaccard index here as a measure of similarity. For an edge $(u, v)\in E$, the Jaccard index~\cite{Liben-Nowell2007} of it is defined as $\frac{\left|\Gamma(u)\cap \Gamma(v)\right|}{\left|\Gamma(u)\cup \Gamma(v)\right|}$, where $\Gamma(u)$ denotes the set of one-hop neighbors of node $u$, and $\left|\Gamma(u) \cap \Gamma(v)\right|$ denotes the number of common neighbors of $u$ and $v$. For edges that connect two trust seeds, we set the prior of it to 0.1 if the edge is an attack edge, and set to 0.9 if the edge is a non-attack edge. For other edges, we compute the corresponding Jaccard index. We scale these indices into the range $[0.1, 0.9]$. These scaled Jaccard scores will then be used as priors.

We can also use other similarity metrics, such as Cosine index~\cite{Lu2011} or Adamic-Adar index~\cite{Liben-Nowell2007}
, and combine them to obtain an overall similarity score. A possible approach is to use these raw similarity scores as features for an edge, and obtain a feature matrix for all edges on the graph. We can then adopt a supervised learning approach by leveraging existing tools, such as Logistic Regression~\cite{hastie01statisticallearning} and Support Vector Machine~\cite{hastie01statisticallearning}, and make probabilistic predictions of each edge being a non-attack/attack edge. These probabilistic outputs can then be used as overall prior scores.

\subsection{Results}
\label{subsec:facebook_result}

\begin{figure}[!t]
\begin{subfigure}[H]{0.24\textwidth}
  \includegraphics[width=\linewidth]{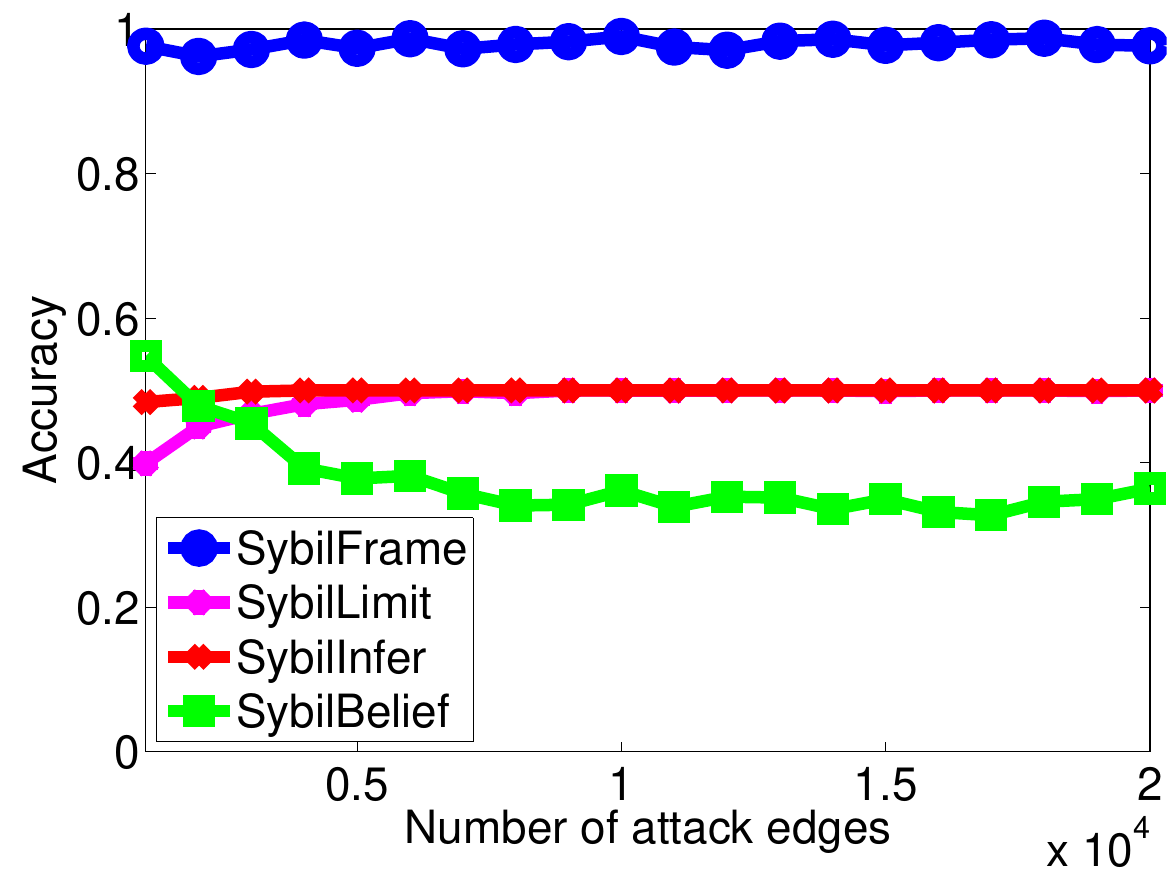}
  \caption{Accuracy}
  \label{fig:facebook_accuracy}
\end{subfigure}%
\begin{subfigure}[H]{0.24\textwidth}
  \includegraphics[width=\linewidth]{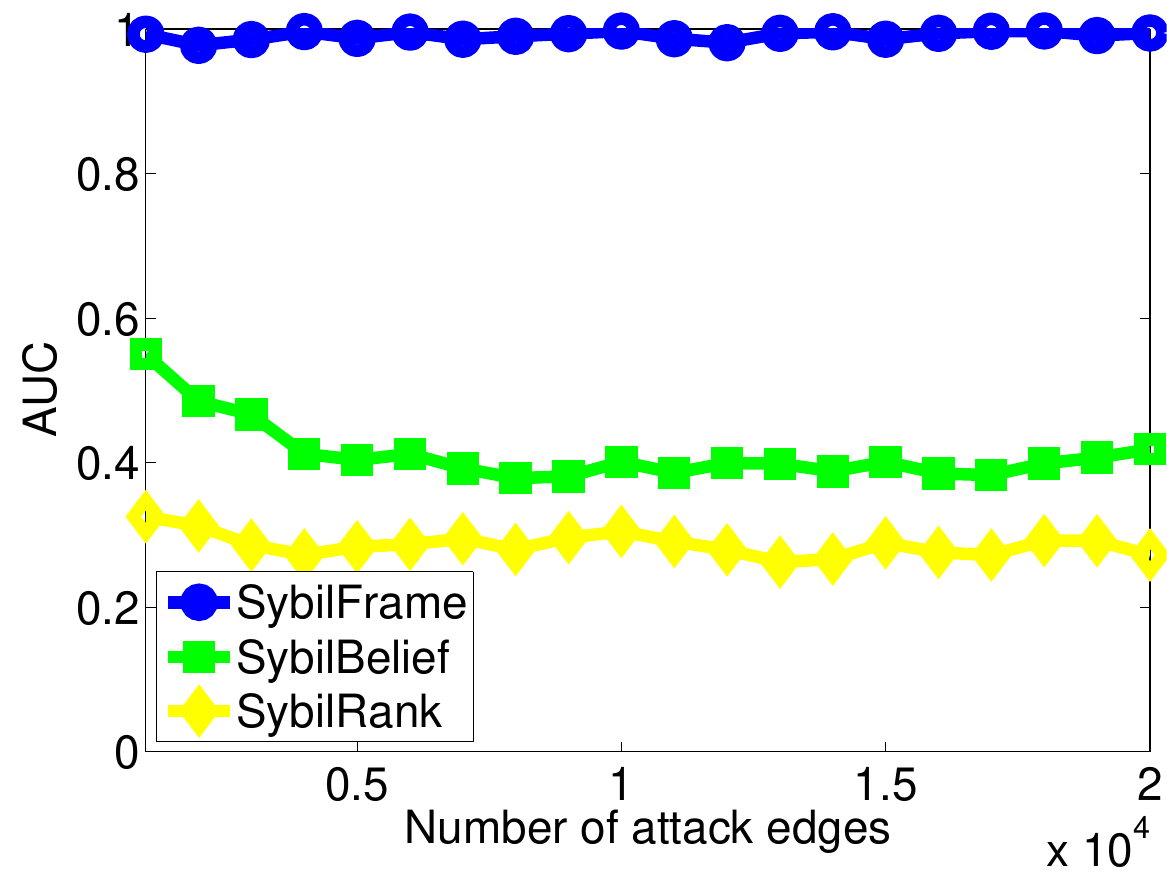}
  \caption{AUC}
  \label{fig:facebook_auc}
\end{subfigure}%

\begin{subfigure}[H]{0.24\textwidth}
  \includegraphics[width=\linewidth]{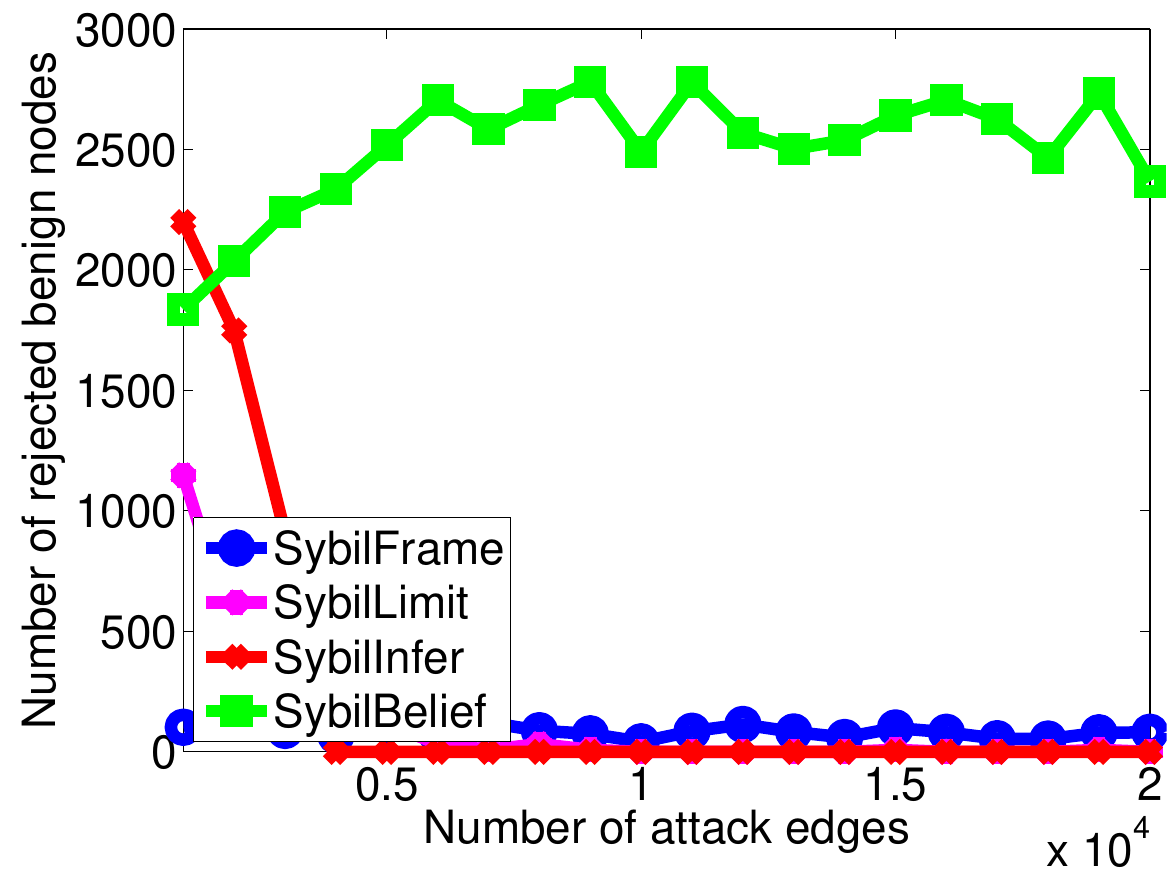}
  \caption{Rejected benign nodes}
  \label{fig:facebook_benign_rej}
\end{subfigure}%
\begin{subfigure}[H]{0.24\textwidth}
  \includegraphics[width=\linewidth]{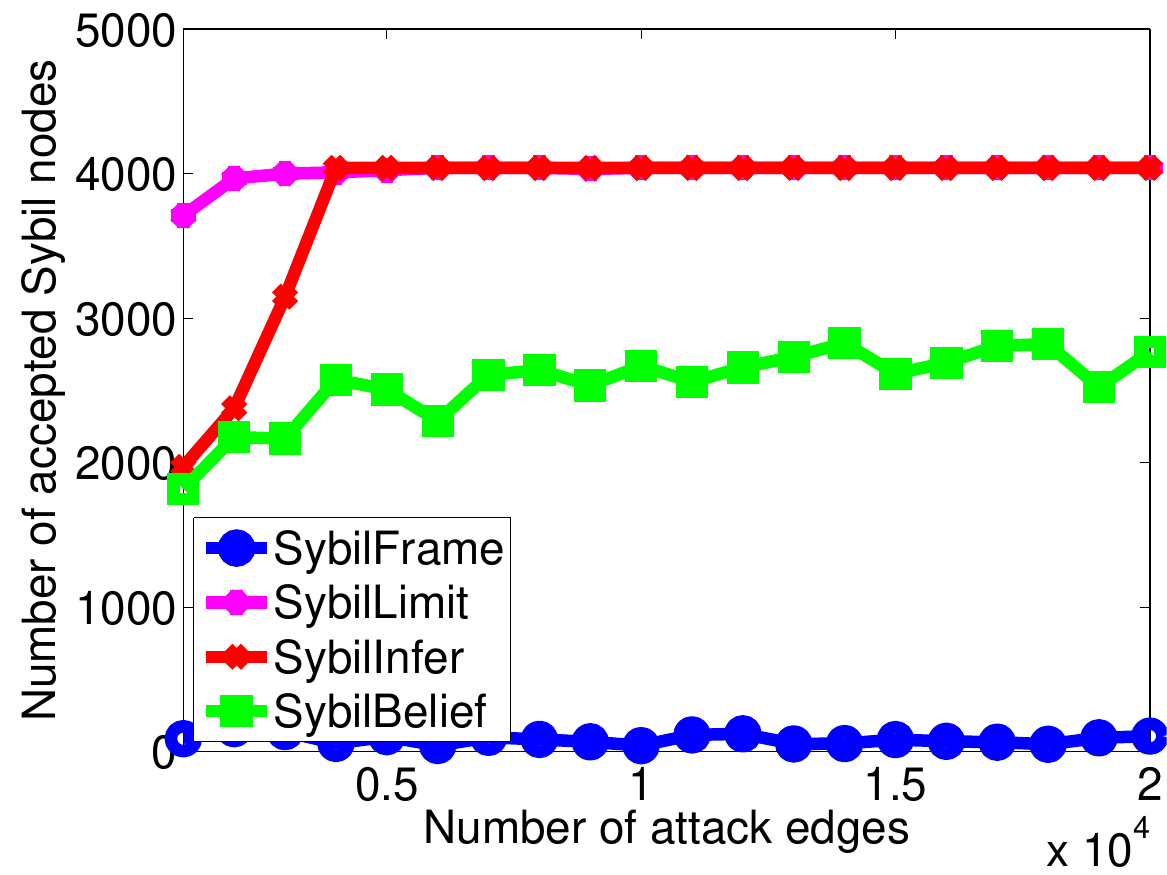}
  \caption{Accepted Sybil nodes}
  \label{fig:facebook_sybil_acc}
\end{subfigure}%
\caption{
Performance comparison on Facebook network}
\label{fig:facebook}
\vspace{-0.6cm}
\end{figure}

Figure~\ref{fig:facebook} shows the performance of SybilFrame and other Sybil defense mechanisms as we vary the number of attack edges from 1000 to 20000. Since SybilRank is a ranking scheme and it is very hard to directly use the degree normalized scores to make predictions, we only compare with SybilRank in terms of AUC. Also, since SybilLimit and SybilInfer output binary predictions rather than belief scores, we do not include them into AUC comparison. We find that: 1) As the number of attack edges increases, the performance of precious methods degrades, with a lower accuracy (SybilLimit, SybilInfer, SybilBelief), and lower AUC (SybilRank). 2) The speed of performance degradation is fast. With more than 3000 attack edges, the detection accuracy of SybilBelief is less than 0.5, worse than a random guess, and SybilLimit and SybilInfer predict all Sybil nodes to be benign, thus losing the detection capability. Thus, SybilBelief, SybilLimit and SybilInfer do not work on weak trust networks with a large number of attack edges. 3) The performance of SybilFrame is stable and near optimal in all cases. By incorporating edge prior information, SybilFrame is able to restrict the amount of message passing across the attack edges. Thus, SybilFrame is able to successfully handle the situation when the number of attack edges is large, and performs orders of magnitudes better than other approaches.

\section{Evaluation on Real-World Large-Scale Twitter Network}
\label{sec:twitter}

In this section, we will evaluate SybilFrame on real-world large-scale Twitter network comprising over $20M$ nodes and $256M$ edges. We will explore ways to compute prior information, and incorporate it to SybilFrame.

\subsection{Collecting Twitter Dataset}
\label{subsec:twitter_analysis}

We obtained a snapshot of the Twitter follower network which was crawled by Kwak et al.~\cite{Kwak10}.  

\myparatight{Pre-processing} Originally, the Twitter network is \textbf{directed}. Since it is easy for attackers to manipulate one-way directed edges, 
we transform this  directed network to an \textbf{undirected} one by retaining an undirected edge between $u$ and $v$ if both directed edges $(u,v)$ and $(v,u)$ exist. Furthermore, we select the largest connected component of the transformed network since all investigated algorithms require the networks to be connected. The largest connected component contains 21,297,772 nodes, and 265,025,545 edges, with average degree 24.9.

We note that some previous works remove nodes with degrees smaller than a threshold from the social networks. For instance,  SybilLimit~\cite{Yu08} removes nodes with degree smaller than 5 and SybilInfer~\cite{Danezis09} removes nodes with degree smaller than 3. Mohaisen et al.~\cite{mohaisen:imc10} found that such pre-processing will prune a large portion of nodes.  Indeed,  social networks often have a long-tail degree distribution (e.g., power-law degree distribution~\cite{Clauset09} and lognormal degree distribution~\cite{Gong12-imc}), in which most nodes have very small degrees. Thus, a large portion of nodes are pruned by such pre-processing. Such pre-processing could result in high FPR or high FNR depending on how the OSN operator treats the pruned nodes.
If the OSN operator treats all the pruned nodes whose degrees are smaller than a threshold as benign nodes, then an attacker can create many malicious nodes with degree smaller than the threshold, resulting in high FNR, otherwise a large fraction of benign nodes will be treated as malicious nodes, resulting in high FPR. Therefore, we do not perform such pre-processing to the Twitter network.

\myparatight{Collecting ground truth} To evaluate the approaches, we need ground truth for the nodes in the Twitter network. The collected Twitter network includes users' Twitter IDs. Therefore, we re-crawled every account using Twitter's API, which tells us the status (i.e., active, suspended or deleted) of each account. 
In summary, we found that 145,156 nodes (i.e., 0.7\%) are suspended, 1,911,482 nodes (i.e., 9.0\%) are deleted, and the rest of the nodes are still active. We take the suspended accounts as Sybil nodes and the active ones as benign nodes.

\subsection{Measuring Twitter Structure}
We find that: 1) Many Sybil nodes are isolated from other Sybils. 2) The number of attack edges is very large. This means that using existing structure-based Sybil detection approaches will achieve limited performance.

\myparatight{No community structure} We adopt \emph{modularity}~\cite{Newman04}, ranging from -0.5 to 1, to quantify if a partition of a network (i.e., the partition in our case consists of the benign and Sybil regions in the Twitter network) can be viewed as two communities. Clauset et al.~\cite{Clauset04} concluded, via a large amount of empirical experiments on real networks, modularity $>$ 0.3 indicates significant community structure. However, we find that the partition consisting of the benign and malicious regions only has modularity 0.0042. 
Thus, the benign and Sybil regions cannot be viewed as two separate communities. Next, we show two reasons: half of the Sybil nodes are isolated and the number of attack edges per Sybil node is high. 
\begin{figure}[!htb]
\centering
{\includegraphics[width=0.4\textwidth]{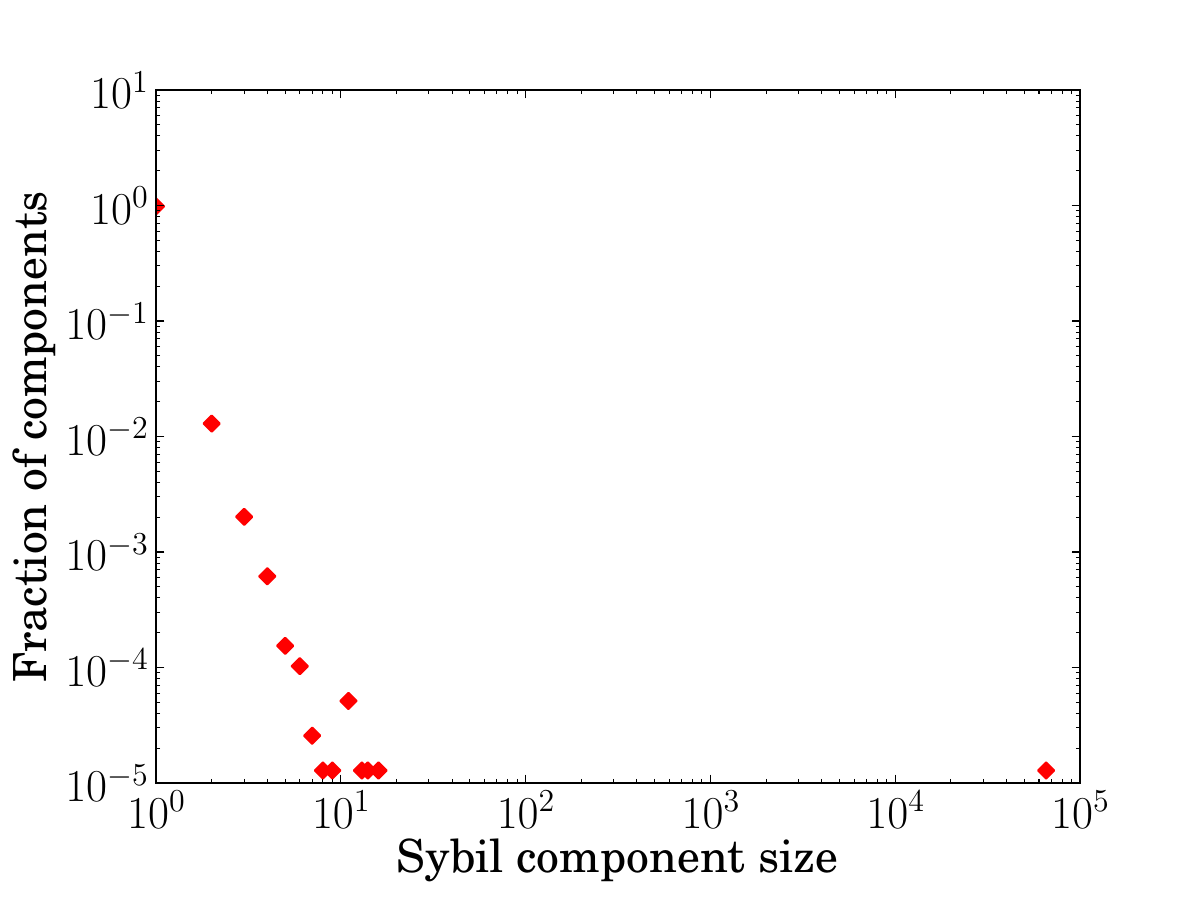}}
\caption{Distribution of connected Sybil component sizes
}
\label{component-size}
\vspace{-0.5cm}
\end{figure}

\begin{figure}
\begin{subfigure}[H]{0.24\textwidth}
  \includegraphics[width=\linewidth]{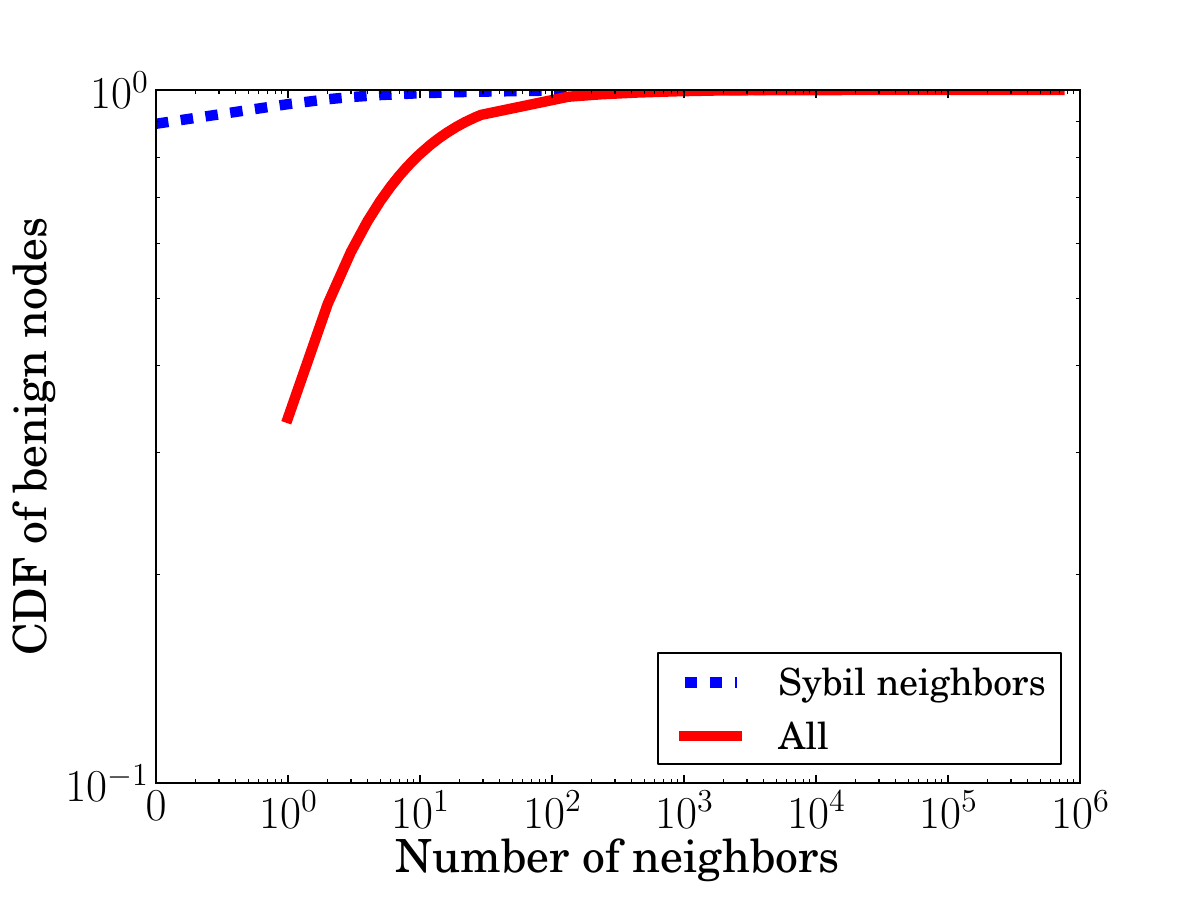}
  \caption{Benign nodes}
  \label{benign-deg}
\end{subfigure}%
\begin{subfigure}[H]{0.24\textwidth}
  \includegraphics[width=\linewidth]{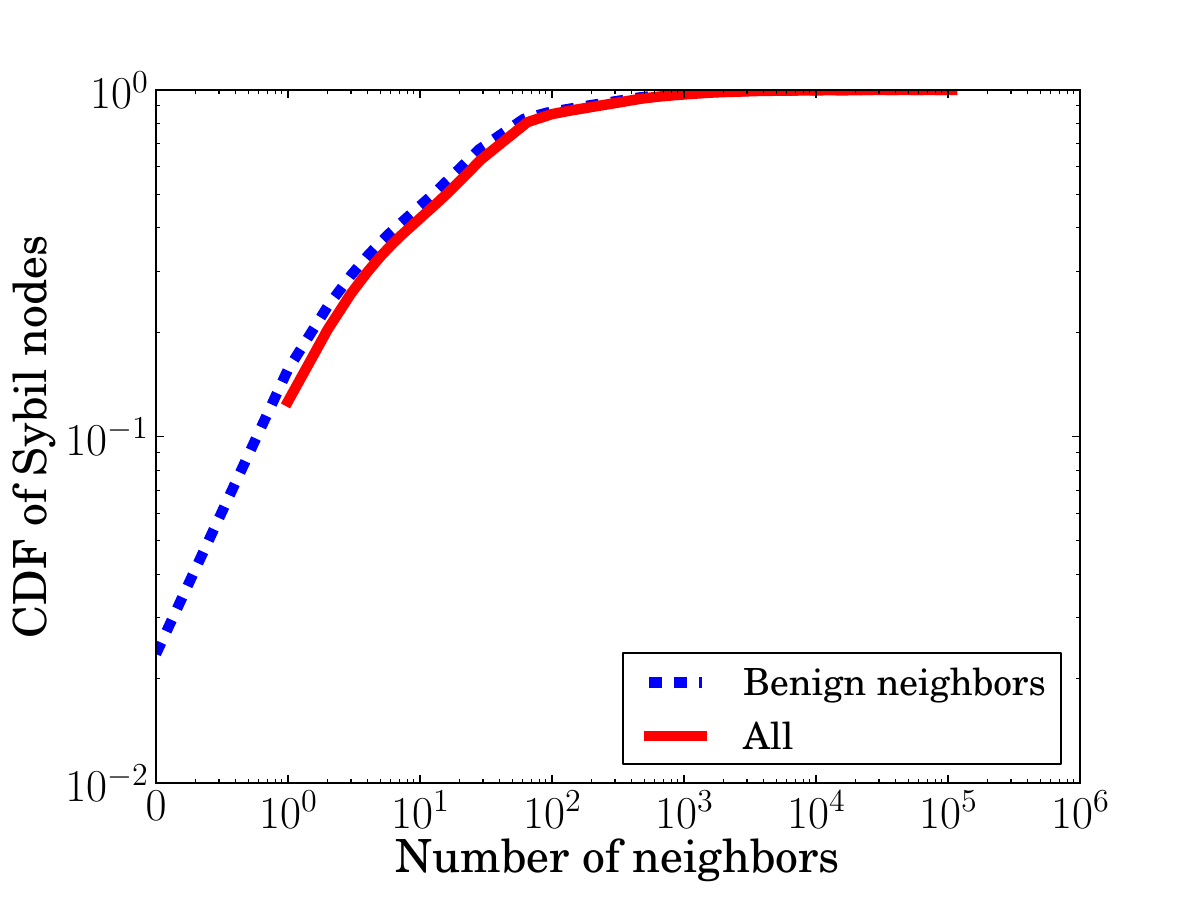}
  \caption{Sybil nodes}
  \label{sybil-deg}
\end{subfigure}%
\caption{CDFs for a) benign and b) Sybil nodes
}
\label{deg-dis}
\vspace{-0.5cm}
\end{figure}

\textbf{1) Half of the Sybil accounts are isolated:} In total, we find 77,917 connected components in the Sybil region (i.e., the subgraph including all malicious nodes and edges between them). Figure~\ref{component-size} shows the distribution of sizes of these components. First, around 50\% of Sybil nodes are \emph{isolated}, i.e., they only link to benign nodes. Second, we find that there exists a large connected component including 45\% of all malicious nodes. Specifically, this component consists of 65,579 nodes and 931,287 edges, resulting in an average degree of 28.40. Thus, the large component is even denser than the benign region whose average degree is 21.62.
We might wonder if this large connected component can be viewed as a community. However,  we find that the modularity of the partition consisting of the benign region and the largest connected component is still only 0.0046, which means that even this large connected component cannot be viewed as a community.  %
Third, the rest of nodes are in connected components whose sizes are less than 20. 

\textbf{2) Large number of attack edges:} We observe that there are 18,414,469 attack edges, which means each Sybil node successfully attacks around 127 benign nodes on average. Figure~\ref{deg-dis} further characterizes how attack edges are distributed among the benign and malicious regions. We can draw several conclusions from this figure.

First, the benign and Sybil regions are structurally similar. Specifically, the number of all neighbors of both benign and Sybil nodes follow long-tail distributions. In fact, such long-tail distributions are also widely observed in other OSNs such as LiveJournal~\cite{Mislove07} and Google+~\cite{Gong12-imc}. We speculate that Sybil nodes are imitating the benign region %
to evade automatic detection.

Second, from Figure~\ref{benign-deg}, we find that around 90\% of benign nodes are not connected to any Sybil node. Moreover, the number of Sybil neighbors of benign nodes also follows a long-tail distribution. This implies that, although around 10\% of benign nodes link to malicious nodes, most attack edges concentrate on a smaller number of benign nodes. For instance, we find that 90\% of attack edges concentrate on only 3\% of benign nodes. We speculate that such nodes are celebrities that tend to follow back to any user who follows them. 

Third, from Figure~\ref{sybil-deg}, we observe that around 2\% of Sybil nodes do not link to any benign node. Again, the number of benign neighbors of Sybil nodes follows a long-tail distribution, which implies that most attack edges are produced by a small portion of Sybil nodes. For instance, we find that 90\% of attack edges are produced by only 16\% of Sybil nodes.  

\begin{figure}[!t]
\centering
{\includegraphics[width=0.35\textwidth]{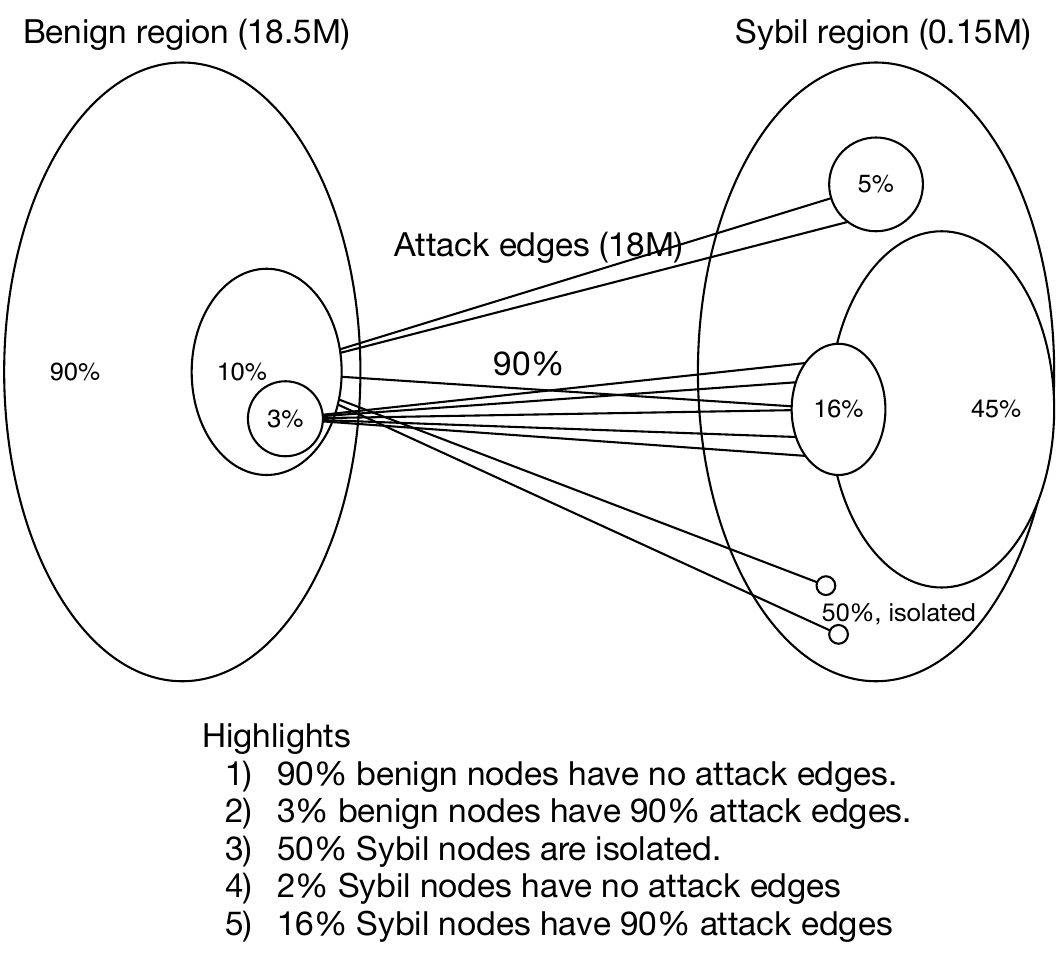}}
\caption{ 
Structure of Twitter network
}
\label{twitter_structure}
\vspace{-0.6cm}
\end{figure}

Note that the structural properties (i.e., many Sybil nodes are isolated and there are a large number of attack edges per Sybil node) of the Sybil nodes in our Twitter dataset match those in another large-scale Twitter network~\cite{Ghosh12} and those in the RenRen social network~\cite{Yang11-sybil}, which indicates the representativeness of our observations. Figure~\ref{twitter_structure} gives a snapshot of the structure.

\myparatight{Summary} We observe that the reason why structure-based Sybil detection approaches fail is that the assumptions they require are not satisfied. Specifically,  the benign and Sybil regions cannot be viewed as two separate communities. One reason is that a significant portion of the Sybil nodes are isolated, and the other reason is that the number of attack edges per Sybil node is high. 

\subsection{Computing Node Priors}
\label{subsec:twitter_nodeprior}

We now discuss ways to compute node priors. The idea is to collect features and train a classifier that outputs probabilistic scores. Since we do not know whether deleted accounts are benign or Sybil, we will not include them in the training, prediction and evaluation process. We just set the priors for them to be 0.5. 

\subsubsection{Collecting node features}
\label{subsubsec:node_feature}

We compute the following three features. We compute Feature 1) and 2) for all nodes on the original directed network, and map to the corresponding nodes on the undirected largest connected component. We directly compute Feature 3) on the undirected topology.

\textbf{1) Incoming requests accepted ratio:} The insight is that a Sybil identity is more likely to accept incoming requests than benign users, in order to quickly propagate spam. Hence on average, Sybil identities shall have a higher incoming requests accepted ratio. Since we only have \textbf{structural} information, we decide to use the incoming and outgoing edges associated with a node to \textbf{model} the ratio. For a node $v$ on the original directed Twitter graph, we denote $\Gamma_{in}(v)$ as the set of all incoming edges of $v$, and denote $\Gamma_{out}(v)$ as the set of all outgoing edges of $v$. Hence $\Gamma_{in}(v) \cap \Gamma_{out}(v)$ is the set of edges that are both incoming and outgoing edges of $v$. The \emph{incoming requests accepted ratio} is modeled as 
\begin{equation}
Req_{in} = \frac{\left| \Gamma_{in}(v) \cap \Gamma_{out}(v) \right|}{\left| \Gamma_{in}(v)\right|}
\end{equation}
\noindent where $\left| \Gamma(v)\right|$ denotes the cardinality of the set $\Gamma(v)$. 

\textbf{2) Outgoing requests accepted ratio:} The insight is that a benign user is more reliable and hence the outgoing friend requests send from him/her are more likely to be accepted. Hence on average, benign users have a higher outgoing requests accepted ratio than Sybil identities. Similarly, we \textbf{model} the \emph{outgoing requests accepted ratio} for a node $v$ as
\begin{equation}
Req_{out} = \frac{\left| \Gamma_{in}(v) \cap \Gamma_{out}(v) \right|}{\left| \Gamma_{out}(v)\right|}
\end{equation}
\noindent where $\left| \Gamma(v)\right|$ denotes the cardinality of the set $\Gamma(v)$. 

\textbf{3) Clustering coefficient:} The clustering coefficient for a vertex is a graph metric that measures how close its neighbors are to being a complete graph. For a node $v$ on the undirected graph $G = (V, E)$, its local \emph{clustering coefficient}~\cite{Gong12-imc} is given as.
\begin{equation}
C = \frac{2\left| \{  (i, j): i, j\in V, (i, j)\in E\} \right|}{k_v(k_v - 1)}
\end{equation}
\noindent where $k_v$ is the degree of $v$, $i$ and $j$ are both friends of $v$. The insight is that benign users tend to have well-connected social cliques, and users in such cliques share some attributes in common and are likely to be friends themselves. Therefore, benign users are likely to have a higher clustering coefficient than Sybil identities. 
\begin{figure}
\begin{subfigure}[H]{0.24\textwidth}
  \includegraphics[width=\linewidth]{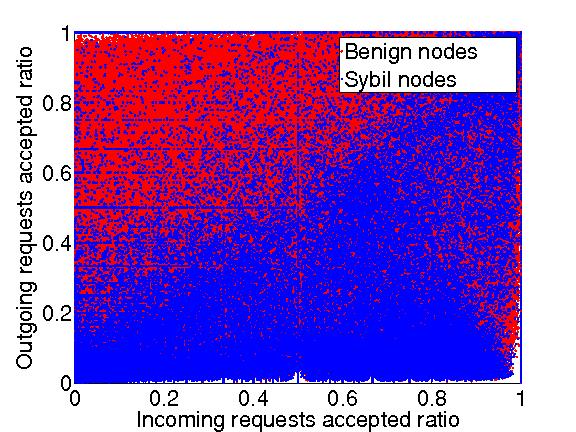}
\caption{Scatter plot}
  \label{fig:twitter_node_feature_scatter}
\end{subfigure}%
\begin{subfigure}[H]{0.24\textwidth}
  \includegraphics[width=\linewidth]{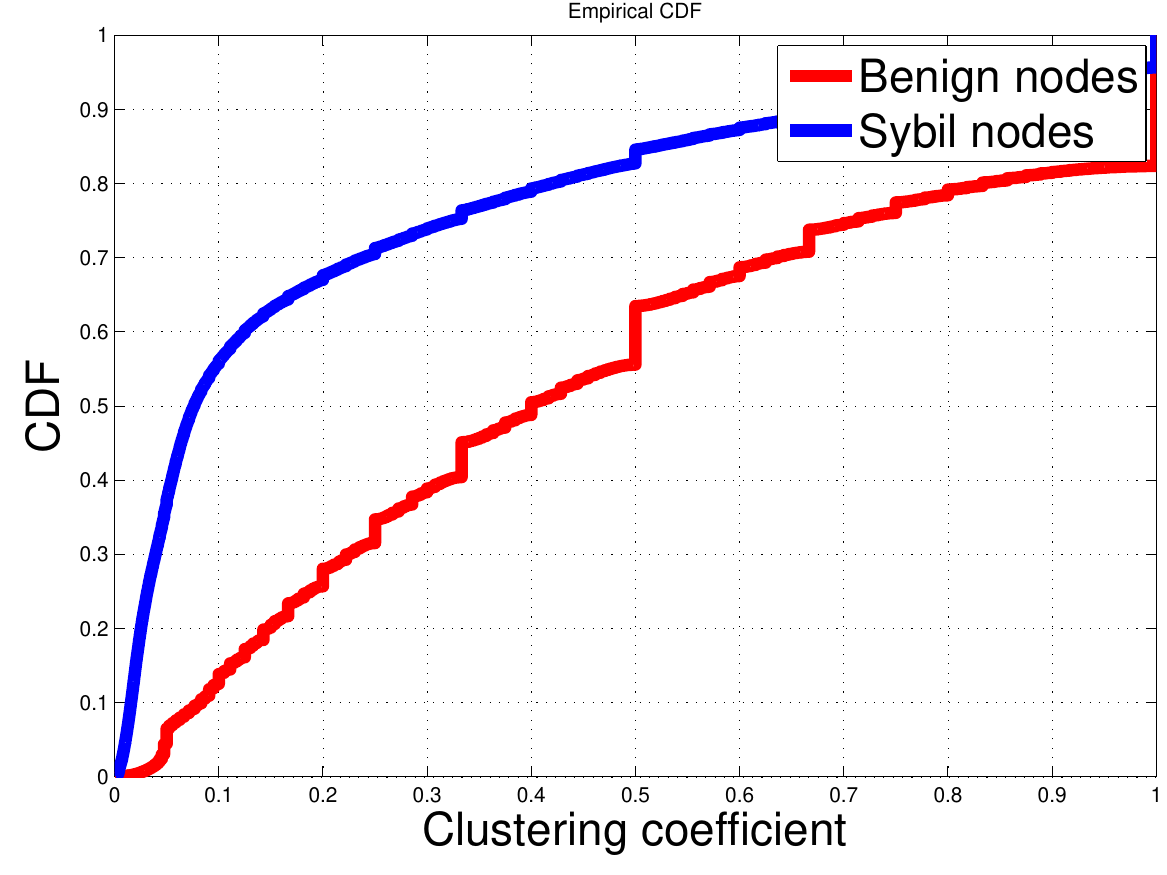}
\caption{CDF}
  \label{fig:twitter_node_feature_cc}
\end{subfigure}%
\caption{Analysis of the node features.}
\label{fig:twitter_node_feature}
\vspace{-0.5cm}
\end{figure}

Figure~\ref{fig:twitter_node_feature} shows the scatter plot of the outgoing requests accepted ratio versus incoming requests accepted ratio, as well as the CDF plot for the clustering coefficient for benign nodes and Sybil nodes. As expected, benign users tend to have a higher outgoing requests accepted ratio, a lower incoming requests accepted ratio, and a higher coefficient. Besides, only using any one of the three features is not able to obtain a clear decision boundary for the classification. Therefore, we need to leverage these three features together by building a machine learning classifier.

\subsubsection{Training a SVM classifier}
\label{subsubsec:svm}

We adopt the \emph{LIBSVM}~\cite{Chang2011} tool to build a Support Vector Machine (SVM)~\cite{hastie01statisticallearning} classifier. We sample a training set comprising 10,000 benign nodes and 10,000 Sybil nodes, and use the remaining nodes as testing. We train a SVM classifier with RBF kernel, whose parameters $c$ and $\gamma$ are obtained from \emph{Grid Search}. The overall prediction accuracy is 90.5\%, with 9.4\% FPR and 31.8\% FNR. Considering the fact that half of Sybil nodes are isolated, it is essentially hard for previous approaches
to detect more than half of total Sybil nodes. Thus, the 68.2\% Sybil nodes detection capability of SybilFrame is impressive. 

Some applications may require a lower FPR. A natural way is to assign a higher penalty term to the benign class and a lower penalty term to Sybil class. In this way, we can reduce the FPR of our node classifier to be 8.5\% with 41.6\% FNR.

\subsubsection{Output node priors}
\label{subsubsec:node_outputs}

To output priors, LIBSVM has an internal scheme to allow for probability estimates by fitting a logistic curve and conducting a cross validation procedure~\cite{Chang2011}. We can use the same parameters obtained from grid search, and enable the probability outputs. These output scores are then used as node priors for SybilFrame. 
\subsection{Computing Edge Priors}
\label{subsec:twitter_edgeprior}

We now explore ways to compute edge priors. As discussed in Section~\ref{subsec:facebook_prior}, we can leverage well-known similarity metrics. For each edge $(u, v)\in E$ on graph $G = (V, E)$, we compute the following similarity metrics: 	

\textbf{Number of Common Neighbors}~\cite{Liben-Nowell2007} $S_{uv} = \left| \Gamma(u) \cap \Gamma(v) \right|$

\textbf{Cosine Similarity Index}~\cite{Lu2011} $S_{uv} = \frac{\left| \Gamma(u)\cap \Gamma(v)\right|}{\sqrt{k_uk_v}}$

\textbf{Jaccard Similarity Index}~\cite{Liben-Nowell2007} $S_{uv} = \frac{\left|\Gamma(u)\cap\Gamma(v)\right|}{\left|\Gamma(u)\cup\Gamma(v)\right|}$

\textbf{Adamic-Adar Similarity Index}~\cite{Liben-Nowell2007} $S_{uv} = \sum_{s\in\Gamma(u)\cap\Gamma(v)}\frac{1}{k_s}$. 

Following a similar procedure, we scale the features and train a RBF-SVM classifier. As a result, we can successfully detect 18\% attack edges, with FPR 10\%. To improve the performance, we may include more complex similarity metrics, i.e. Katz Index~\cite{Lu2011} and Leicht-Holme-Newman Index~\cite{Lu2011}, which may cost a longer time to compute. Another way is to use node priors to infer edge priors. Generally, for an edge whose end nodes have different predicted labels, we can assign a lower score to indicate a higher possibility to be an attack edge; otherwise, we set the score to be default 0.9 to model homophily. Since our node priors work much better than edge priors, we adopt this inference procedure. 

\subsection{Scalable Implementation}
\label{subsec:bp_parallel}

We adopt the \emph{GraphLab} parallel framework~\cite{graphlab} to implement Loopy Belief Propagation in parallel. The parallel framework distributes nodes to multiple processors, and each processor passes and updates messages for the nodes that are assigned to. Essentially, computing node priors and edge priors is also parallelizable.

\subsection{Results}
\label{subsec:results}

We now present our experimental results. We compare SybilFrame with SybilBelief in terms of detection accuracy, FPR and FNR. We compare with SybilBelief and SybilRank in terms of relative ranking of Sybil nodes. If Sybil nodes tend to rank before benign nodes, OSN operators can leverage crowdsourcing (i.e. Amazon Mechanical Turks~\cite{Wang13}) to manually screen and label suspicious accounts. Since SybilLimit and SybilInfer do not scale to large datasets, we do not compare with them.

\myparatight{Detection results}
\label{subsubsec:induce_results}
We randomly select 1000 benign and Sybil seeds, and run SybilBelief and SybilFrame. Table~\ref{tab:detection_results} shows the results. We draw several conclusions: 1) Due to large number of attack edges, SybilBelief predicts all nodes to be Sybil thus completely losing detection capabilities. (We validated the implementation and results with the authors of SybilBelief.) 2) Node prior classifier of SybilFrame has certain detection power, which is able to detect 68.2\% Sybil nodes at maximum. By assigning different penalty terms, the FPR can be reduced to 8.5\% (Node classifier - II). 3) Incorporating node priors into SybilFrame can reduce FPR to 4.2\%. (SybilFrame - II), and achieve a better accuracy 95.4\%. 4) Since we label our ground truth based on whether the accounts are suspended by Twitter, it is possible that Twitter fails to detect some Sybil accounts, which are labeled by SybilFrame as positive examples. Thus, the true FPR should be \emph{lower} than our estimates. 5) Considering that half of Sybil nodes are isolated, the detection capability of 68.2\% Sybil accounts is impressive.
\begin{table}[!h]
\caption{Detection Performance on Twitter}\label{tab:detection_results}
\centering
\begin{tabular}{|l|c|c|c|}
\hline
~				&Accuracy		&FPR	&FNR\\\hline
SybilBelief			&0.7\%		&99.3\%	&0.00\\\hline
Node classifier		&90.5\%		&9.4\%	&31.8\%\\\hline
SybilFrame		&91.8\%		&8.0\%	&33.5\%\\\hline
Node classifier - II	&91.2\%		&8.5\%	&41.6\%\\\hline
SybilFrame - II		& 95.4\%		&4.2\%	&48.9\%\\\hline
\end{tabular}
\vspace{-0.2cm}
\end{table}

\myparatight{Ranking results}
\label{subsubsec:rank_results}
We rank the posteriors score generated from SybilFrame, as well as scores of SybilBelief and SybilRank, in ascending order. We then compute the portion of Sybil identities in the 1K, 10K, 50K, 100K, 1M and 10M lowest-ranked users. 
\begin{figure}[!htp]
\centering
\includegraphics[width=0.4\textwidth]{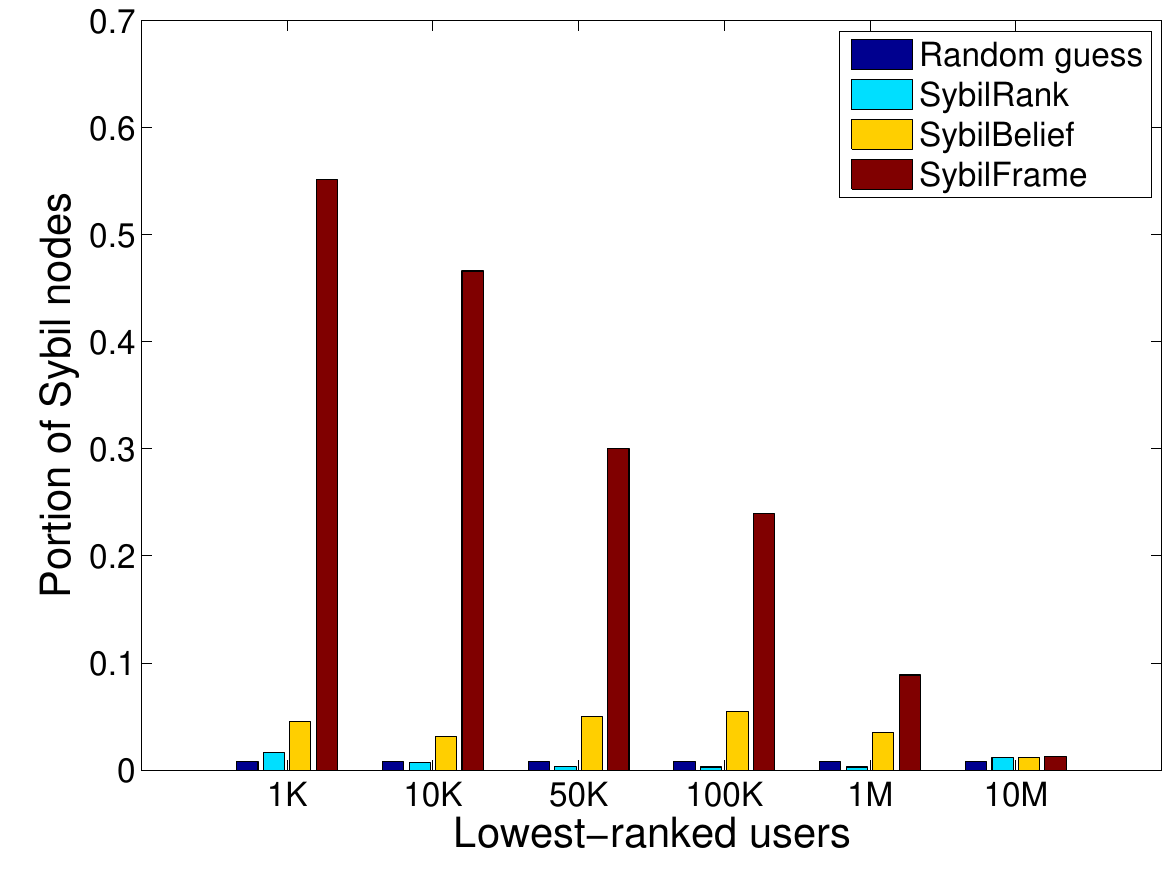}
\caption{Portion of Sybil identities} 
\label{fig:twitter_sybil_portion}
\vspace{-0.4cm}
\end{figure}
Figure~\ref{fig:twitter_sybil_portion} shows the results of four schemes: random guess, SybilRank, SybilBelief and SybilFrame. We draw several conclusions: 1) In the first 1K users, SybilFrame is able to rank over 500 Sybil accounts, 12 times better than SybilBelief, 35 times better than SybilRank and 72 times better than random guess. 2) There exists a significant descending trend of portions in SybilFrame, while SybilRank and SybilBelief do not have such trend. This means that SybilFrame has much more power to rank Sybil nodes in the top part of the ranking list, while SybilRank and SybilBelief roughly distribute the Sybil accounts evenly. 3) Given the same amount of time and human resource, OSN operators can use SybilFrame to detect more Sybil nodes than using SybilRank or SybilBelief.

\myparatight{Problem with Twitter's detection policy} 
\label{subsubsec:top_100}
Recall that we obtained our ground truth based on whether the account was active or suspended by Twitter. Thus, it is possible that some accounts are actually Sybil but evade Twitter's detection policy. To test this, first we re-crawl the top 1K accounts, and find that 7 additional accounts have been suspended by Twitter since our first crawl. Next, we manually examine the top 100 accounts, of which 71 are suspended and 29 are active. We examine the profile of each of the 29 active accounts, and find that only 3 accounts are likely to be real, with a long timeline and diverse tweets. Besides, 24 accounts are highly likely to be fake, with common characteristics such as same account images and few tweets. Furthermore, most of their tweets are published around 7/5/2009, and they all contain URLs and are about making money. 
Thus, we suspect that these 24 accounts were created by attackers and belong to the Sybil category.
The remaining 2 accounts have less than two tweets and a protected profile, which are marked as suspicious. We give a complete list of these 29 active accounts in Appendix~\ref{subsec:29_active}. 

From the above analysis, we conclude that: 1) Twitter's Sybil detection policy is not optimal. 2) SybilFrame is able to uncover a large fraction (24/29) of suspicious accounts that Twitter fails to detect. Hence, the true FPR of SybilFrame should be lower than our estimates.

\subsection{Summary}
\label{subsec:twitter_summary}

In this section, we discussed ways to compute priors and implemented SybilFrame in parallel. We evaluated SybilFrame on real-world, large-scale Twitter network, and have following observations:

\textbf{1)} In terms of detection performance, SybilFrame performs orders of magnitudes better than SybilBelief. Even when the dataset is noisy and the number of attack edges is large, SybilFrame can detect 68.2\% Sybil nodes at maximum. By tuning parameters, SybilFrame can achieve 4.2\% FPR with 51\% detection rate.

\textbf{2)} In terms of ranking performance, SybilFrame performs orders of magnitudes better than SybilBelief and SybilRank. Among the first 1K users, SybilFrame is able to successfully rank 552 Sybil accounts, 12 times better than SybilBelief and 35 times better than SybilRank.

\textbf{3)} SybilFrame is able to uncover large fraction of suspicious accounts that Twitter fails to detect.

\section{Discussion}
\label{sec:discussion}

\myparatight{Defense-in-depth} We discuss the resilience of our approach to attackers that aim to mimic the features we use in \emph{Stage 1}.
Specifically for Twitter experiment, if the attacker wants to mimic the features and let more Sybils bypass the node classifier, he/she needs to control Sybil identities to establish more connections between themselves and form Sybil clusters, in order to have a lower $Req_{in}$, a higher $Req_{out}$ and a higher clustering coefficient. As a result, Sybils will be much more densely connected, and the edge classifier in \emph{Stage 1} together with LBP in \emph{Stage 2} will be more effective to detect them. This is the basic idea of SybilFrame's multi-layered protection and defense-in-depth. 
Also, it remains to be discussed whether the attacker wants to spend time in performing such a complex strategy, which consumes both a lot of time and resource. In our Twitter experiment, we found that Sybil identities were often less intelligent (e.g. half of them are isolated and they share common characteristics as discussed in Section~\ref{subsubsec:top_100}), which makes it very easy for a human expert to identify them. However, Twitters fails to detect a significant fraction of Sybils but SybilFrame is able to uncover them.

Furthermore, we recall that although we collected structural features only to evaluate SybilFrame, SybilFrame is an open framework that is able to incorporate content information. Similarly, we can extract content features of each node and edge, and build a content-based classifier, or even combine structural and content features together to build a more powerful general classifier.

\myparatight{Lower FPR} Our experiment considers suspended accounts in Twitter as a ground truth for Sybil identities. Correspondingly, accounts that were not suspended were labeled as benign accounts. We note that this evaluation is conservative: our analysis considers accounts that are labeled as malicious by SybilFrame, but not suspended by Twitter as false positives. It is very well possible that these labeled false positives are actually malicious, however not detected by Twitter. As experimented in Section~\ref{subsubsec:top_100}, Twitter's current detection policy is far from optimal, and there is a large fraction of malicious accounts that Twitter fails to suspend. Therefore, our $4.2\%$ FPR should be essentially lowered. 

\myparatight{Sybil detection capability in Twitter} Recall that SybilFrame was able to detect 66.5\% Sybil identities (Table~\ref{tab:detection_results}, SybilFrame). By tuning parameters, SybilFrame was able to reduce FPR to 4.2\% while still detecting 51\% Sybil identities (Table~\ref{tab:detection_results}, SybilFrame - II). We believe that this result is close to the optimal that any structure-based approach could achieve. On the Twitter graph, half of the Sybil identities form a connected component, and the remaining half are isolated and only connect to benign nodes. Since previous structure-based approaches are mostly based on detecting local communities, they are limited in their ability to detect those isolated Sybil nodes.

\myparatight{Resilience against social churn}
\cite{temporal-sybil} proposed that existing Sybil defenses such as SybilInfer~\cite{Danezis09} and SybilRank~\cite{sybilrank} are vulnerable to the churn in social graphs, in which the attacker gradually moves the attack edges closer to the trusted seeds. The success of these temporal attacks requires that the locaiton of the trusted seeds is known to the attacker and remains steady for a certain period of time. Unlike other Sybil defenses, SybilFrame does not propagate from trusted seeds. Instead, SybilFuse computes local scores via local classifiers and propagate these scores for all nodes. This mechanism is able to mitigate such temporal attacks since the attacker does not have a direction for moving the attack edges gradually. Nevertheless, we recommend that the system operator frequently change the trusted seeds and rerun the propagation.

\myparatight{Broader applicability}  Our approach of defense-in-depth, and using a multi-stage classification framework that is able to incorporate prior information about nodes and edges has broad applicability for network security. For example, the area of botnet detection can benefit from similar techniques that combine host-level information with network structure-based information.

\section{Conclusion}
\label{sec:conclusion}

In this paper, we proposed SybilFrame, a defense-in-depth framework, for structure-based Sybil detection in online social systems. SybilFrame uses a multi-stage classification mechanism, which is able to incorporate heterogeneous sources and types of information about the social network. By leveraging the fine grained local information about users and edges, SybilFrame transforms local information into beliefs of labels, and then propagate those beliefs to make collective inferences.

We experimentally evaluated the accuracy of our approach using both synthetic and real-world social network topologies. We evaluated SybilFrame on a large-scale Twitter dataset. Our results demonstrate that SybilFrame is resilient to high number of attack edges, and performs an order of magnitude better than previous structure-based approaches.

Future work includes collecting and incorporating local content information, and enforcing more fine grained control on the belief propagation rules.

\bibliographystyle{IEEEtran}
\bibliography{refs}

\section{Appendix}
\label{sec:appendix}

\subsection{Prior Generator}
\label{subsec:prior_generator}

\myparatight{Node prior generator} Algorithm~\ref{alg:node_prior} gives our \emph{Node Prior Generator} for experiments on synthetic graphs in Section~\ref{subsec:node_prior}. We randomly generate node priors based on FPR and FNR.

\begin{algorithm}
\DontPrintSemicolon
\KwData{nodes with true labels, set of trust seeds, FPR and FNR}
\KwResult{priors of all nodes}
\For{each node}{
	\lIf{$v$ is a benign trust seed}{ set $Prior_v = 0.9$}
	\lElseIf{$v$ is a Sybil trust seed} { set $Prior_v = 0.1$ }
	\uElse{
		\eIf{the true label of $v$ is benign}{
			$i = randDouble(0, 1)$ \;
			\eIf{$i < FPR$} {
				$Prior_v = randDouble(0.1, 0.5)$ \;
			}{
				$Prior_v = randDouble(0.5, 0.9)$ \;
			}
		}{
			$i = randDouble(0, 1)$ \;
			\eIf{$i < FNR$} {
				$Prior_v = randDouble(0.5, 0.9)$ \;
			}{
				$Prior_v = randDouble(0.1, 0.5)$ \;
			}
		}
	}
}
\caption{Node Prior Generator}\label{alg:node_prior}
\end{algorithm}

\myparatight{Edge prior generator}
Algorithm~\ref{alg:edge_prior} gives our \emph{Edge Prior Generator} for experiments on synthetic graphs in Section~\ref{subsec:edge_prior}. We randomly generate edge priors based on FPR and FNR.
\begin{algorithm}
\DontPrintSemicolon
\KwData{nodes with true labels, set of trust seeds, FPR and FNR}
\KwResult{priors of all edges}
\For{each edge $(u, v)$}{
	\eIf{both $u$ and $v$ are trust seeds}{
		\eIf{$u$ and $v$ have different labels}{
			Set $Prior_{u, v} = 0.1$\;
		}{
			Set $Prior_{u, v} = 0.9$\;
		}
	}{
		\eIf{$u$ and $v$ have different labels} {
			$i = randDouble(0, 1)$ \;
			\eIf{$i < FNR$} {
				$Prior_v = randDouble(0.5, 0.9)$ \;
			}{
				$Prior_v = randDouble(0.1, 0.5)$ \;
			}
		}{
			$i = randDouble(0, 1)$ \;
			\eIf{$i < FPR$} {
				$Prior_v = randDouble(0.1, 0.5)$ \;
			}{
				$Prior_v = randDouble(0.5, 0.9)$ \;
			}
		}
	}
}
\caption{Edge Prior Generator}\label{alg:edge_prior}
\end{algorithm}

\subsection{Experiments on the Influence of Edge Priors}
\label{syn_edge_fpr0.1fpr}
\begin{figure}[!htb]
\begin{subfigure}[H]{0.24\textwidth}
  \includegraphics[width=\linewidth]{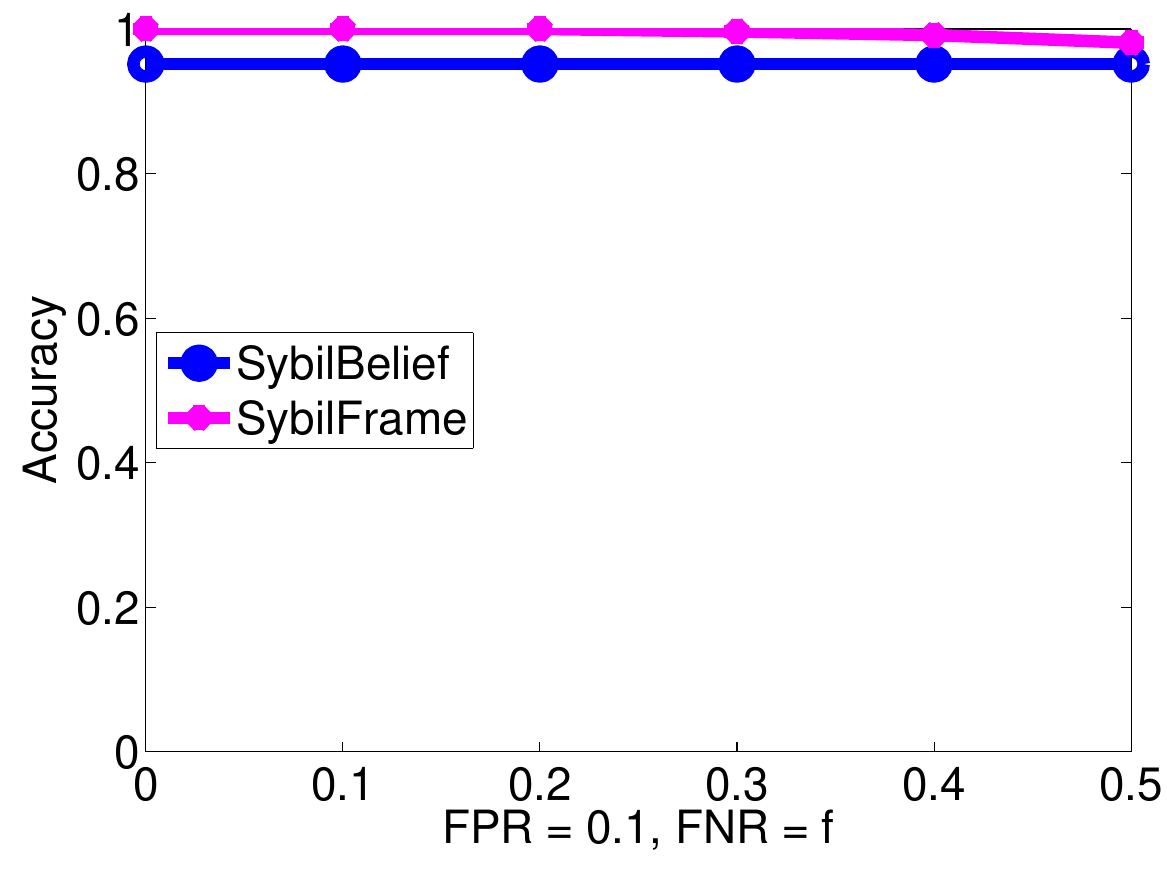}
  \caption{Accuracy}
  \label{fig:edge_prior_accuracy_1_2}
\end{subfigure}%
\begin{subfigure}[H]{0.24\textwidth}
  \includegraphics[width=\linewidth]{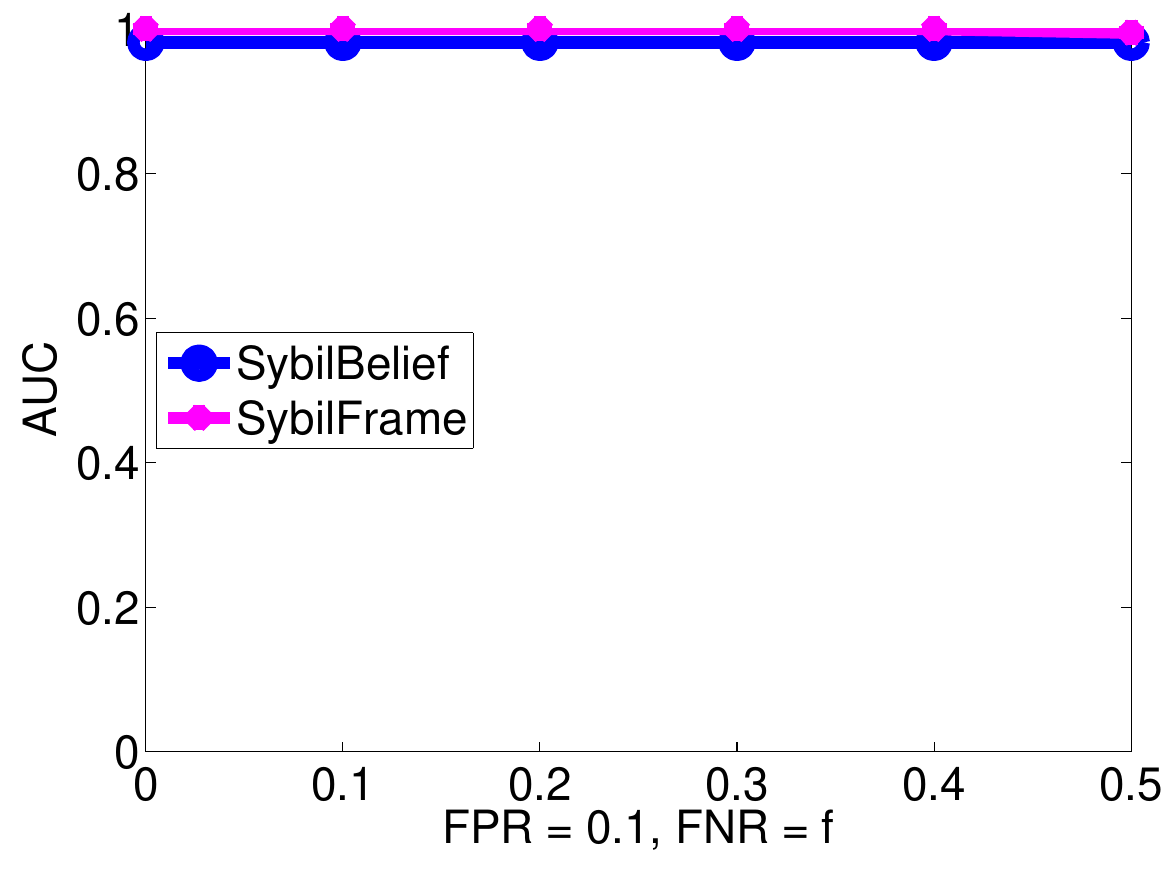}
  \caption{AUC}
  \label{fig:edge_prior_auc_1_2}
\end{subfigure}%

\begin{subfigure}[H]{0.24\textwidth}
  \includegraphics[width=\linewidth]{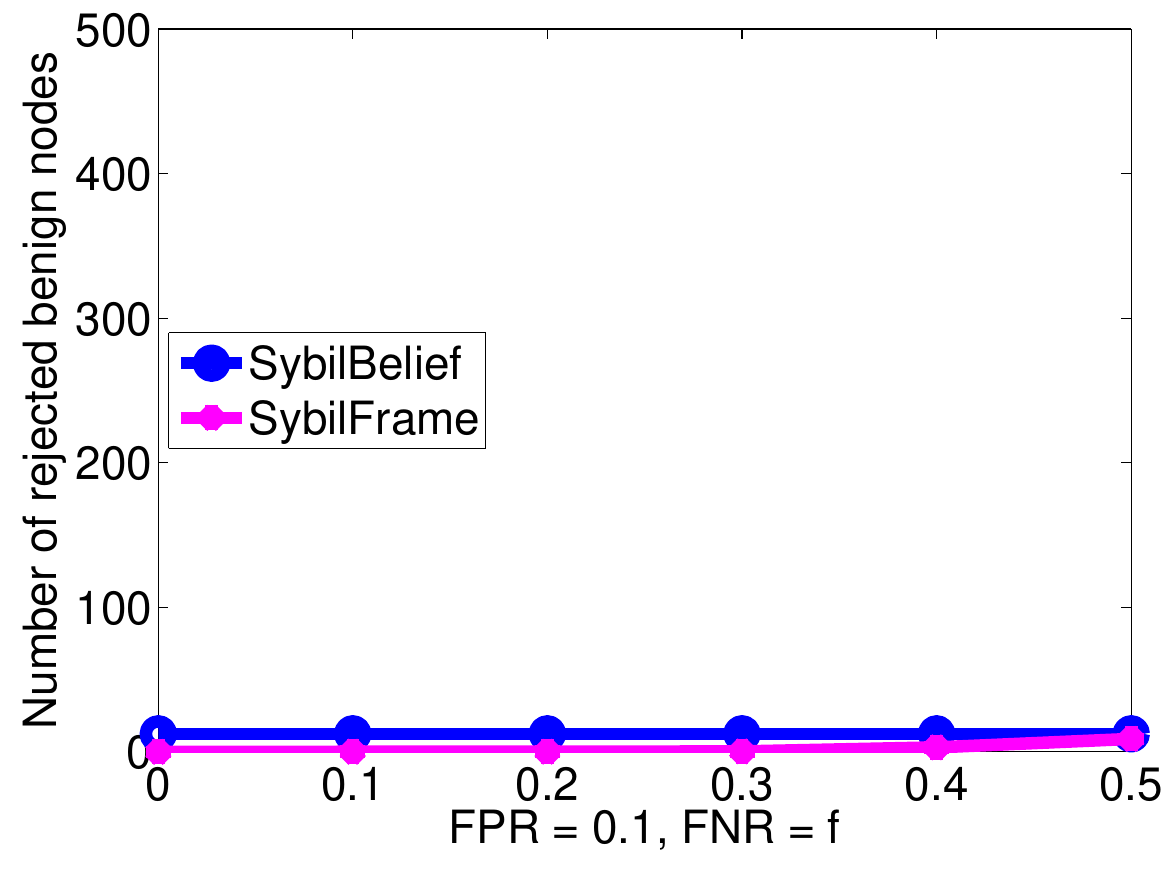}
  \caption{Rejected benign nodes}
  \label{fig:edge_prior_benign_rej_1_2}
\end{subfigure}%
\begin{subfigure}[H]{0.24\textwidth}
  \includegraphics[width=\linewidth]{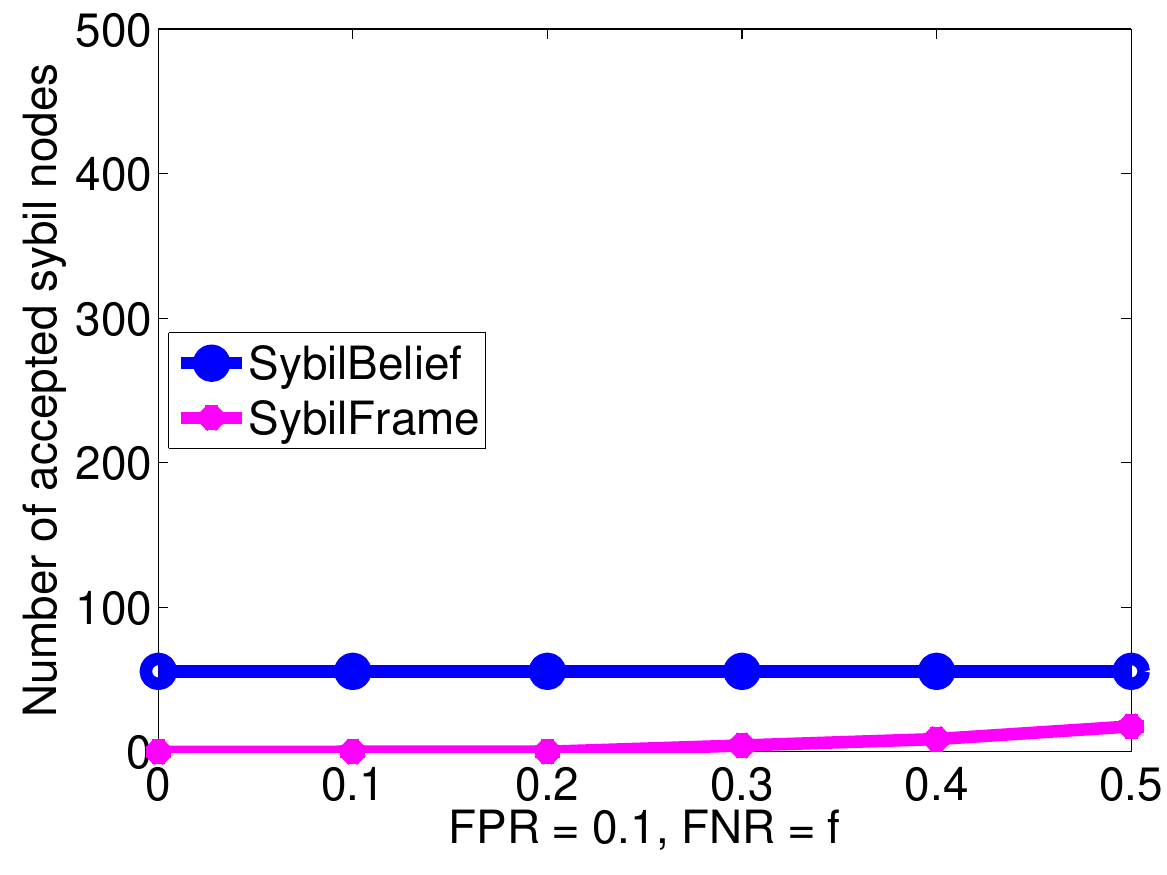}
  \caption{Accepted Sybil nodes}
  \label{fig:edge_prior_sybil_acc_1_2}
\end{subfigure}%
\caption{
Set FPR=0.1 and vary FNR (edge prior)}
\label{fig:edge_prior_1_2}
\vspace{-0.6cm}
\end{figure}

As in Section~\ref{subsubsec:edge_fpr_fnr_f}, we \emph{set $FPR=0.1$ and tune $FNR$ from 0 to 0.5}, and fix everything else in the basic setup. Figure~\ref{fig:edge_prior_1_2} shows the results of SybilFrame and SybilBelief. We can see that SybilFrame has good performance and outperforms SybilBelief even when \emph{FNR} is 0.5.

\begin{figure}[!htb]
\begin{subfigure}[H]{0.24\textwidth}
  \includegraphics[width=\linewidth]{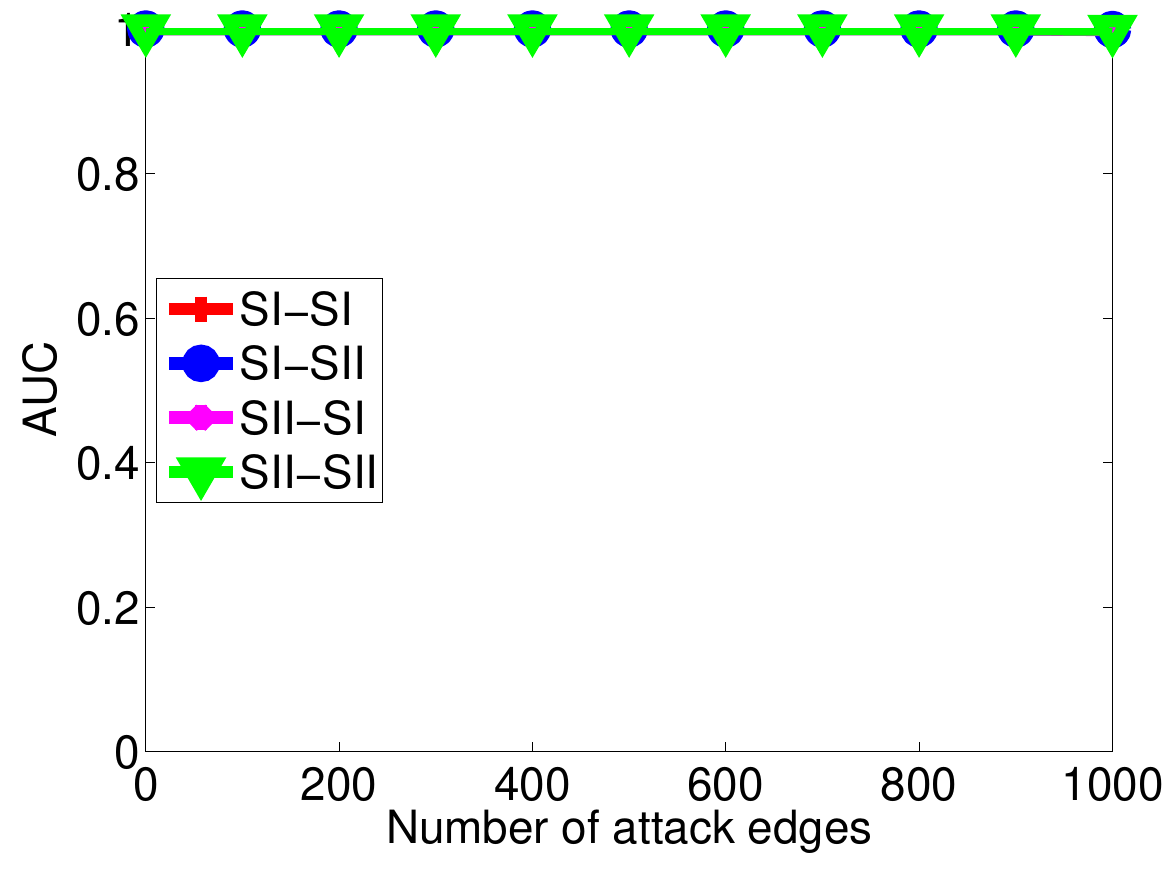}
  \caption{AUC (node prior)}
  \label{fig:targeted_node_auc}
\end{subfigure}%
\begin{subfigure}[H]{0.24\textwidth}
  \includegraphics[width=\linewidth]{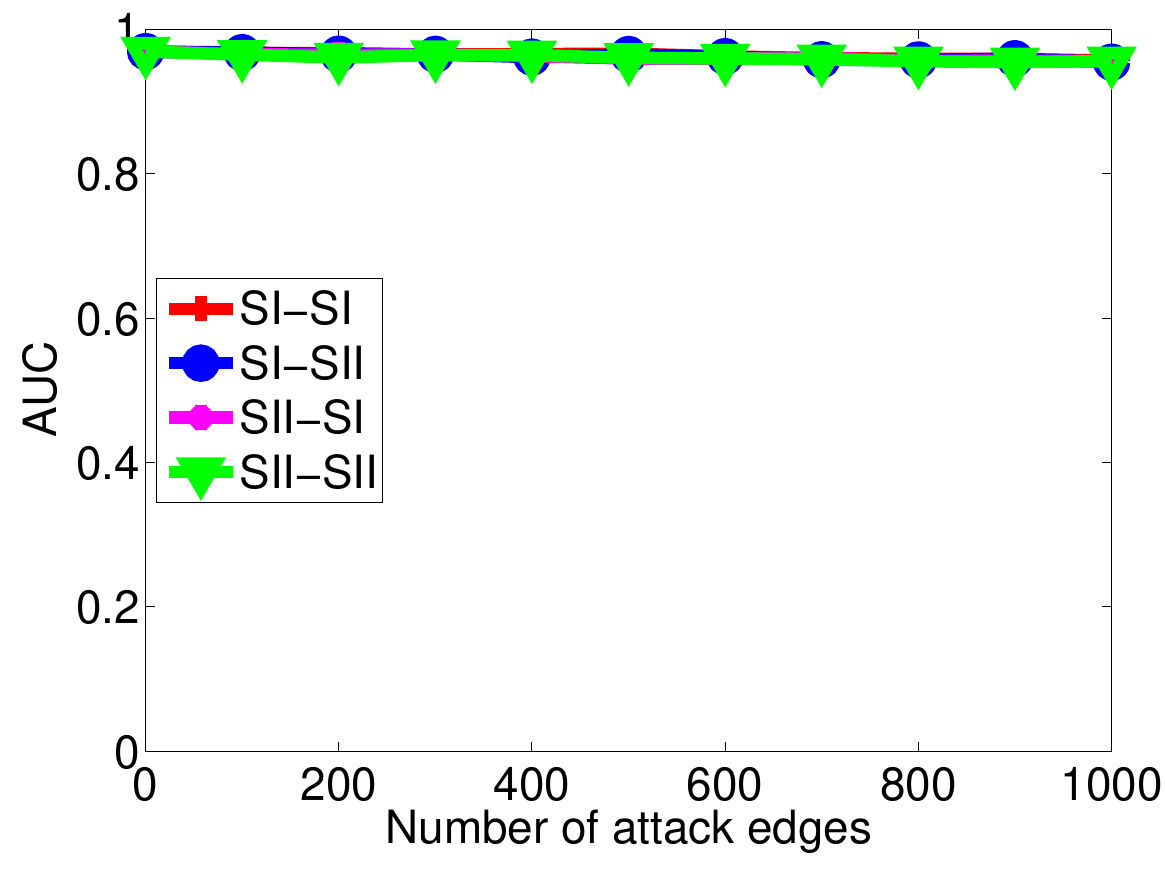}
  \caption{AUC (edge prior)}
  \label{fig:targeted_edge_auc}
\end{subfigure}%
\caption{
AUC of SybilFrame under seed targeting attacks. (a) Given node priors. (b) Given edge priors.}
\label{fig:targeted_auc}
\end{figure}

\begin{figure}[!htb]
\begin{subfigure}[H]{0.24\textwidth}
  \includegraphics[width=\linewidth]{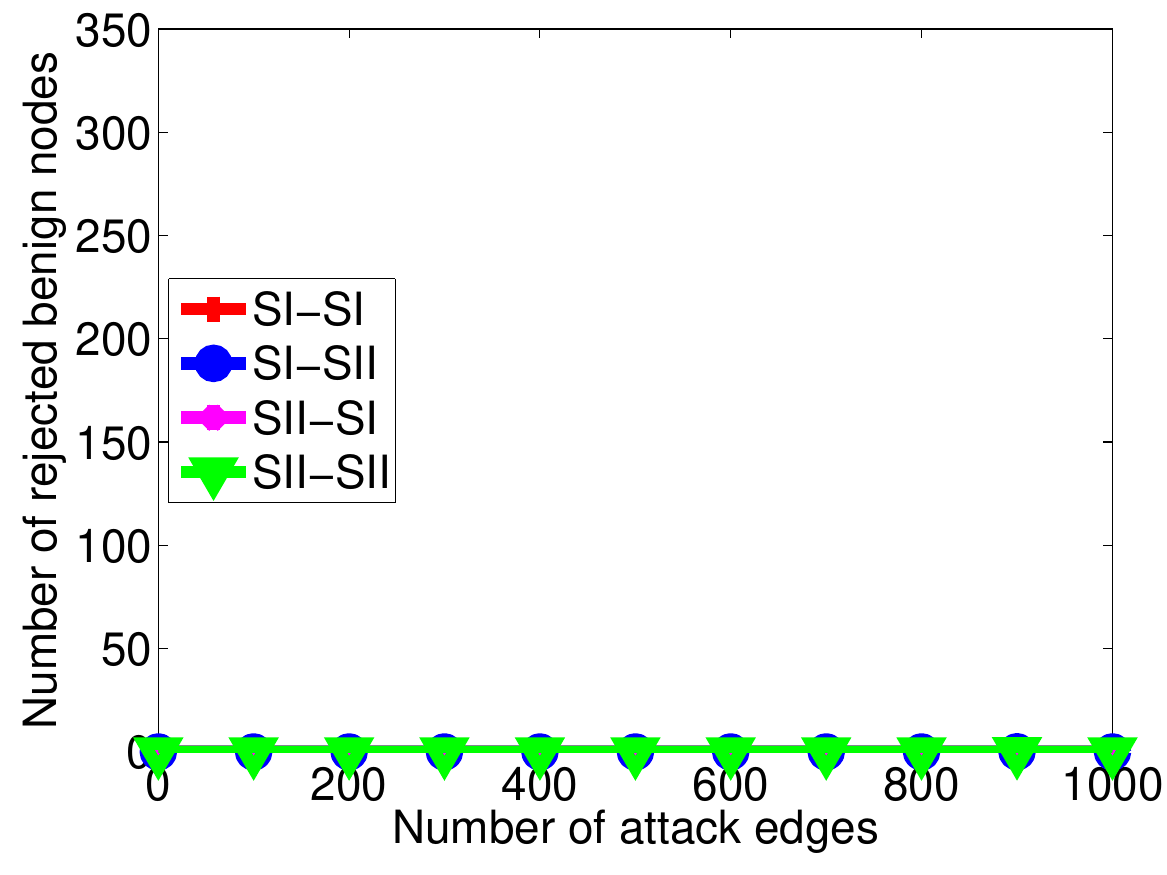}
  \caption{Rejected benign nodes (node prior)}
  \label{fig:targeted_node_fp}
\end{subfigure}%
\begin{subfigure}[H]{0.24\textwidth}
  \includegraphics[width=\linewidth]{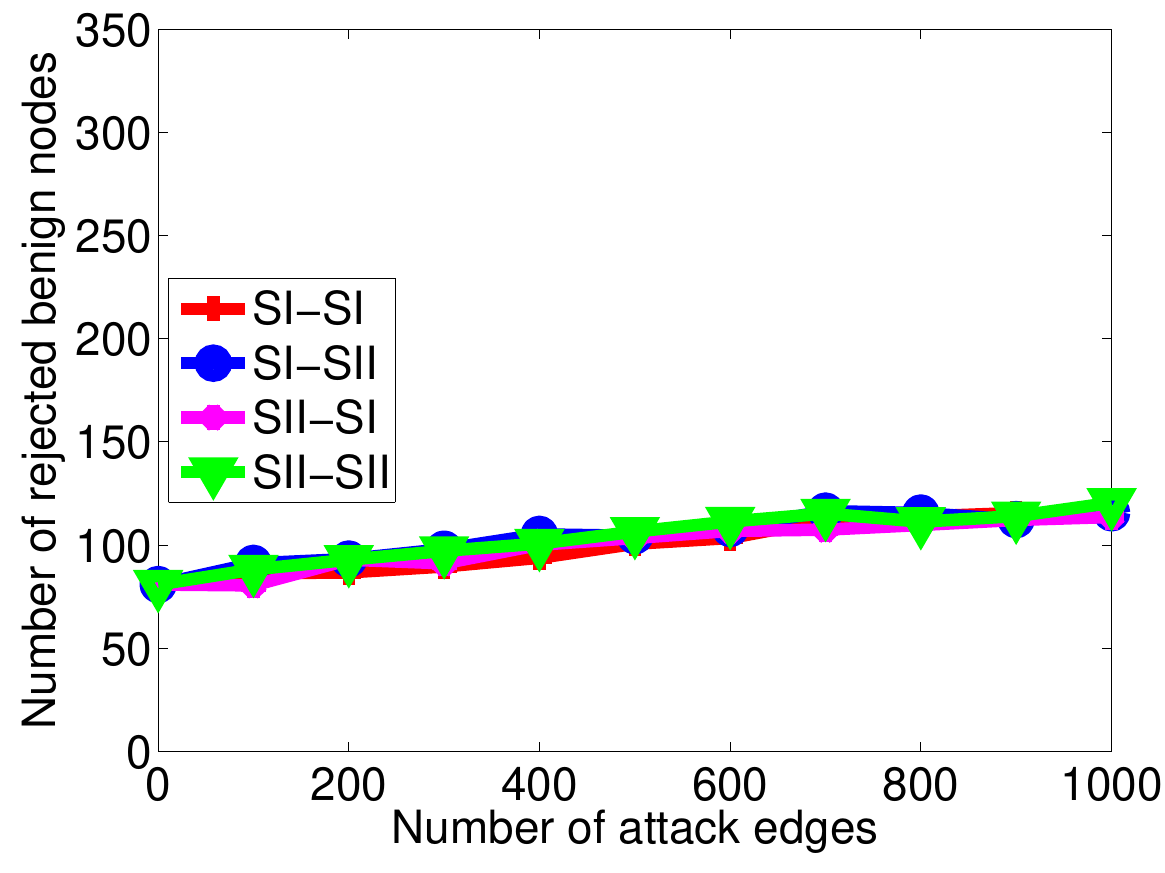}
  \caption{Rejected benign nodes  (edge prior)}
  \label{fig:targeted_edge_fp}
\end{subfigure}%

\begin{subfigure}[H]{0.24\textwidth}
  \includegraphics[width=\linewidth]{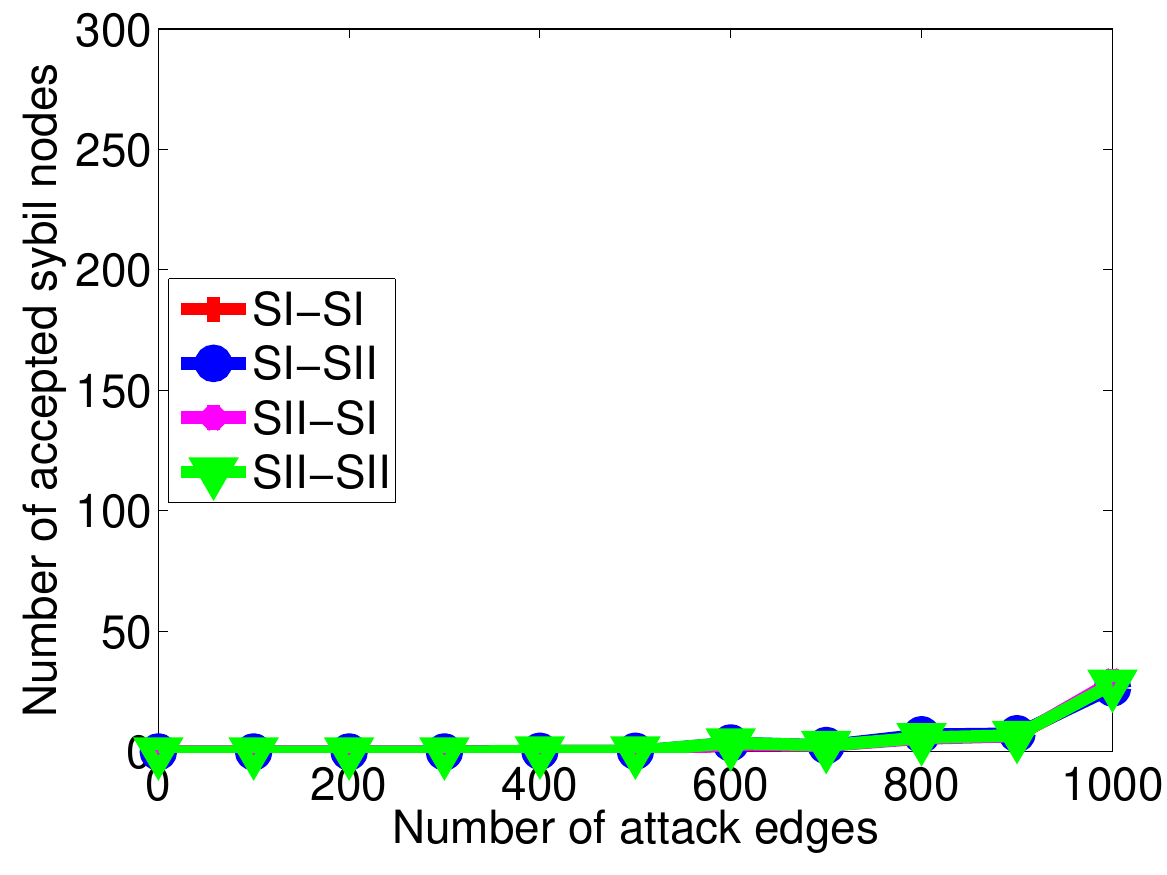}
  \caption{Accepted Sybil nodes (node prior)}
  \label{fig:targeted_node_fn}
\end{subfigure}%
\begin{subfigure}[H]{0.24\textwidth}
  \includegraphics[width=\linewidth]{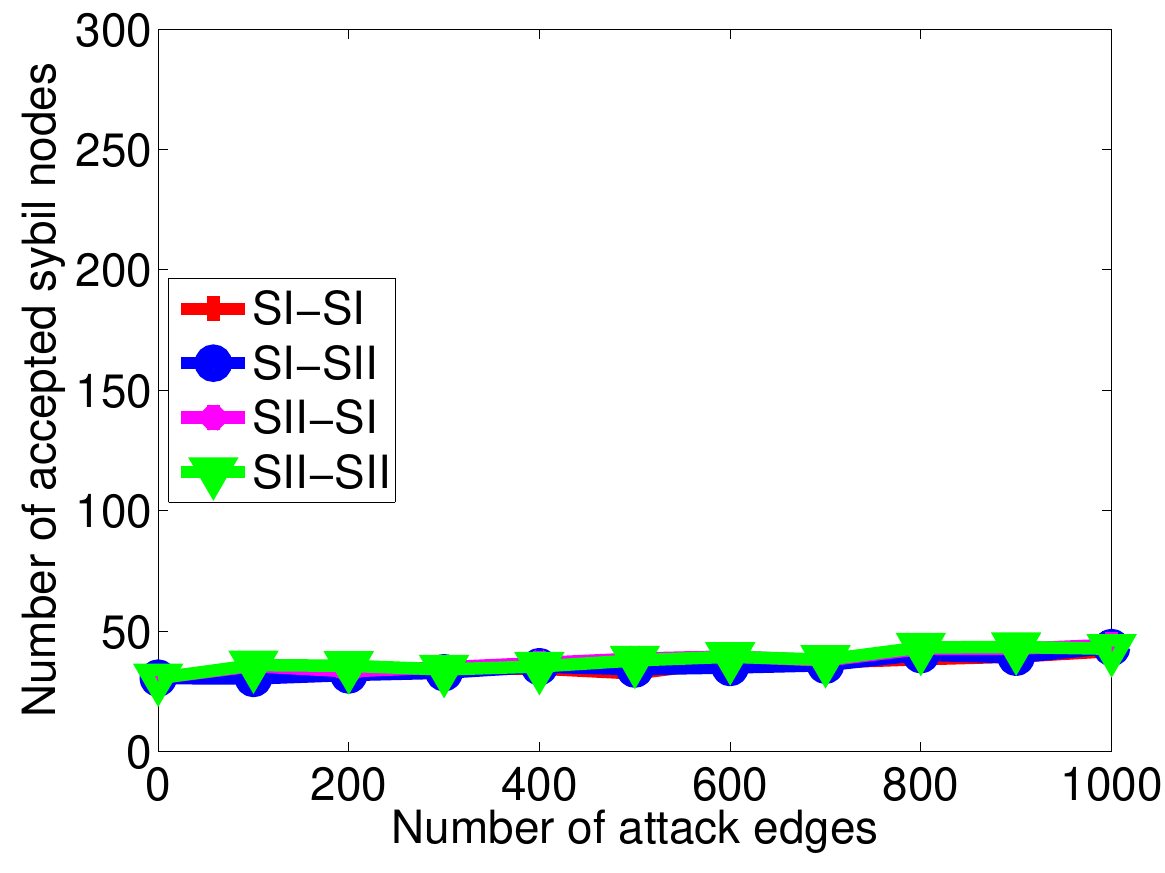}
  \caption{Accepted Sybil nodes (edge prior)}
  \label{fig:targeted_edge_fn}
\end{subfigure}%
\caption{
Performance of SybilFrame under seed targeting attacks.}
\label{fig:targeted_fpfn}
\end{figure}

\subsection {Experiments on Seed Targeting Attacks}
\label{subsec:seed_targeted_appendix}

As in Section~\ref{subsec:targeted_attack}, we evaluate SybilFrame under seed targeting attacks. Figure~\ref{fig:targeted_auc} shows the AUC as a function of the number of attack edges for four scenario combinations of trust seeds, in the node prior experiment (FPR=FNR=0.3) and edge prior experiment (FPR=0.1, FNR=0.5). Figure~\ref{fig:targeted_fpfn} shows results for the number of rejected benign nodes and the number of accepted Sybil nodes.

\subsection{Complete List of 29 Active Accounts}
\label{subsec:29_active}
Table~\ref{tab:29active} gives a complete list of 29 active accounts among the top 100 ranked accounts. For simplicity, we use pseudo names for account images; accounts sharing the same image name have the same account image. We have the following observations:

1) Among 29 active accounts, 3 are benign users, 2 are suspicious and 24 are Sybil identities.

2) Benign users tend to have a long timeline with diverse tweets.

3) Sybil identities have much more following users than their followers. Besides, Sybil identities have few tweets (e.g. mostly less than 5) and a short timeline (e.g. around 7/5/09). Most of their tweets are about making money and contain URLs. Furthermore, Sybil identities tend to share common account images.

\begin{sidewaystable*}[htbp]
  \centering
  \caption{Complete List of 29 Active Accounts among Top 100 Ranked Accounts by SybilFrame}
    \begin{tabular}{rrrrrrrrrr}
	\hline
    Ranking & Username & Tweets & Following & Followers & Active Time & Account Image & Topic & Most Tweets has URL & Identity \\
    \hline
    1     & boomBOXbul88741 & 3     & 914   & 81    & 7/4/09 - 7/8/09 & None  & Money & Y     & Sybil \\
    2     & LuvGirl162 & 6     & 619   & 76    & 7/28/09 - 7/28/09 & None  & Learn French & Y     & Sybil \\
    3     & Christina\_78661 & 3     & 554   & 41    & 7/4/09 - 7/8/09 & Img1  & Money & Y     & Sybil \\
    4     & candy3456 & 0     & 1     & 0     & 9/1/09 & None  & Money & None & Sybil \\
    5     & lynnodonne37181 & 3     & 650   & 69    & 7/4/09 - 7/8/09 & Img2  & Money & Y     & Sybil \\
    6     & sarasexmeu57808 & 1     & 520   & 42    & 7/4/09 & Img1  & ``Picture" & Y     & Sybil \\
    7     & sarasexmeu2375 & 1     & 535   & 33    & 7/4/09 & Img1  & ``Picture" & Y     & Sybil \\
    11    & Ceqlyq23 & 38    & 885   & 123   & Up to 8/23/13 & Customized & Politics & N     & Benign \\
    15    & subslavest14109 & 2     & 366   & 43    & 7/5/09 & Img3  & Money & Y     & Sybil \\
    19    & chrischamb46593 & 4     & 336   & 34    & 7/3/09 - 7/7/09 & Img4  & Money & Y     & Sybil \\
    24    & mismarygra16029 & 2     & 289   & 20    & 7/5/09 & Img1  & Money & Y     & Sybil \\
    32    & romyzz & 0     & 0     & 0     & 10/1/09 & None  & No tweets & None & Sybil \\
    33    & JudithDupo70244 & 2     & 365   & 41    & 7/5/09 & Img5  & Money & Y     & Sybil \\
    40    & Cheap\_cray18037 & 4     & 291   & 17    & 7/3/09 - 7/7/09 & Img1  & Money & Y     & Sybil \\
    43    & basketbras94491 & 2     & 362   & 38    & 7/5/09 & None  & Money & Y     & Sybil \\
    45    & kolektor\_blog & 0     & 399   & 1     & 6/1/09 & None  & None & Protected & Suspicious \\
    48    & MsLuvULong68914 & 2     & 315   & 25    & 7/5/09 & Img6  & Money & Y     & Sybil \\
    50    & SarahMulli25962 & 2     & 326   & 38    & 7/5/09 & Img7  & Money & Y     & Sybil \\
    52    & karataylor59102 & 4     & 329   & 39    & 7/3/09 - 7/7/09 & Img2  & Money & Y     & Sybil \\
    54    & PrincessLd76238 & 2     & 292   & 15    & 7/5/09 & Img1  & Money & Y     & Sybil \\
    57    & mismarygra69078 & 2     & 325   & 28    & 7/5/09 & Img5  & Money & Y     & Sybil \\
    59    & Mom3CuteBo44691 & 2     & 292   & 30    & 7/5/09 & Img8  & Money & Y     & Sybil \\
    61    & pri\_oftran39924 & 2     & 314   & 31    & 7/5/09 & Img7  & Money & Y     & Sybil \\
    65    & julianasar75283 & 3     & 579   & 66    & 7/4/09 - 7/8/09 & Img6  & Money & Y     & Sybil \\
    69    & DawnSellsS11353 & 2     & 281   & 12    & 7/5/09 & Img1  & Money & Y     & Sybil \\
    76    & blackmore138719 & 2     & 286   & 16    & 7/5/09 & Img1  & Money & Y     & Sybil \\
    80    & muppetleung & 1507  & 317   & 32    & Still active & Customized & Diverse & N     & Benign \\
    88    & singlessalsa & 35    & 417   & 752   & Up to 2/13/13 & Customized & Dance & Y     & Benign \\
    93    & sobradinhobar & 1     & 257   & 1     & 7/1/09 & Customized & Protected &  Protected & Suspicious \\
\hline
    \end{tabular}%
  \label{tab:29active}%
\end{sidewaystable*}%

\end{document}